\documentclass[11pt,a4paper]{article}
\usepackage{jheppub}
\usepackage{amsmath}
\usepackage{multicol,bbm}
\usepackage{enumerate}
\usepackage{bibentry}
\usepackage{epstopdf}

\graphicspath{{./Figs/}} 
\newcommand{\be}{\begin{eqnarray}}
\newcommand{\ee}{\end{eqnarray}}
\newcommand{\Z}{\mathcal{Z}}
\newcommand{\la}{\langle}
\newcommand{\ra}{\rangle}
\newcommand{\A}{\hat{A}}

\newcommand{\4}{\hat{4}}

\def\nl{\nonumber\\}

\def\ZZ{\mathcal{Z}}

\def\l{\langle}
\def\r{\rangle}

\title{The Amplituhedron from Momentum Twistor Diagrams}
\author[a]{Yuntao Bai,}
\author[b,c]{Song He}
\affiliation[a]{Department of Physics, Princeton University, Princeton, NJ 08544 } 
\affiliation[b]{School of Natural Sciences, Institute for Advanced
Study, Princeton, NJ 08540, USA} 
\affiliation[c]{Perimeter Institute for Theoretical Physics, Waterloo, ON N2L 2Y5,
Canada}
\emailAdd{ytbai@princeton.edu, songhe@ias.edu} 

\abstract{We propose a new diagrammatic formulation of the all-loop scattering amplitudes/Wilson loops in planar $\mathcal{N}=4$ SYM, dubbed the ``momentum-twistor diagrams". These are on-shell-diagrams obtained by gluing trivalent black and white vertices defined in momentum twistor space, which, in the reduced diagram case, are known to be related to diagrams in the original twistor space. The new diagrams are manifestly Yangian invariant, and they naturally represent factorization and forward-limit contributions in the all-loop BCFW recursion relations in momentum twistor space, in a fashion that is completely different from those in momentum space. We show how to construct and evaluate momentum-twistor diagrams, and how to use them to obtain tree-level amplitudes and loop-level integrands; in particular for the latter we identify an isolated bubble-structure for each loop variable, arising from a forward limit, or entangled removal of particles. From a given diagram one can directly read off the  $C$, $D$ matrices via a generalized ``boundary measurement"; this in turn determines a cell in the amplituhedron associated with the amplitude, and our diagrammatic representations of the amplitude can provide triangulations of the amplituhedron with generally very intricate geometries. To demonstrate the computational power of the formalism, we give explicit results for general two-loop integrands, and the cells of the complete amplituhedron for two-loop MHV amplitudes. }

\begin{document}
\maketitle 
\section{Introduction and Motivations}
One of the most fundamental objects in quantum field theory is the S-matrix. In the past decades, unexpected simplicities and rich structures have been discovered for scattering amplitudes in gauge theories and gravity, especially in planar $\mathcal{N}=4$ supersymmetric Yang-Mills theory (SYM). One notable example is the Grassmannian/on-shell-diagram program~\cite{Grassmannian, posGrassmannian}, which provides a reformulation for all-loop scattering amplitudes in planar $\mathcal{N}=4$ SYM, without referring to Feynman diagrams, Lagrangian or spacetime. The planar integrand consists of diagrams constructed by gluing together fundamental three-point on-shell amplitudes, which admit representations via Grassmannian integrals, and the way these diagrams are assembled together can be determined by BCFW-like recursion relations to all loop orders~\cite{all loop}. The on-shell diagrams are extremely interesting both from mathematical and physical points of view; in particular, through the recursion they combine to exhibit the correct behavior of the all-loop integrand, including factorizations at physical poles and forward-limit behavior at the so-called single cuts. 

Another remarkable property of scattering amplitudes in planar $\mathcal{N}=4$ SYM is that they are dual to null polygonal Wilson loops in a dual spacetime~\cite{aldaymalda,Drummond:2007aua,WL1,WLDrummond,Bern:2008ap,WLDrummond2,maldacenaberkovits}.
The duality was originally discovered for bosonic Wilson loops/MHV amplitudes, and later generalized to super Wilson loops/super-amplitudes (which contain all helicity-amplitudes)
\cite{masonskinner,simonWL}; they enjoy a hidden, dual superconformal symmetry~\cite{aldaymalda,magic}. The dual symmetry can be understood as the symmetry of the dual Wilson loops, which, together with the ordinary superconformal symmetry, generate an infinite-dimensional Yangian symmetry~\cite{Yangian}. An advantage of the Grassmannian/on-shell diagram formulation is that the original superconformal symmetry is made manifest; although the Grassmannian form can be rewritten in the dual space-time to make dual symmetry manifest~\cite{Mason:2009qx, dualGrassmannian, posGrassmannian}, an important goal yet to be fulfilled is to directly understand the dual conformal symmetry in a diagrammatic formulation in momentum-twistor space, similar to on-shell diagrams in the original space.

The most compact form of amplitudes/Wilson loops are given in momentum twistor variables, which are twistors of the dual space-time, introduced by Hodges~\cite{hodges}. These variables manifest the dual symmetry, and trivialize momentum conservation and massless on-shell conditions simutaneously. Very recently, a direct reformulation of amplitudes even without referring to recursion relations or on-shell diagrams, the ``amplituhedron", was proposed~\cite{amplituhedron}; in terms of the amplituhedron, the  planar integrand in momentum-twsitor space, at any loop order, is reformulated as forms in an auxiliary space. Despite significant progress, the geometry of the amplituhedron inside this space has not been fully understood, and there is rich mathematical structure to be explored. It is highly desirable to derive triangulations of the amplituhedron from all-loop BCFW representations, thus a systematic study of the recursion will be of great importance for understanding the structure of the amplituhedron.  

In this paper, we propose a new type of diagrams called ``momentum twistor diagrams", which as we will argue play a key role in the study of amplitudes/Wilson loops along all the aforementioned directions. These diagrams are manifestly Yangian invariant, and serve as the building blocks for the all-loop recursion relations, and thus for the amplituhedron, in momentum twistor space. Although the diagrams have formally the same ingredients as those in the original space, their meanings are completely different, and instead they exhibit the behavior of amplitudes/Wilson loops at singularities in momentum twistor space. Already at tree level it is interesting to see how the reduced diagrams combine into the tree amplitude, according to the factorization term of the recursion. More importantly, we will generalize the construction to all loops, which requires a systematic way of dealing with iterated forward limit terms of the recursion, and by doing so we find direct connections to the amplituhedron geometry. The new diagrams, as dictated by the all-loop recursion, become not only conceptually interesting as providing triangulations of the amplituhedron, but also practically powerful for explicit computations of multi-loop integrands. 

After a brief review of momentum twistor variables, all-loop recursion relations, and the amplituhedron, we present the definition of the new on-shell diagrams in momentum-twistor space in section \ref{diagram}. We study the dual Grassmannian formulation of the diagrams, which is parallel to that of the original on-shell diagrams; in addition, we give several examples, such as the R-invariant, and operations on the diagrams, including BCFW bridge, fusing, adding and removing particles. We proceed to representing general factorizations of amplitudes using momentum-twistor diagrams in section \ref{tree}, which is given by gluing two sub-amplitudes by the R-invariant; this completes our diagrammatic representation of all tree amplitudes, and we write down explicitly NMHV and N${}^2$MHV examples. In section \ref{loop}, we apply the diagrams to loop level, where one can see the full strength of the formalism. We find that the pair of particles in the forward limit can be represented by an isolated, bubble-like structure, and it is much more efficient for producing loop integrands than the original diagrams. After giving the Kermit representation for all one-loop amplitudes, by iterating the procedure of taking forward limits, we show how to obtain the diagrams for higher-loop integrands. To demonstrate this, we present the full two-loop integrand, for MHV and for general cases. The diagrams makes it possible to systematically determine cells of the amplituhedron, without actually understanding the intricate geometries, and in particular we obtain the cells for the amplituhedron of two-loop MHV.

\paragraph{Amplitudes in momentum-twistor space}

Let us begin with a brief review of amplitudes/Wilson loops in momentum-twistor space. Denote the $n$-point, N${}^k$MHV, $L$-loop amplitude as $\mathcal{A}_{n,k}^{(L)}$, and throughout the paper we will consider the amplitude, $A_{n,k}^{(L)}$, with MHV tree stripped off, 
\be
\mathcal{A}_{n,k}^{(L)}= \frac{\delta^{2\times (2|4)} (\sum_{i=1}^n \lambda^{\alpha}_i (\tilde\lambda^{\dot{\alpha}}_i |\eta^A_i) )}{\l 1 2\r \ldots\l n{-}1 n\r\l n 1\r} A_{n,k}^{(L)}\,, 
\ee
where $\alpha, \dot\alpha=1,2$ are SU(2) indices of spinors $\lambda_i$ and
their conjugates $\tilde\lambda_i$ encoding the null momenta of $n$
particles,  and $A=1,...,4$ is the SU(4) index of Grassmann variables
$\eta_i$ describing their helicity states. The Wilson loop dual to the $n$-point amplitude is formulated along a $n$-sided null polygon in a chiral
superspace with coordinates $(x,\theta)$; for $i=1,\ldots,n$, we have \be x_i^{\alpha
\dot\alpha}-x_{i{-}1}^{\alpha
\dot\alpha}=\lambda_i^{\alpha}\tilde\lambda_i^{\dot\alpha},\qquad
\theta_i^{\alpha A}-\theta_{i{-}1}^{\alpha
A}=\lambda_i^{\alpha}\eta_i^A\,, \ee  

The (super) momentum twistors are in the fundamental representation of the superconformal group of this dual space;  explicitly 
\be \Z_i=
(Z_i^a | \chi_i^A)=(\lambda_{i\alpha}, \mu_i^{\dot\alpha} |\chi_i^A)\equiv (\lambda_{i\alpha},x_i^{\alpha \dot\alpha}
\lambda_{i \alpha} | \theta_i^{\alpha A} \lambda_{i \alpha})\,.
\ee 
The momentum twistors are unconstrained and they determine $\tilde\lambda,\eta$ via,
\be
(\tilde\lambda |\eta)_i=\frac{\l i{-}1\,i\r (\mu|\chi)_{i{+}1}+\l i{+}1\,i{-}1\r (\mu|\chi)_{i}+\l i\,i{+}1\r (\mu|\chi)_{i{-}1}}{\l i{-}1\, i\r\l i\,i{+}1\r}
\ee
We further define the totally antisymmetric contraction of four bosonic twistors $\l i j k l\r\equiv \varepsilon_{abcd}Z^a_iZ^b_jZ^c_kZ^d_l$. The factorization poles $x^2_{i,j}=0$ , with $x_{i,j}\equiv x_i-x_j$, can be written in these variables as $\l i{-}1\,i\,j{-}1\,j\r=0$.  In addition, we have the basic R-invariant of five super-twistors, \be
[i,j,k,l,m]\equiv \frac{\delta^{0|4}(\l\l i\,j\,k\,l\,m\r\r)}{\l i j
k l\r \l j k l m\r\l k l m i\r\l l m i j \r\l m i j k\r},\ee where
the argument of Grassmann delta function is $\l\l
i\,j\,k\,l\,m\r\r^A\equiv \chi_i^A\l j k l m\r+\textrm{cyclic}$. 

The central object we will study in this paper is the {\it integrand} of amplitudes/Wilson loops in momentum twistor space. We will denote the integrand for $A_{n,k}^{(L)}$ as $Y_{n,k}^{(L)}$, which is a form of degree $4L$ in the $L$ loop variables denoted as $\ell$'s. Formally we have

\be
A^{(L)}_{n,k}=\int_{\rm reg} Y^{(L)}_{n,k}(\Z_1,\ldots,\Z_n; \{\ell_1,\ldots,\ell_L\})=\int_{\rm reg} \prod_{m=1}^L d^4 \ell_m\,I_{n,k}^{(L)} (\Z_1,\ldots,\Z_n; \{\ell_1,\ldots,\ell_L\})\,,\nl
\ee
where ``reg" means regularizations which are needed for the loop integrals, and by writing the integral measure explicitly, the remaining part of $Y^{(L)}_{n,k}$, as a rational function, is denoted as $I^{(L)}_{n,k}$. Note that both $Y$ and $I$ are cyclic in external twistors, $\Z_1,\ldots,\Z_n$, which will be denoted as $1,\ldots, n$ and completely symmetrized in loop variables, $\ell_1,\ldots,\ell_L$, which will be suppressed when possible.

The loop variables $\ell$'s correspond to points in dual space (for computing Wilson loops, they are positions of Lagrangian insertions, see~\cite{simonWL}). Accordingly, they are lines in momentum-twistor space, and we will always represent $\ell$'s by bi-twistors: $\ell_m\equiv (A_m B_m)\equiv(A B)_m$, for $m=1,\ldots,L$. The loop integral measure in momentum-twistor space is defined as
\be\label{loopmeas}
d^4 \ell  \equiv \l A B d^2 A\r\l A B d^2 B\r = \frac{d^4 Z_A d^4 Z_B}{{\rm vol}\; GL(2)}
\ee
where the factors of $\left<AB\right>$ always drop out because the integrand is dual conformal invariant, so we have neglected writing them in eq.~(\ref{loopmeas});  the integral over the line $(AB)$ is given by that over a pair of points (twistors) $Z_A$ and $Z_B$, divided by the GL$(2)$ redundancies labeling their positions on the line~\cite{all loop}. The integrand has, in addition to factorization poles, poles from a propagator involving loop variables going on shell, e.g. the so-called single cut corresponds to poles of the form $\l A_m B_m\, i{-}1\,i\r=0$, for the loop variable $\ell_m$. 

\paragraph{All-loop recursion relations} The integrand $Y_{n,k}^{(L)}$ can be determined by BCFW recursion~\cite{all loop}. By applying a shift of the form $\hat{\Z}_n=\Z_n+w \Z_{n-1}$, we obtain contributions from three different types of poles in $w$:
\be\label{recursion}
Y_{n,k}^{(L)}(1,\ldots, n)=\text{B}+\text{FAC}+\text{FL}
\ee
where for simplicity we have used indices $1,2,...,n$ to denote super-twistors $\mathcal{Z}_1,\mathcal{Z}_2,...,\mathcal{Z}_n$.

Here B represents the boundary contribution from $w\rightarrow \infty$, given by removing $\mathcal{Z}_n$.
\be
\text{B} = Y_{n-1,k}^{(L)}(1,\ldots,n-1)\,.
\ee
Note that although B originates from a different pole than those in FAC below, it can be regarded as a special factorization term, as is clear from the momentum-space point of view: it is the factorization into $Y_{n{-}1,k}$ and the three-point conjugate-MHV amplitude, ${\cal A}_{3,-1}^{(0)}$,  which as it stands does not have a momentum-twistor representation since no MHV tree can be stripped off from it (it would be something like a $Y_{3,-1}$ with  $k$-charge $-1$). 

The FAC term represents contributions from factorization poles:
\be
\text{FAC} =\frac {1}{L!} \sum_{i=3}^{n-2}\sum_{k',L'}\sum_{\sigma(\ell)} [i{-}1,i,n{-}1,n,1]Y_{i,k'}^{(L')}(1,\ldots,i{-}1,\hat{i}) Y_{n{+}2{-}i, k{-}1{-}k'}^{(L{-}L')}(\hat{i}, i,\ldots, n{-}1, \hat{n}_i)\,,\nl
\ee
where we sum over all the poles of the form $\l i{-}1\,i\,\hat{n}\,1\r=0$, and at each pole the internal leg and the shifted leg are given by $\hat{i}\equiv (i-1,i)\cap(1,n-1,n)$, $\hat{n}_i=(n-1,n)\cap (1,i-1,i)$, with $(a,b)\cap (c,d,e)\equiv Z_a\l b\, c\,d\,e\r-Z_b\l a\,c\,d\,e\r$ defined as the intersection of the line $(a,b)$ with the plane $(c,d,e)$; in addition, we sum over $k'=0,\ldots, k{-}1$, $L'=0,\ldots,L$, and over distributions of $\ell_1,\ldots,\ell_L$ into the two subsets, with $L'$ and $L{-}L'$ variables, which explains the overall symmetrization factor $1/L!$. 
 
The FL term represents the forward-limit contributions which come from single cuts:

\be
\text{FL} = 
\frac 1 L\,\sum_{m=1}^L \int \frac{d^{3|4}\ZZ_A d^{3|4}\ZZ_B}{{\rm vol\; GL(2)}}\int_{GL(2)} [A_m,B_m,n{-}1,n,1]Y_{n{+}2,k{+}1}^{(L{-}1)}(1,...,n{-}1,\hat{n}_{\ell_m},A_m, B_m)\,,\nl
\ee
where we sum over $L$ loop variables $\ell_m=(AB)_m$ (with a symmetrization factor $1/L$), and each term comes from the pole $\left<A_m B_m \hat{n} 1\right>=0$, with $\hat{n}_{\ell}= (n-1, n)\cap (A, B, 1)$. The $\int_{GL(2)}$ integral is defined as follows. We first set $Z_A\rightarrow Z_A+\alpha Z_B\equiv Z_A'$ and $Z_B\rightarrow Z_B+\beta Z_A\equiv Z_B'$ for parameters $\alpha,\beta$, which is equivalent to moving the two points $Z_A,Z_B$ without changing the line they span. Then, we take a double residue in $\alpha,\beta$ such that $\left<A',1,n{-}1,n\right>,\left<B',1,n{-} 1,n\right>\rightarrow 0$, which is equivalent to taking $A',B'$ to lie on the plane $(1,n{-}1,n)$. Formally, we have

\be
\int_{GL(2)} \equiv \int_{\left<A',1,n{-} 1,n\right>\rightarrow 0}d\alpha \;\int_{\left<B',1,n{-} 1,n\right>\rightarrow 0} d\beta \; (1-\alpha\beta)^2
\ee
This residue is equivalent to setting $Z_A',Z_B'\rightarrow (A,B)\cap(1,n{-}1,n)$. The $(1-\alpha\beta)^2$ is a determinant factor that makes the poles in $\alpha,\beta$ simple.\\

\paragraph{The amplituhedron} Here we briefly review the definition of the amplituhedron~\cite{amplituhedron}. At tree level, one can extract the super-amplitude from the ``volume", or the form, of the tree amplituhedron.  The kinematic data is given by $z^I_i=(Z^a_i, \phi_{1,A} \chi_i^A,\ldots,\phi_{k,A} \chi_i^A) $ for $i=1,\ldots, n$  and $I=1,\ldots,k{+}4$, which are bosonic variables associated with super momentum-twistors, and $\phi_1,\ldots,\phi_k$ are auxiliary Grassmann parameters. We restrict to $z\in M_+(k{+}4,n)$,  where the space $M_+(k{+}4,n)$ is defined as the set of $(k+4)\times n$ matrices with all ordered-minors positive: $ \l z_{i_1}\,\ldots z_{i_{k{+}4}}\r>0$ for $i_1<\cdots<i_{k{+}4}$. The tree amplituhedron, $A(n,k,0)$, is defined as a subspace of $G(k,k{+}4)$, determined by ``positive" linear combinations of the positive data, 
\be A(n,k,0)\equiv\left\{y\in G(k,k{+}4): y^I_\alpha=C_{\alpha i} z^I_i,\; C\in G_+(k,n)\right\}\,,\ee
where $G_+(k,n)$ is the positive Grassmannian ($k$-plane in $n$-dimensional space)with all ordered minors $\l i_1 \ldots i_k\r>0$ for $i_1<\cdots<i_k$.  One then defines the canonical form $\Omega_{n,k}(y;z)$ of the tree amplituhedron to have logarithmic singularities on all boundaries of $A(n,k,0)$. Given a particular top-dimensional cell $\Gamma$ of the tree amplituhedron parametrized by positive coordinates $(\alpha_1,\ldots,\alpha_{4k})^\Gamma$, the form with logarithmic singularities on the boundaries of the cell is given by
\be
\Omega_{n,k}^\Gamma(y;z)=\frac{d \alpha_1^\Gamma}{\alpha_1^\Gamma}\ldots \frac{d \alpha_{4k}^\Gamma}{\alpha_{4k}^\Gamma}.
\ee 
Given a set of cells that triangulate the amplituhedron, the canonical form on the full amplituhedron is given by the sum of the forms associated with each cell. The logarithmic singularities that live on the boundary between any two adjacent cells are not true singularities of canonical form, and in fact they cancel in pairs in the sum.

The super-amplitude $Y_{n,k}$ is extracted from $\Omega_{n,k}(y;z)$ by localizing it to a special point $y_0=(0_{4\times k}| I_{k\times k})$ (note the four-dimensional space, complementary to $y_0$, can be thought of as the bosonic momentum-twistor space):
\be\label{extract amp}
Y_{n,k}(\Z)&=&\int d^4\phi_1\ldots d^4\phi_k\int \Omega_{n,k}(y;z)\delta^{4k}(y;y_0)\nl
&=&\sum_{\Gamma} \prod_{a=1}^{4k} \frac{d \alpha_a^\Gamma}{\alpha_a^\Gamma}\prod_{\alpha=1}^k \delta^{4|4}(\sum_{i=1}^n C_{\alpha i}(\alpha^{\Gamma}) \Z_i )\,.
\ee
where $C_{\alpha i}(\alpha^\Gamma_1,\ldots,\alpha^\Gamma_{4k})$ are coordinates of a dimension-$4k$ cell in $G_+(k,n)$. 

At loop level, in addition to the $k\times n$ $C$-matrix, we have $L$ $2\times n$ matrices $D_m=(D^{(A)}_m,D^{(B)}_m)$ with $m=1,\ldots L$ which live in the $(n{-}k)$ dimensional complement of $C$. The amplituhedron $A(n,k,L)$ is the subspace of all $y$'s and $\ell$'s
\be
y^I_\alpha=C_{\alpha, i} z^I_i\,,\quad \ell^I_m=(A_m,B_m)^I=(D^{(A)}_m, D^{(B)}_m)_i\,z^I_i \,,
\ee
with $C\in G_+(k,n)$ and $D$'s satisfy the positivity condition that all the ordered $(k{+}2l)\times (k{+}2l)$ minors of the $(k{+}2l)\times n$ matrix $(D_{m_1},\ldots, D_{m_l}, C)$ are positive, for any $l=1,...,L$ and any $\{m_1,...,m_l\}\subset\{1,...,n\}$; the space of $(D_1,\ldots,D_L; C)$ with these conditions are dubbed as the space $G_+(k,n; L)$. Note we have used $\ell=(A,B)$ to denote $(k{+}4)$-dimensional vectors  and $(D^{(A)}, D^{(B)})$ to denote the two rows of $D$, for which we mod out the GL$(2)$ redundancy.

The canonical form, $\Omega_{n,k,L}(y,\ell;z)$, is again defined to have logarithmic singularities at all boundaries of $A(n,k,L)$. Given positive coordinates $(\alpha_1,\ldots,\alpha_{4(k{+}L)})^\Gamma$ for a cell $\Gamma$, again the form is the product of the $d \log$'s, and one extracts the integrand for super amplitudes, $Y_{n,k}^{(L)}(\Z,\ell)$, exactly the same as in eq.~(\ref{extract amp}):
\be
\Omega^\Gamma_{n,k,L}(y,\ell;z)=\prod_{a=1}^{4(k{+}L)} \frac {d \alpha_a^\Gamma}{\alpha_a^\Gamma},\quad Y^{(L)}_{n,k}(\Z,\ell)=\sum_{\Gamma}\Omega^\Gamma_{n,k,L}(y,\ell;z)\times\prod_{\alpha=1}^k \delta^{4|4}(\sum_{i=1}^n C_{\alpha i}(\alpha^{\Gamma}) \Z_i )\,.\nl
\ee

The systematic study of cell decompositions, and more concretely, finding positive coordinates such that $(D_1,\ldots,D_L; C)\in G_+(k,n;L)$, remains an extremely interesting open question. In the following, by exploiting the on-shell-diagrams in momentum-twistor space, and the all-loop recursion relations, (\ref{recursion}), we will provide a prescription for finding the BCFW cell decomposition and positive coordinates for the amplituhedron.

\section{New on-shell diagrams in momentum-twistor space}\label{diagram}

We start by presenting fundamental ingredients for on-shell diagrams in momentum-twistor space.  A generic on-shell diagram consists of trivalent white and black vertices connected by external and internal edges, all drawn on a disk. Note that we do not assume that the diagrams are planar. In fact, as we will discuss later, non-planarity is a surprising feature that only appears in forward limit terms at loop level, despite the fact that they compute planar loop amplitudes.
\begin{itemize}

\item The external edges (suppose there are $n$ of them) of a diagram are connected to the boundary of a disk. They represent the $n$ color-ordered external states (here we always work with the canonical ordering $1,2,\ldots,n$), and are associated with momentum twistors $\ZZ_1,\ZZ_2,\ldots,\ZZ_n$. 

\item Each internal edge of the diagram is associated with a momentum twistor $\ZZ$ which is then integrated over with the measure 
\be
d^{3|4} \ZZ = \frac{d^{4|4}\ZZ}{{\rm vol\,GL} (1)}
\ee 

\item Each white vertex with three (internal or external) twistors $\ZZ_a,\ZZ_b,\ZZ_c$ is represented by an integral over $C\in G(1,3)$ (a $1\times 3$ matrix with GL$(1)$ redundancies) \be
W(a,b,c)=\vcenter{\hbox{\includegraphics[width=2cm]{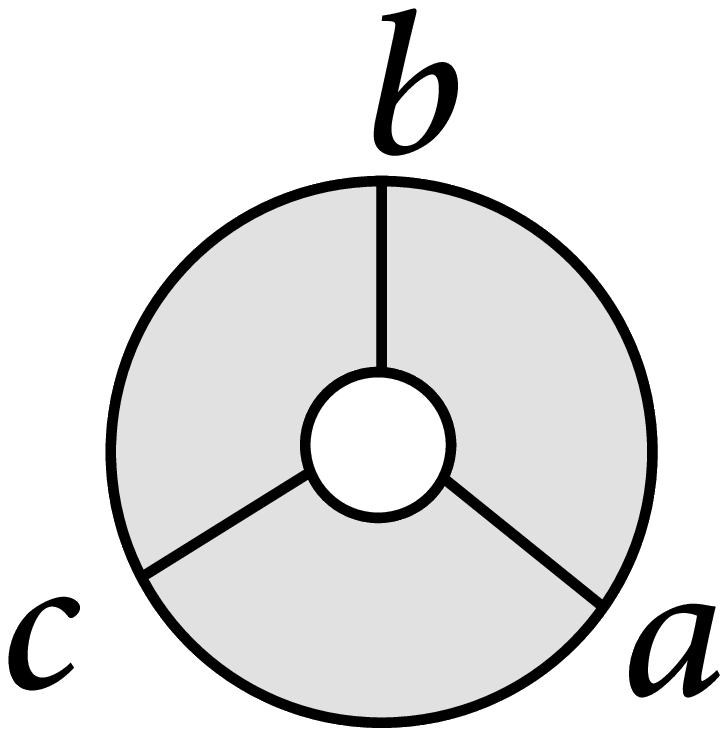}}}=\int \frac 1{{\rm vol\,GL} (1)}\frac{d^{1\times 3} C}{(a) (b) (c)}\delta^{4|4} (C_a \ZZ_a{+} C_b \ZZ_b{+} C_c\ZZ_c)\,.
\ee
where $(i)=C_i$ for $i=a,b,c$. The white vertex thus enforces $\ZZ_a,\ZZ_b,\ZZ_c$ to be on the same projective line.

\item Each black vertex  with three (internal or external) twistors $\ZZ_a,\ZZ_b,\ZZ_c$ is represented by an integral over $C\in G(2,3)$ (a $2\times 3$ matrix up to GL$(2)$ redundancies, with minors defined as $(i\,j)\equiv C_{1,i} C_{2,j}-C_{2,i} C_{1,j}$),
\be
B(a,b,c)=\vcenter{\hbox{\includegraphics[width=2cm]{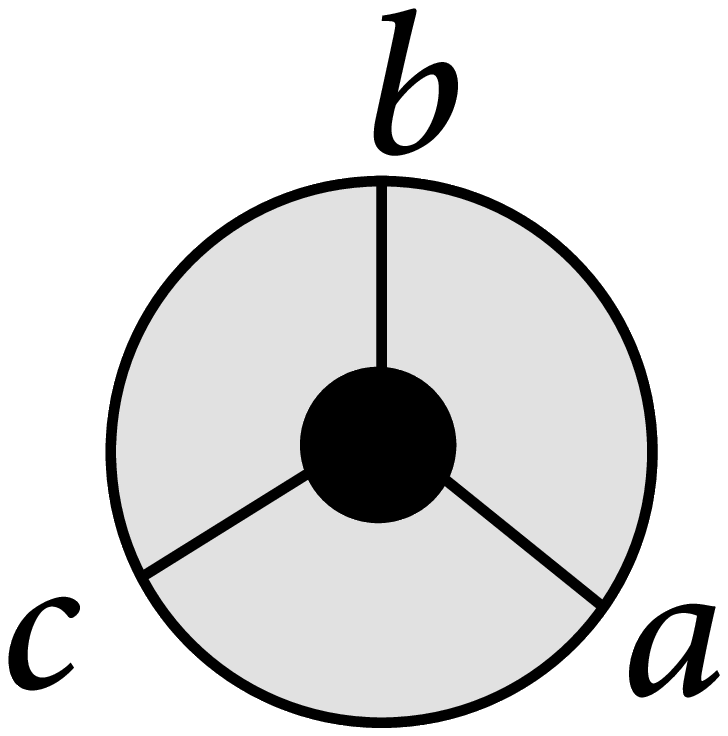}}}=\int \frac 1{{\rm vol\,GL} (2)}\frac{d^{2\times 3} C}{(ab) (bc) (ca)}\prod_{\alpha=1}^2 \delta^{4|4} (C_{\alpha,a} \ZZ_a{+} C_{\alpha,b} \ZZ_b{+} C_{\alpha,c}\ZZ_c)\,.\nl
\ee
The black vertex thus identifies $\ZZ_a,\ZZ_b,\ZZ_c$ projectively. There are degenerate cases: a black vertex with two edges can be deleted from the diagram, with the two edges identified to be one edge (the two twistors are identified), and a black vertex connected to the boundary by one external edge can also be deleted, making the diagram independent of the corresponding external twistor. 

\end{itemize}

\subsection{The Grassmannian representation of momentum-twistor diagrams}

These are all the necessary ingredients for evaluating on-shell diagrams in momentum twistor space.  Formally they are identical to the vertices and edges in the original space, when written in terms of the original twistor variables. However, as we will see shortly, using these diagrams, (MHV-tree stripped) amplitudes/Wilson loops are expressed in a completely different fashion from the way amplitudes are written in terms of the original diagrams. In this section we focus on reduced diagrams, and present their representation using Grassmannian $G(k,n)$ in momentum-twistor space. Note that the $k$-charge here (the Grassmann degree is defined as $4k$) is related to the $k$-charge in the original space (the MHV degree of the full amplitude) by $k_{\rm here}=k_{\rm original}{-}2$. 

The $k$-charge of a diagram is easy to determine: each trivalent white vertex has $k=1$ and each trivalent black vertex has $k=2$; for each internal edge the $k$ is reduced by one, thus the total $k$-charge is 
\be
k=n_W{+}2 n_B{-}n_I\,,
\ee
where $n_W, n_B$ and $n_I$ are the number of trivalent white vertices, black vertices and internal edges. Note that this is the counting after deleting degenerate black vertices. Let us denote the number of relevant external edges as $m$ (note $k+4\leq m\leq n$; each of the remaining $n{-}m$ edges is connected to a monovalent black vertex and thus can be deleted), then another useful relation is 
\be
m=3(n_W+ n_B)-2 n_I\,.
\ee

The derivation for the momentum-twistor Grasssmannian representation of our diagram is parallel to that in section 4 of~\cite{posGrassmannian}, and we will not repeat it.  Roughly speaking, we perform all integrals over internal twistors, and obtain the delta functions given by amalgamation of those from fundamental vertices, and in the end an on-shell diagram can be represented by ``$d \log$" integrals over edge variables, 
\be
\int \prod_{v} \frac 1{\rm{vol\,GL} (1)} \prod_e \frac{d\alpha_e}{\alpha_e} \prod_{I=1}^k\,\delta^{4|4}\big(\sum_{a=1}^n C_{I,a}(\alpha) \ZZ_i\big)\,, 
\ee
where $v,e$ runs over all vertices and edges respectively. Here $C\in G(k,n)$ is the amalgamation of $G(1,3)$'s and $G(2,3)$'s, and it can be put in a GL$(k)$ gauge-fixed form associated with a perfect orientation of the diagram: choosing $k$ external edges as incoming sources, labeled by $A$ (or sometimes $B$), and the remaining $n{-}k$ sinks labeled by $a$, then we can determine a perfect orientation from left-right-path of the diagram. 

The gauge-fixed form of $C$ has an identity-matrix part, $C_{A,B}=\delta^A_B$, and the remaining, non-trivial part of matrix $C$ is determined by the``boundary measurements": the weight for each path $\Gamma$ from $A$ to $a$ is given by the product of all edge variables on the path, and we sum over all such paths, 
\be
C_{A,a}=-\sum_{\Gamma\in \{ A\to a\}} \prod_{e \in \Gamma} \alpha_e\,.
\ee Equivalently, one can define the boundary measurements in terms of face variables,  
\be
C_{A,a}=-\sum_{\Gamma\in \{ A\to a\}} \prod_{f\in \hat\Gamma} (-f)\,
\ee
where $\hat{\Gamma}$ is the set of faces enclosed by the counterclockwise completion of $\Gamma$, and the value of the graph is given by, 
\be
\int \frac 1{\rm{vol\,GL} (1)} \prod_{f} \frac{d f} f\prod_{I=1}^k\,\delta^{4|4} \big(\sum_{a=1}^n C_{I,a}(f) \ZZ_i\big)\,. 
\ee
We note that there is an overall GL$(1)$ redundancy for the face variables since the product of all face variables is unity, $\prod_i f_i=1$, so when evaluating the graph we include all but one face variable.

In practice, we find it convenient to merge all trivalent black vertices connected to each other (without passing through white vertices) into a single one. For any region of the graph with exclusively black vertices, all momentum twistors associated with the edges are identified, thus effectively we have a unique twistor, $\ZZ$, for the region. In the end all black vertices in the entire region evaluate to unity, and, when no external twistors are involved, one simply integrates over the internal twistor with $\int d^{3|4}\ZZ$. 

To see this, it is enough to look at a single trivalent black vertex, attached by $a,b,c$, and we represent the remaining part of the diagram by a projective function $f$:
\be\label{blackmerge}
\int d^{3|4} \ZZ_a\,d^{3|4} \ZZ_b\,d^{3|4} \ZZ_c\,B(a,b,c)\,f(\ZZ_a,\ZZ_b,\ZZ_c)&=&\int d^{3|4} \ZZ_I\,f(\ZZ_a=\ZZ_I,\ZZ_b=\ZZ_I, \ZZ_c=\ZZ_I)\,,\nl
\int d^{3|4} \ZZ_a\, d^{3|4}\ZZ_b\,B(a,b,c)\,f(\ZZ_a,\ZZ_b,\ZZ_c)&=&f(\ZZ_a=\ZZ_c,\ZZ_b=\ZZ_c, \ZZ_c)\,,
\ee 
where we have included two cases: when $a,b,c$ are all internal edges or when one of them, say $c$, is external. Thus in general, as we will see repeatedly in the following, one can evaluate on-shell diagrams by computing the result from those parts with white vertices and using black vertices for connecting them and identifying twistors. 


As studied in details in~\cite{posGrassmannian} any reduced on-shell diagram from can be mapped to a decorated permutation, $\sigma: i\to i\leq \sigma(i)<i{+}n$ with the $k$-charge given by $k_{\rm original}=\frac 1 n \sum_{i=1}^n (\sigma(i){-}i)$. One can decompose the decorated permutation into a series of adjacent transpositions, and the diagram is constructed by the composition of BCFW bridges, which are in one-to-one correspondence to the transpositions. 

Given a decorated permutation with $\sigma(i)>i{+}1$, we can define the dual decorated permutation as $\sigma': \sigma'(i{+}1)=\sigma(i){-}1$. It turns out that, all the reduced diagrams in momentum-twistor space, are the on-shell-diagram associated with the permutation $\sigma'$, i.e. they are constructed by  the composition of BCFW bridges corresponding to adjacent transpositions of $\sigma'$. Although the interpretation of the BCFW bridge is different from that in the original space, this prescription indeed gives the correct diagrams in momentum twistor space. In addition, it also explains why our on-shell-diagrams are computing MHV-tree stripped amplitudes, or equivalently Wilson loops. The reason is that those original diagrams associated with the MHV tree amplitude are exactly those with $\sigma(i)\leq i{+}1$. We will not pursue this connection to the original diagrams any further, but turn to the study of our on-shell-diagrams independently, as building blocks for tree and loop amplitudes. 
 
\subsection{Examples and operations on the diagrams}

Given the general prescriptions, we first study a few simple examples, which will be useful for defining some basic operations acting on the momentum-twistor diagrams. 

The so-called ``lollipop diagrams" are those with all external edges connected to monovalent black vertices.  As we have discussed, one can delete such vertices in which case the diagram becomes the trivial diagram, which is the unique $k=0$ reduced diagram. They give unity, which is the MHV tree-amplitude in momentum-twistor space. Here we draw a 6 point example
\be\vcenter{\hbox{\includegraphics[width=4cm]{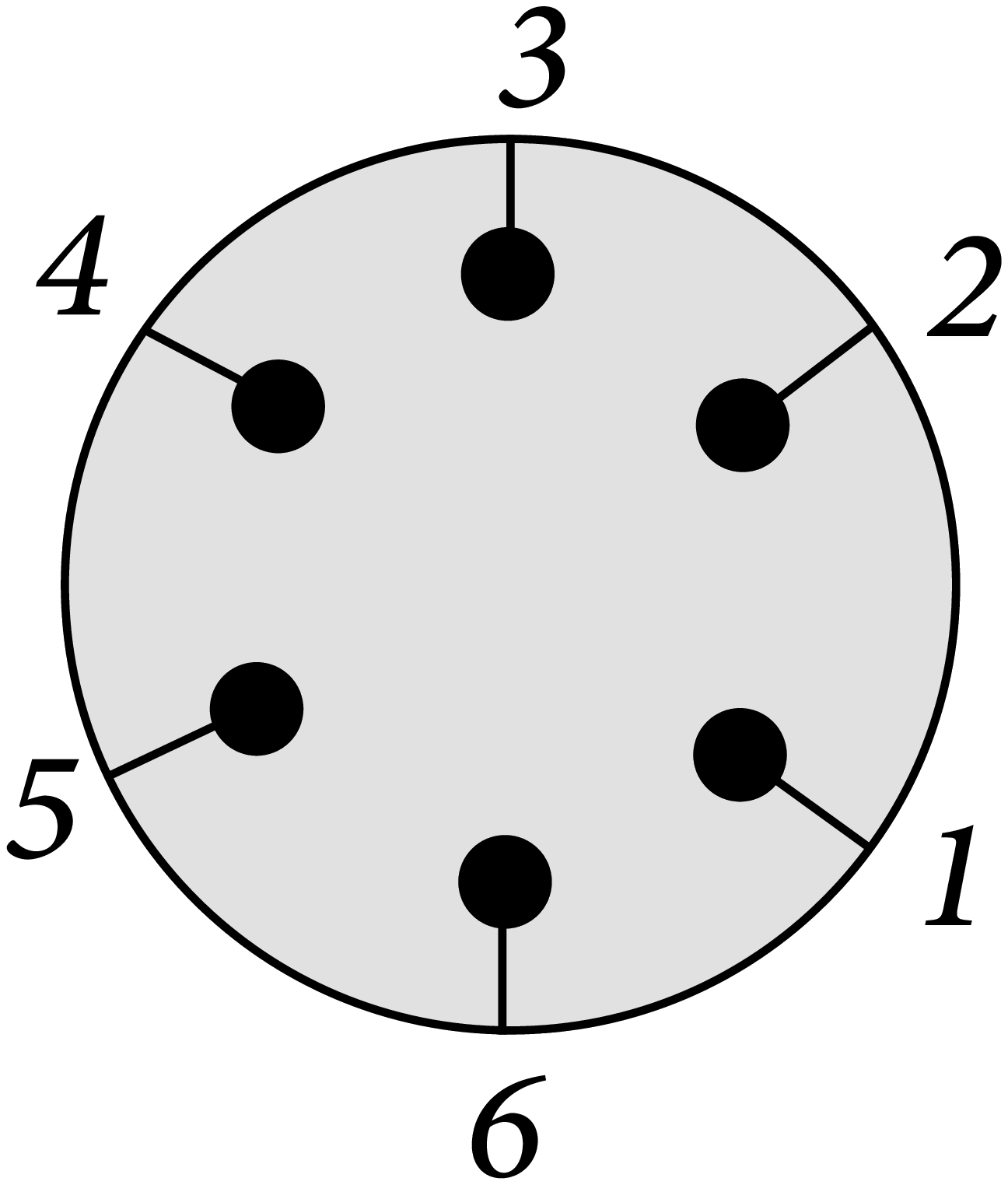}}}\,.\nonumber \ee

The simplest non-trivial diagram is given by two white vertex connected by an external edge, with $n=4, k=1$
\be\label{fac_Rinv}
&&\int d^{3|4} \ZZ_I\,W(a,b, I) W(I, c,d)=\int\frac 1{{\rm vol\,GL} (1)} \frac{d\alpha}{\alpha}\,\frac{d\beta}{\beta}\,\frac{d\gamma}{\gamma}\,\frac{d\delta}{\delta} \delta^{4|4} (\alpha\ZZ_a{+}\beta\ZZ_b{+}\gamma\ZZ_c{+}\delta \ZZ_d)\nl
&&\equiv W(a,b,c,d)=\int d^{3|4} \ZZ_I\,W(b,c,I) W(I,d,a)\,.
\ee
where on the second line we have seen that the two collinear constraints together enforce $\ZZ_a,\ZZ_b,\ZZ_c,\ZZ_d$ to be on a projective plane; this is manifestly cyclic, thus we can merge it as a single white-vertex with four edges attached to it, which we denote as $W(a,b,c,d)$, and then reexpand it to the other channel (which imposes the same kinematic constraints), as shown by the following diagrammatic identity
\be  \vcenter{\hbox{\includegraphics[width=10cm]{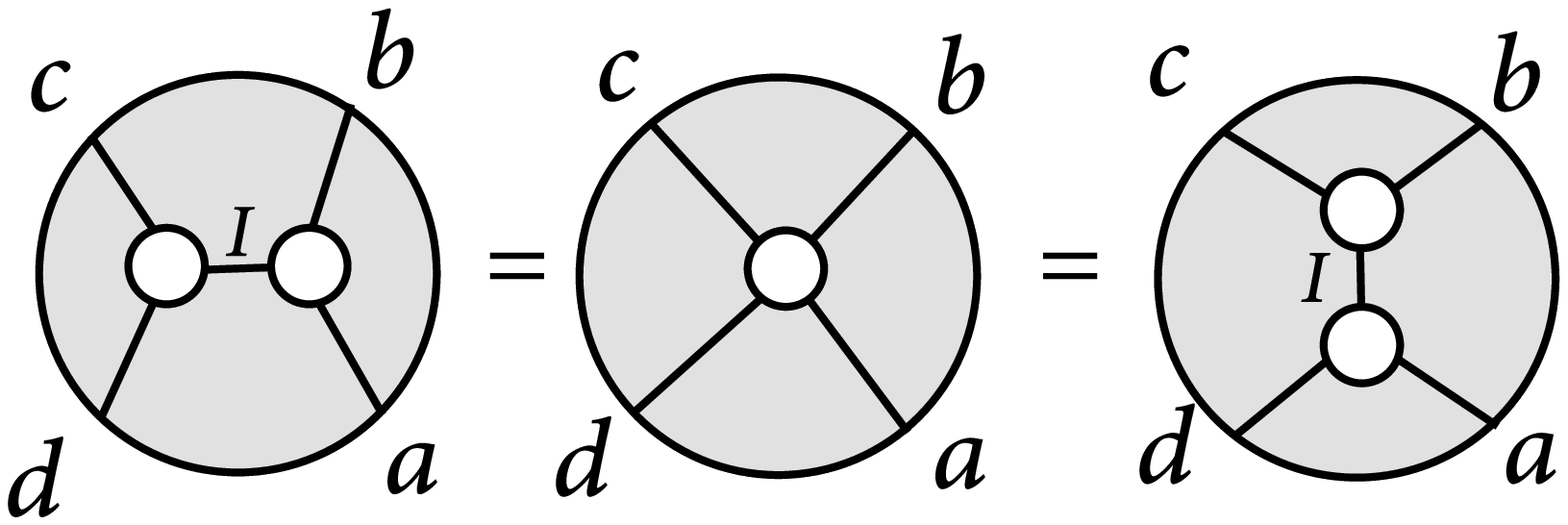}}}\,. \nonumber\ee

This is nothing but the simplest factorization diagram (i.e. with propagator put on-shell) one can access in momentum twistor space. Note that the two-particle channels of the MHV amplitude is invisible since it is stripped off, thus the simplest factorization would be that of a NMHV ($k=1$) amplitude, divided by MHV amplitude. Consider the factorization pole $\l i{-}1\,i\,j{-}1\,j\r=0$, then we can perform the integrals in eq.~(\ref{fac_Rinv}) using a reference twistor $\ZZ_*$ (the result is independent of $*$):
\be~\label{simplestfac}
W(i{-}1,i,j{-}1,j)=\delta(\l i{-}1\,i\,j{-}1\,j\r)\frac{\delta^{0|4}(\l\l *, i{-}1, i, j{-}1, j\r\r)}{\l *\,i{-}1\,i\,j{-}1\r\l *\,i{-}1\,i\,j\r\l *\,j{-}1\,j\,i{-}1\r\l\ *\,j{-}1\,j\,i \r}\,,\nl
\ee
which is indeed the residue at the factorization pole $\l i{-}1\,i\,j{-}1\,j\r=0$ for any NMHV R-invariant with this pole, $[*, i{-}1,i,j{-}1,j]$.

Given the factorization diagram that depends on four twistors, we can obtain the full R-invariant, which depends on five twistors, by adding a \textit{BCFW bridge}. The operation is very simple: it attaches a bridge with a black vertex and a white vertex to two adjacent, external edges, respectively. Denoting the original diagram by $Y(1,\ldots, n)$, then adding the bridge $br(n,1)$, with white and black vertex attached to $n$ and $1$ respectively, amounts to
\be
Y'(1,\ldots, n)=br(n,1)\cdot Y(1,\ldots,n)\equiv \int \frac{d c} c\,Y(1,\ldots, \hat{n})
\ee
where $\hat{\Z}_n=\Z_n+ c\,\Z_1$ and $c$ is the edge variable associated with the bridge; diagrammatically
\be br(n,1)\cdot Y(1,\ldots,n)=\vcenter{\hbox{\includegraphics[width=3cm]{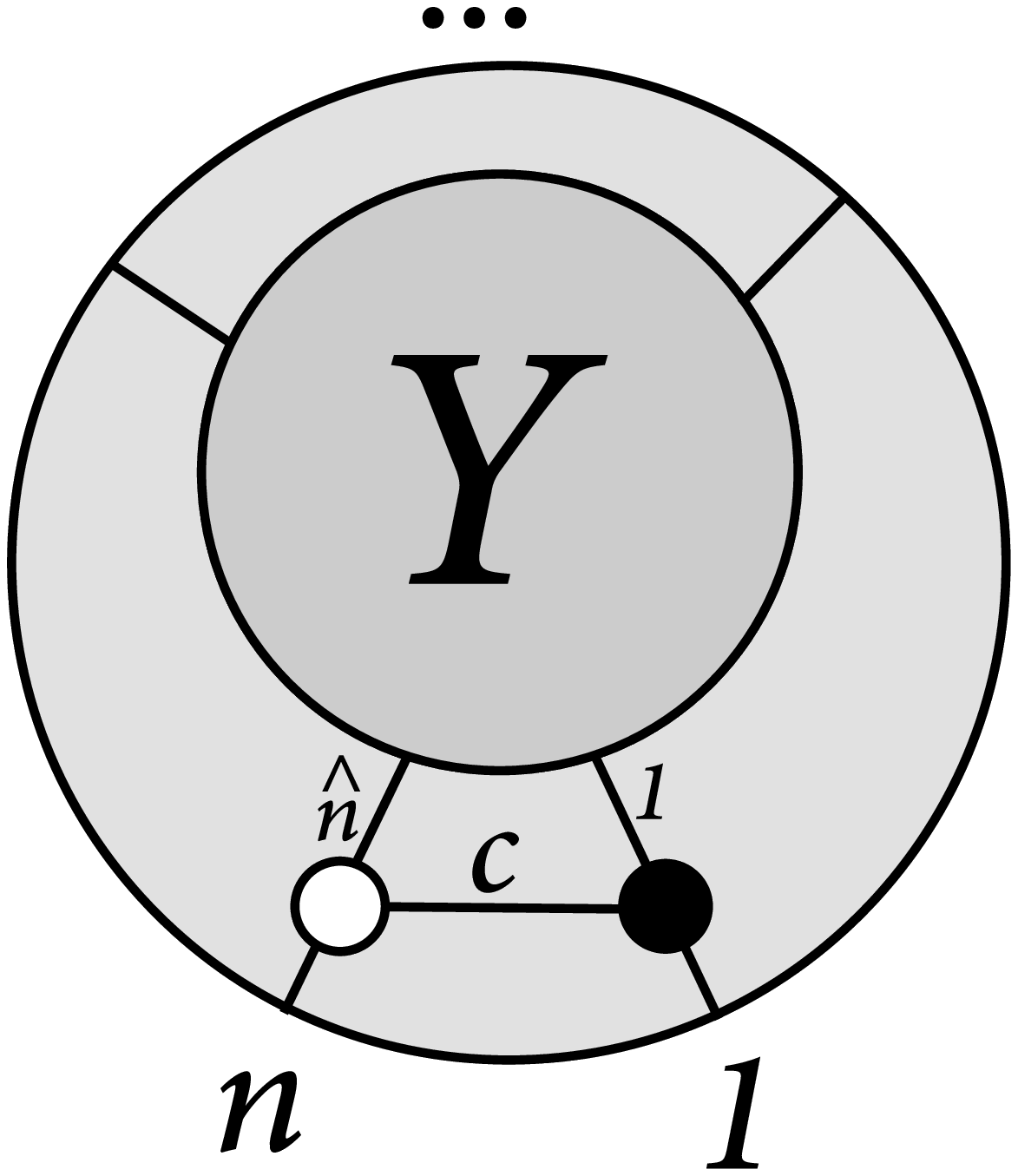}}}~.\nonumber\ee
By adding the BCFW bridge $br(d,e)$ to the diagram from \textit{fusing} $W(a,b,c,d)$ and a lollipop diagram with a black vertex attached by $e$,  $B(e)$, we obtain~\footnote{Fusing two diagrams, denoted by $Y_1 \otimes Y_2$, trivially gives the product of the two diagrams, $Y'=Y_1 Y_2$. This special case, where $Y_2$ is a lollipop diagram with a single external leg, corresponds to adding a particle, as we will see below.} 
\be\label{Rinv}
br(d,e)\cdot (W(a,b,c,d) \otimes B(e))=\int \frac{d c} c\,W(a,b,c,\hat{d})=[a,b,c,d,e]\ee
\be\vcenter{\hbox{\includegraphics[width=6.7cm]{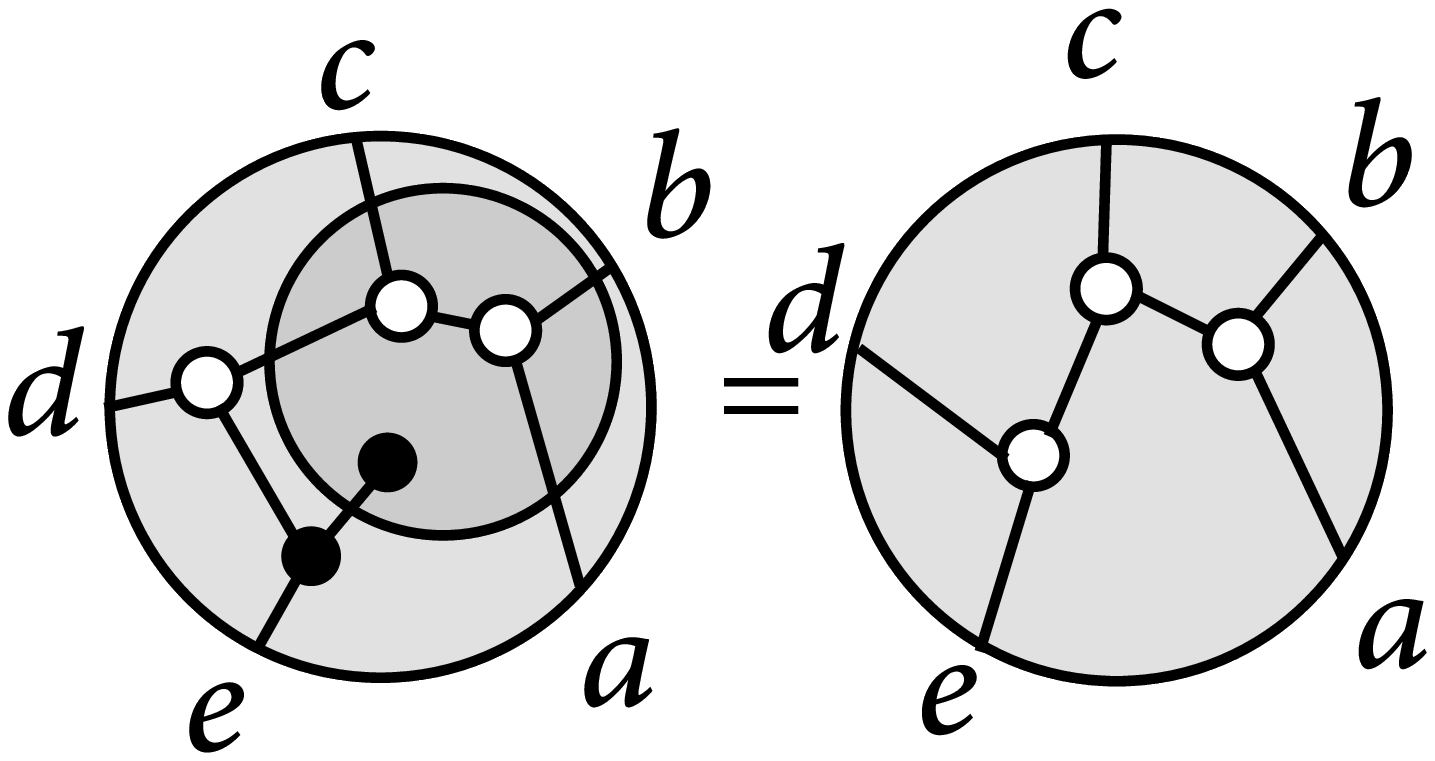}}} \nonumber \ee
which is the on-shell diagram for R-invariant that manifests the factorization channel $W(a,b,c,d)$. We can merge and re-expand the white vertices and obtain various different representations of the same R-invariant:
\be \label{Rinv2} \vcenter{\hbox{\includegraphics[width=10cm]{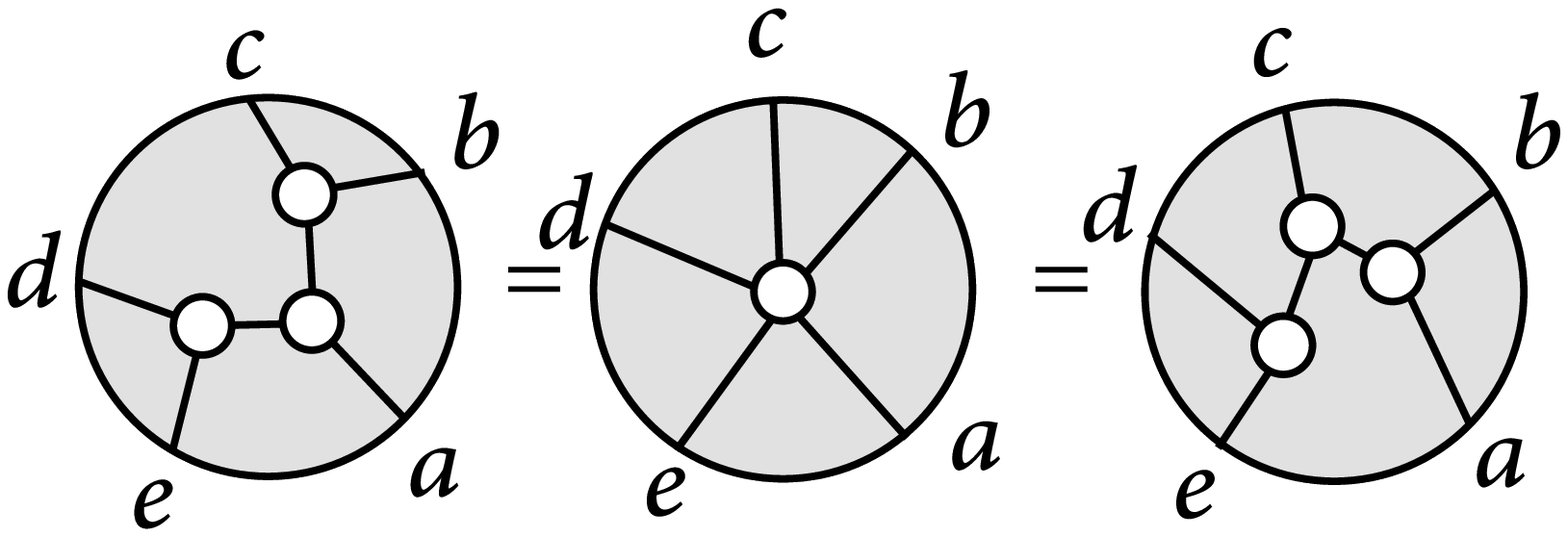}}} \nonumber \ee

In addition to BCFW bridges, we can have operations that add or remove particles for on-shell diagrams. For a generic diagram with external particles $1,\ldots,n{-}1$, one can add an additional particle, $n$, which produces a diagram with $n$ external particles. This corresponds to the ``inverse soft limit", which has two cases,  the \textit{$k$-preserving} and \textit{$k$-increasing} operations. 

The $k$-preserving operation simply adds a lollipop; the external edge $n$ is attached to a monovalent black vertex, and the before and after diagrams evaluate to identical results,
\be\label{softkpres}
Y' (1,\ldots,n)=\vcenter{\hbox{\includegraphics[width=3.5cm]{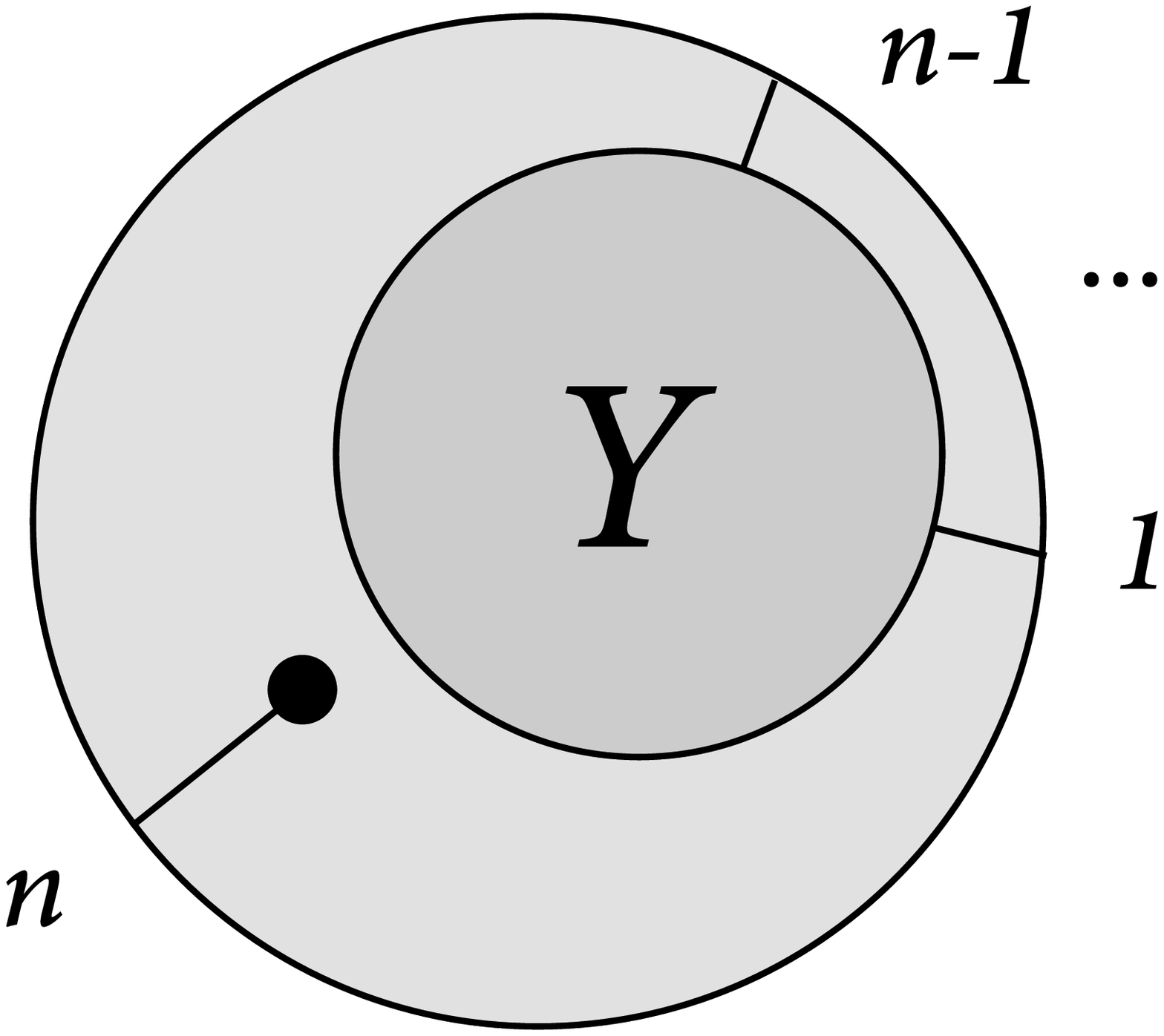}}}=Y(1,\ldots,n{-}1)\otimes B(n)=Y(1,\ldots,n{-}1)~.
\ee

The $k$-increasing operation is more interesting. It ``adds" an R-invariant (thus increase its $k$-charge by $1$) which involves the additional leg, say $n$ and four neighboring legs, $n{-}2,n{-}1,1,2$ to the original diagram, and it also involves shifting the two legs $n{-}1,1$
\be
Y'(1,\ldots, n)= \vcenter{\hbox{\includegraphics[width=4.5cm]{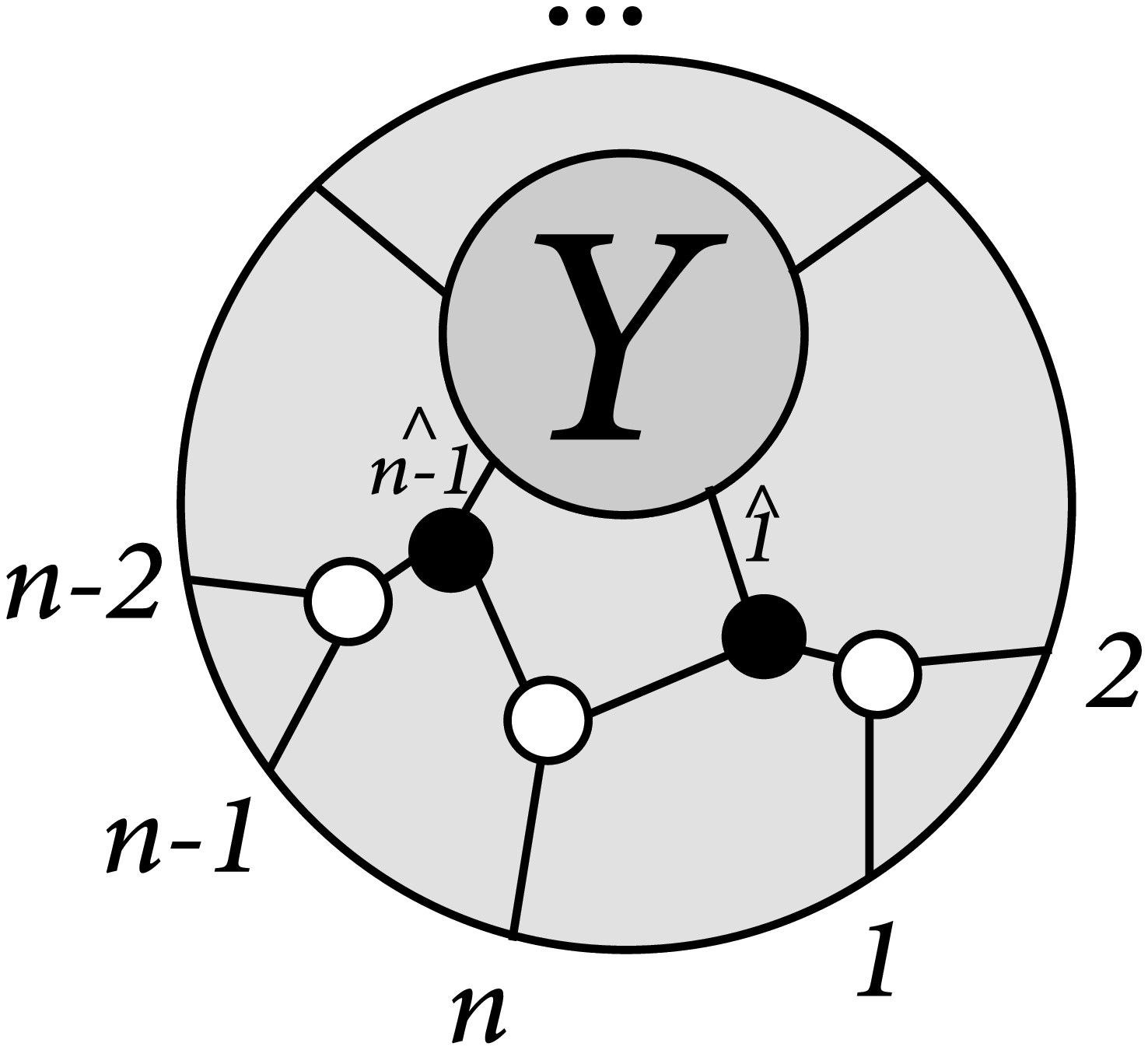}}}=[n{-}2,n{-}1,n,1,2]\,Y(\hat{1},\ldots,\widehat{n{-}1})~,
\ee
where $\widehat{n{-}1}=(n{-}2, n{-}1)\cap (n, 1, 2)$ and $\hat{1}=(1,2)\cap (n{-}2, n{-}1, n)$. 

The opposite operations are those that remove a particle from the diagrams of the form above. Correspondingly they are \textit{$k$-preserving} and \textit{$k$-decreasing soft limits}:
\be
Y'(1,\ldots, n{-}1)=Y(1,\ldots, n)\,,\nl
Y'(1,\ldots, n{-}1)=\int d^{3|4} \ZZ_n Y(1,\ldots, n)\,.
\ee

\section{The amplituhedron from momentum-twistor diagrams: tree level}\label{tree}

In this section we study how to represent factorizations of amplitudes by our diagrams directly in momentum-twistor space. The result will suffice to yield all tree-level amplitudes/Wilson loops, and it generalizes to the factorization terms of all-loop integrand.    

\subsection{Momentum-twistor diagrams for factorizations}

Here we re-derive all factorization contributions for amplitudes in terms of momentum twistor diagrams, including the aforementioned B and FAC terms. The B term is the residue at $w\to \infty$, or $\Z_n\to \Z_{n{-}1}$ projectively. This is nothing but the $k$-preserving soft limit,  eq.~(\ref{softkpres}), given by a lower point diagram with a lollipop on particle $n$,

\be
\text{B} = \;\;\; \vcenter{\hbox{\includegraphics[width=5cm]{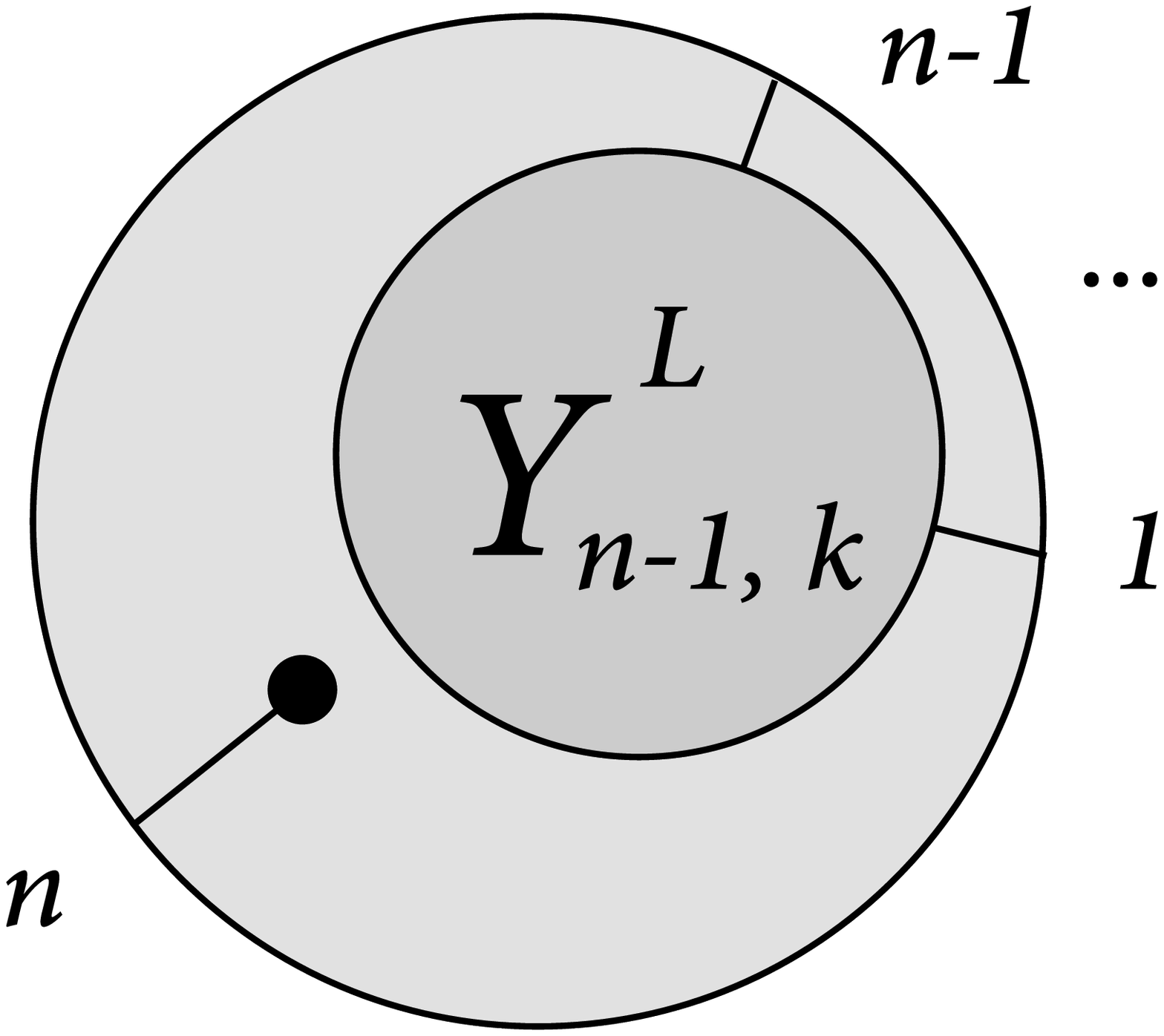}}}
\ee\\

Now we move to the contributions from poles of the form $\l j{-}1\,j\,\hat{n}\,1\r=0$. Recall that we have worked out the simplest cases with $k=1$ in eq.~(\ref{simplestfac}), where both left and right part are unity MHV amplitudes. In general, by connecting the left and right amplitudes with $W(j{-}1,j,n,1)$ we obtain the factorization limit: 
\be
&&Y_L(\hat{j},j,\ldots,n)\,W[j{-}1, j, n,1]\,Y_R(1,\ldots, j{-}1,\hat{j})\nl
=&&\frac{\delta(\l j{-}1\,j\,n\,1\r)\,\delta^{0|4}(\l\l *, i{-}1, i, j{-}1, j\r\r)}{\l *\,i{-}1\,i\,j{-}1\r\l *\,i{-}1\,i\,j\r\l *\,j{-}1\,j\,i{-}1\r\l\ *\,j{-}1\,j\,i \r}\,Y_L(\hat{j},j,\ldots,n)\,Y_R(1,\ldots, j{-}1,\hat{j})\nonumber
\ee
where $\hat{j}=(j{-}1 j)\cap (* n 1)=(n 1)\cap (* j{-}1 j)$ is exactly the twistor corresponding to the intersection of the lines $(j{-}1 j)$ and $(n 1)$, as shown in the diagram of~(\ref{fac}). The diagram is clearly independent of the reference $*$, and note that at this stage it is symmetric under the exchange of left and right amplitude (including $(j{-}1,j) \leftrightarrow (n,1)$). 

As explained in eq.~(\ref{Rinv}), each contribution in the FAC term can be obtained by attaching the BCFW bridge, $br(n,n{-}1)$, to the factorization limit,
\be
&&br(n,n{-}1) \cdot (Y_L(\hat{j},j,\ldots,n)\,W[j{-}1, j, n,1]\,Y_R(1,\ldots, j{-}1,\hat{j}))\nl
=&&[j{-}1, j, n{-}1, n,1] Y_L(\hat{j},j,\ldots,\hat{n}_j)\,Y_R(1,\ldots, j{-}1,\hat{j})
\ee
where we have $\hat{j}=(j{-}1, j)\cap(n{-}1,n,1)$ and $\hat{n}_j=(n{-}1 n)\cap (1 j{-}1 j)$, which are the two intersection points in the diagram. By summing over these, we have the FAC term:
\be\label{fac}
\text{FAC} = \sum_{j=3}^{n-2}\;\;\; \vcenter{\hbox{\includegraphics[width=7cm]{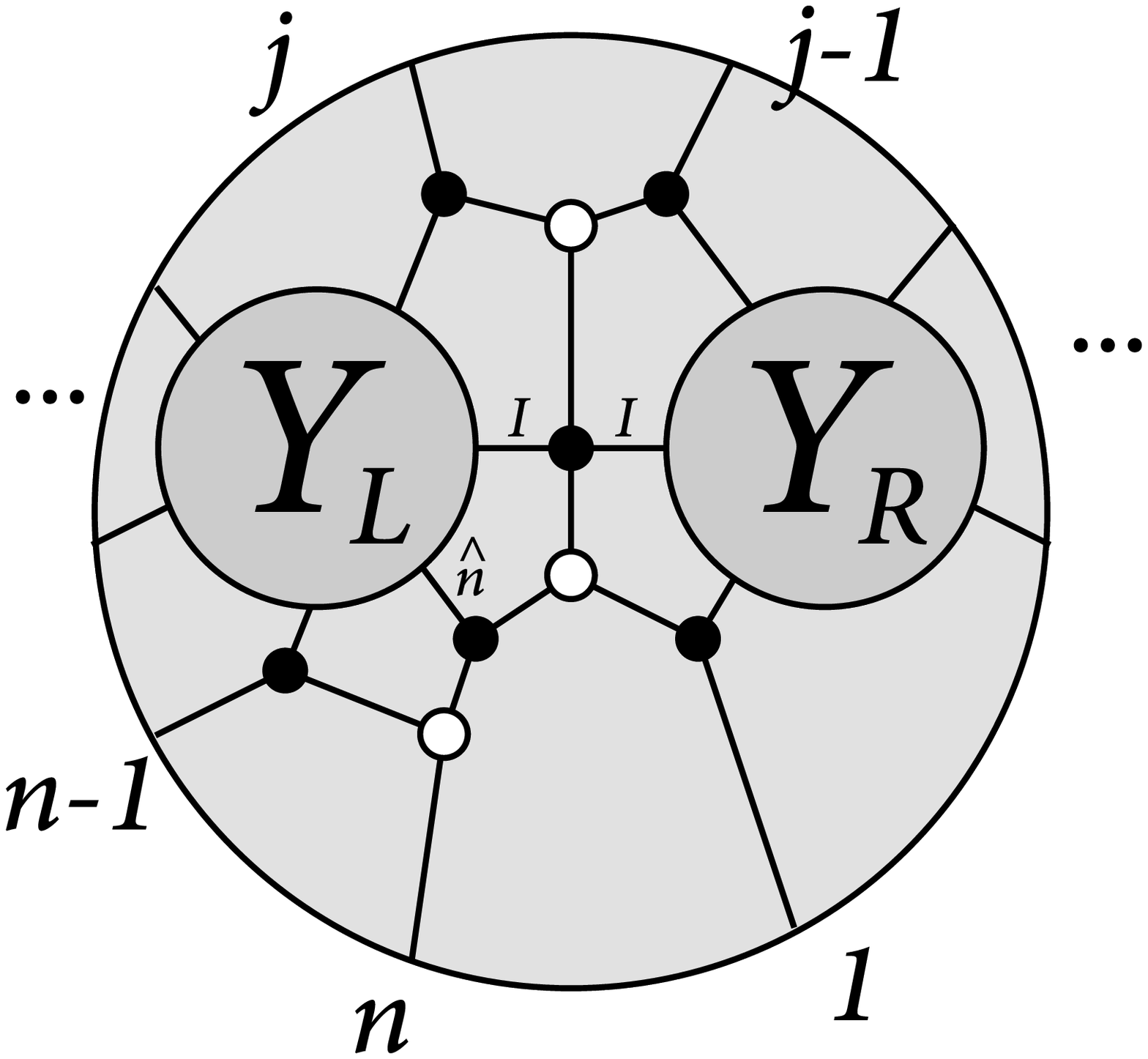}}}
\ee\\

Given that this is our first full-fledged example, we find it necessary to perform the computation directly from the final diagram in details, which also serves as a good example to demonstrate the way we evaluate these diagrams. Explicitly, one can use eq.~(\ref{blackmerge}) to identify those internal twistors attached to the same black vertex, and then only keep the delta functions from white vertices:

\be
\vcenter{\hbox{\includegraphics[width=8cm]{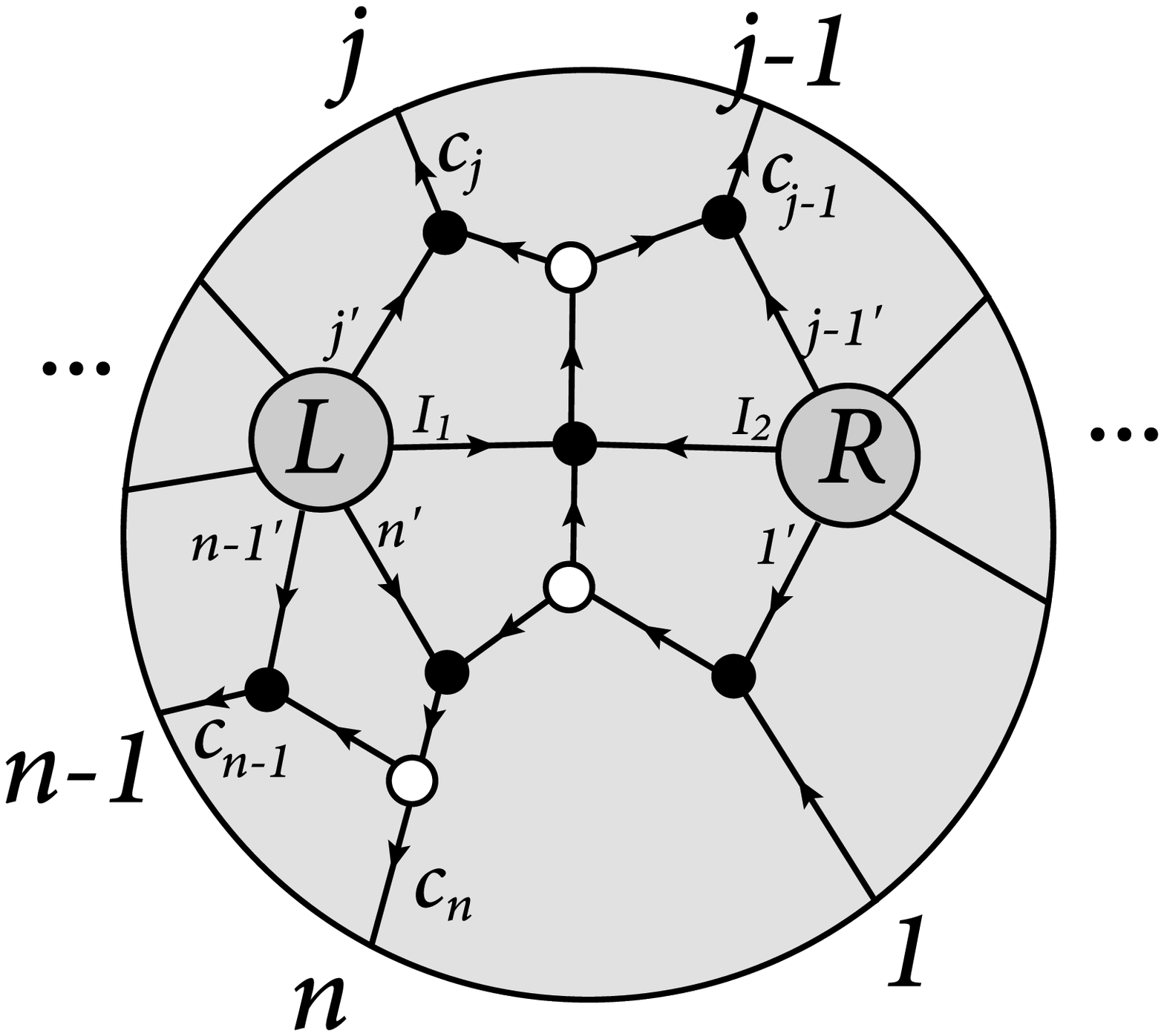}}}\
\ee
\be
&=&\int d^{3|4}\Z_I\,d^{3|4}\Z_n'\,\frac{d c_{j-1}\,d c_j\,d c_{n-1}\,d c_n\,d c_I\,d c_{n'}}{c_{j{-}1}\,c_{j}\,c_{n{-}1}\,c_n\,c_I\,c_{n'}}\nl
&&\times \delta^{4|4}(c_I \Z_I-c_{j-1}\Z_{j{-}1}-c_j\Z_j)\,\delta^{4|4}(c_{n'} \Z_n'-c_{n{-}1}\Z_{n{-}1}-c_n\Z_n)\,\delta^{4|4}(\Z_1{-}c_I \Z_I{-} c_{n'} \Z_{n'})\nl
&&\times Y_{n_L,k_L}^{L_L}(I_1,j',...,(n{-}1)',n')\,Y_{n_R,k_R}^{L_R}(1',2,...,(j{-}1)',I_2)\,.
\ee
Now from the delta functions we can see the following constraints,
\be
\Z_I\sim c_{j-1}\Z_{j-1}+c_j \Z_j &\sim& (j-1,j)\cap(1,n-1,n)\equiv \hat{\Z}_j\nl
\Z_{n'}\sim c_{n-1}\Z_{n-1}+c_{n}\Z_n &\sim& (n-1,n)\cap(1,j-1,j) \equiv \hat{Z}_n\,,
\ee
where we used $\sim$ since the twistors are defined projectively.  The same can be directly seen from the geometry of the diagram, e.g. consider the two lines labeled $I$. The white vertex above the two lines impose that $\Z_I$ lies on the line $(j-1,j)$, while the two white vertices below say that $\Z_I$ lies on the plane $(1,n-1,n)$. It follows that $\Z_I = (j-1,j)\cap (1,n-1,n)$, exactly as required by BCFW recursion. A similar argument shows that $\hat{\Z}_n$ has the correct shift.  

Now we can perform the integrals over $\Z_I, c_I$ and $\Z_{n'}, c_{n'}$ (which effectively combine into $d^{4|4}\Z_I$ and $d^{4|4}\Z_{n'}$) using the first two delta functions. On their support, the argument of the last delta function becomes $\Z_1{-}c_{j{-}1}\Z_{j{-}1}{-}c_j \Z_j{-}c_{n{-}1}\Z_{n{-}1}{-}c_n\Z_n$, and the result is\\
\be
&&\int \frac{dc_{j{-}1}\,dc_j\,dc_{n{-}1}\,dc_n}{c_{j{-}1}\,c_{j}\,c_{n{-}1}\,c_n}\,\delta^{4|4}(\Z_1{-}c_{j{-}1}\Z_{j{-}1}{-}c_j \Z_j{-}c_{n{-}1}\Z_{n{-}1}{-}c_n\Z_n)\,Y_L(I,\ldots,n')Y_R(1,\ldots,I)\nl
&=&[1,j{-}1,j,n{-}1n]\,Y_L(\hat{j},j,...,{n-1},\hat{n})Y_R(1,2,...,j{-}1,\hat{j})
\ee\\
where the integration over the remaining $c$ variables gives the R-invariant $[j{-}1,j,n{-}1,n,1]$, and in the left and right amplitudes, we made the replacement $I\to \hat{j}$, $n'\to \hat{n}$. 

\subsection{Examples}

Thus all tree amplitudes can be determined by solving the recursion with only the factorization term. Here we present two examples, NMHV and N${}^2$MHV, which is enough to illustrate the result. On the other hand, we know a prior what the resulting diagrams are: they are given by the on-shell diagrams with permutations $\sigma'$ satisfying $\sigma'(i{+}1)=\sigma(i)-1$, for those $\sigma$'s associated with diagrams in the original space. It is straightforward to check the following result from this perspective. 

\subsubsection{NMHV trees}

We now apply our diagrams to obtain the Yangian invariant $Y_{n,k=1}(\Z_1,..,\Z_n)$ for NMHV trees, where we have suppressed writing the $L=0$ superscript. Recall that NMHV trees factorize as the product of two MHV trees. By representing MHV amplitudes as a series of lollipops, the factorization term becomes

\be
\text{FAC} = \sum_{j=3}^{n-2}\;\;\; \vcenter{\hbox{\includegraphics[width=4cm]{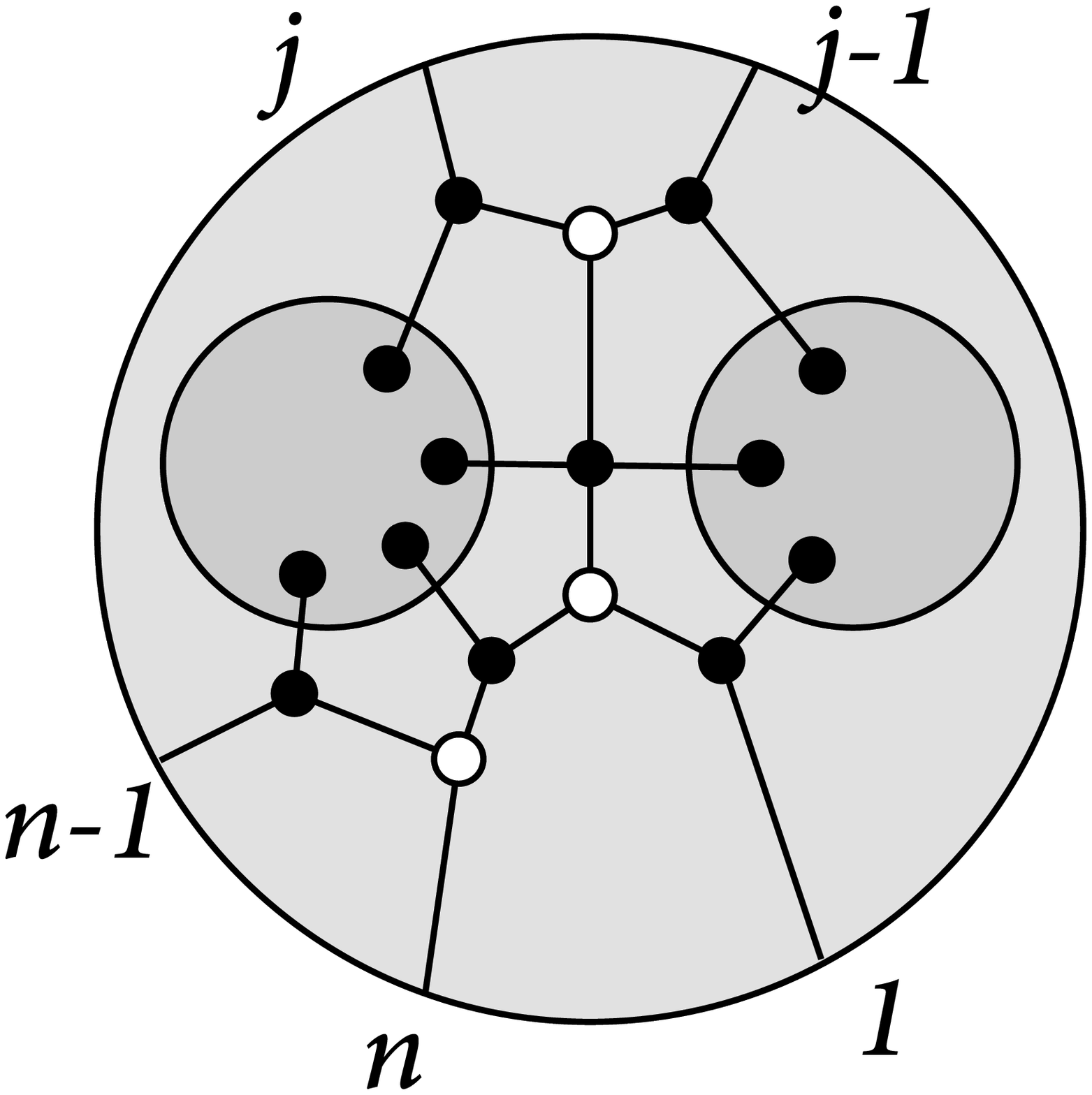}}}\;\;\; = \;\;\;\sum_{j=3}^{n-2}\;\;\; \vcenter{\hbox{\includegraphics[width=4cm]{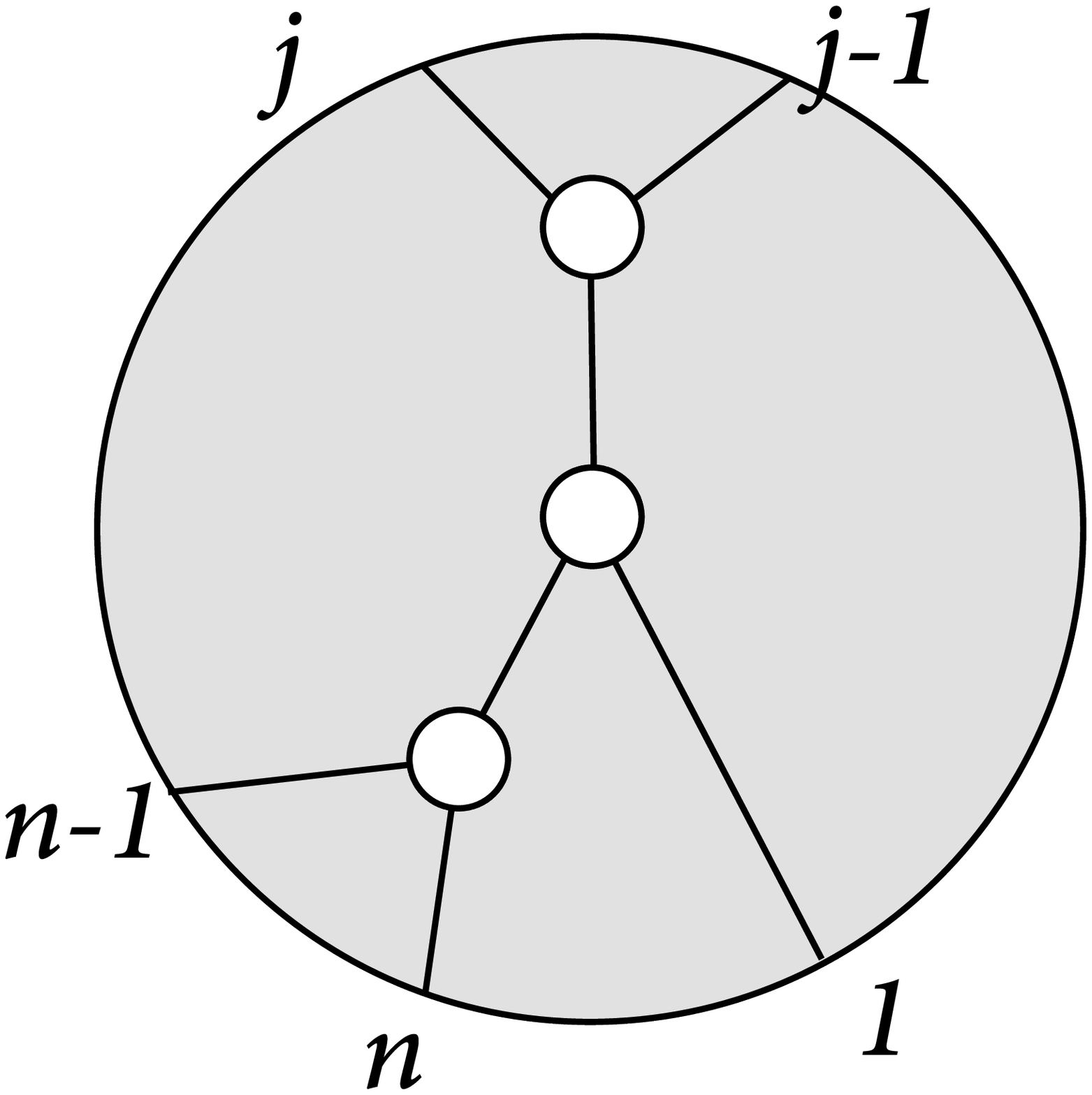}}}
\ee\\

In going from the first diagram to the second, we deleted any lollipops attached to internal lines, and any vertex attached to only two lines.\\

Recall from our earlier discussion that this diagram is just the R-invariant $[1,j{-}1,j,n{-}1,n]$. The BCFW recursion is therefore
\be
Y_{n,k=1}(1,...,n) = Y_{n-1,k=1}(1,...,n-1)+\sum_{j=3}^{n-2}[1,j{-}1,j,n{-}1,n]
\ee
As is well known, the following closed form expression for $Y_{n,k=1}$ satisfies the recursion relation, for which we have a diagrammatic representation now,
\be
Y_{n,k=1}(1,...,n) =\sum_{i<j} [1,i{-}1,i,j{-}1,j]~.
\ee

\subsubsection{$\text{N}^2\text{MHV}$ trees}

We can also apply our diagrams to $\text{N}^2\text{MHV}$ trees, which factorize as the product of MHV and NMHV. Consider for example the 6 point case where the B term vanishes. A moment's thought reveals that there is only one FAC diagram that contributes, which contains 5 point NMHV tree on the left and 3-point MHV tree on the right.
\be \label{N2MHV}
\text{FAC} = \;\;\; \vcenter{\hbox{\includegraphics[width=4.9cm]{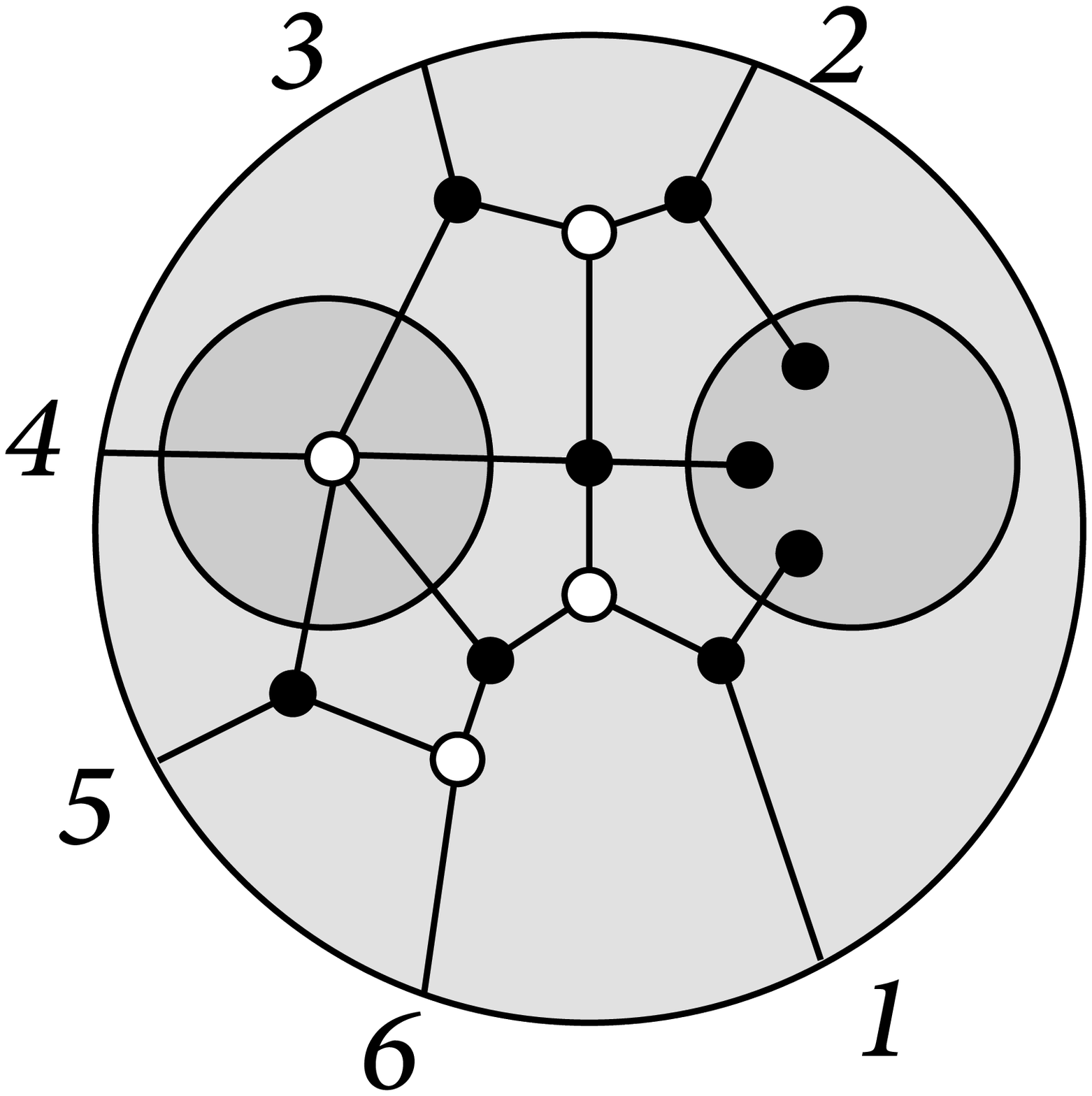}}} \;\;\;= \;\;\; \vcenter{\hbox{\includegraphics[width=5.5cm]{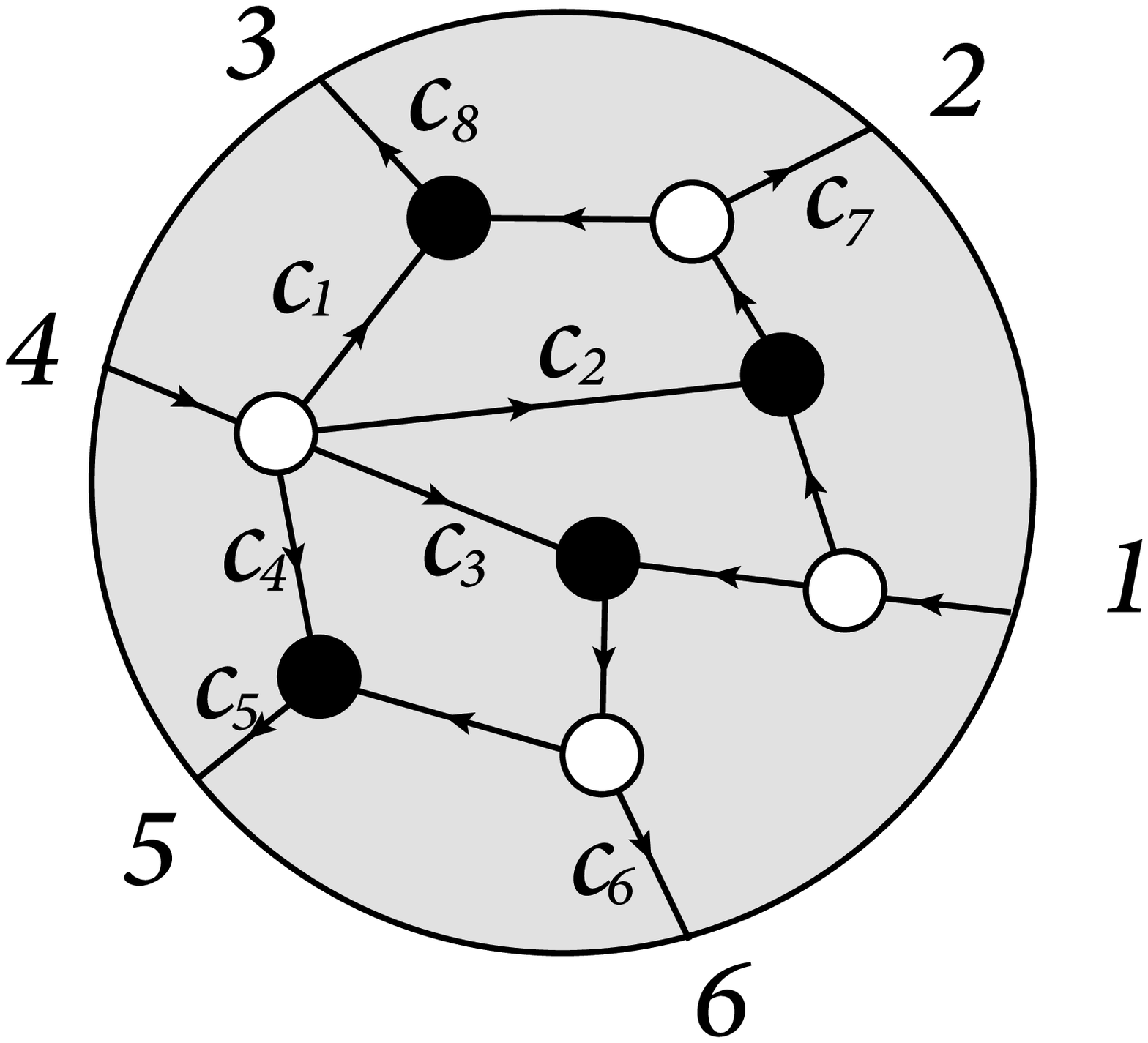}}}
\ee\\

We can then compute this diagram by performing boundary measurements. Since the diagram contains $n_F = 9$ faces, the number of integration variables must be $n_F-1 = 8$, so we can gauge fix some of the bridge variables until only 8 are left. The diagram above shows one particular choice of leftover bridge variables. The explicit formula for this diagram is thus given by
\be
\text{FAC} =\int \frac{dc_1 \; ... \; dc_8}{c_1 \; ... \; c_8}&&\delta^{4|4}(\Z_1-c_5 \Z_5-c_6 \Z_6-c_7 \Z_2-c_8 \Z_3)\nl &\times&\delta^{4|4}(\Z_4 -c_1 c_8 \Z_3 -c_2(c_7 \Z_2+c_8 \Z_3)- c_3(c_5 \Z_5+ c_6 \Z_6)-c_4c_5 \Z_5)\nl
\ee
On the support of the first delta function, it is easy to see that
\be
c_5 \Z_5 + c_6 \Z_6 &\sim& (56)\cap(123) \equiv \Z_5'\nl
c_7 \Z_2 + c_8 \Z_3 &\sim& (23)\cap(156) \equiv \Z_2'
\ee

Substituting these into the second delta function and rescaling the integration variables appropriately gives
\be
\text{FAC} = \int \frac{dc_1 \; ... \; dc_8}{c_1 \; ... \; c_8}\delta^{4|4}(\Z_1-c_5 \Z_5-c_6 \Z_6-c_7 \Z_2-c_8 \Z_3)\nl
\times\delta^{4|4}(\Z_4-c_1 \Z_3-c_2 \Z_2'- c_3 \Z_5'-c_4 \Z_5 )\,.
\ee

The integral is now trivial to perform. It just gives us two R-invariants:
\be
Y_{6,k=2}(\Z_1,...,\Z_6) &=& [3,4,5,2',5'][1,2,3,5,6] \nl
&=& [3,4,5,(23)\cap(156),(56)\cap(123)][1,2,3,5,6]
\ee

As mentioned above, in practice it is usually not productive to work out all the boundary measurements step by step and identify all proper shifts like $\ZZ_2'$ and $\ZZ_5'$ on the support of the delta functions; the shifts can be identified more quickly by looking at the diagram and remembering the role of the black and white vertices. From our general rules and examples, it is straightforward to work out (reduced) momentum-twistor diagrams for all tree-level amplitudes/Wilson loops.

\section{The amplituhedron from momentum-twistor diagrams: loop level}\label{loop}

In this section we turn to loop level. Since B and FAC of any loop integrand are identical to those at tree level, we focus on the final contribution to its BCFW expansion,  the forward limit term FL. It comes from a loop propagator going on shell. In other words, given the BCFW deformation $\hat{\mathcal{Z}}_n=\mathcal{Z}_n+w \mathcal{Z}_{n-1}$, we expect to find poles when $\left<AB1\hat{n}\right>\rightarrow 0$. Each of these poles contributes to the FL term.

\subsection{Momentum-twistor diagrams for forward-limit contributions}

The FL term from $Y^{(L{-}1)}_{n{+}2,k{+}1}(1,\ldots,n, A,B; \{\ell\}/(AB))$, which contributes to $Y^{(L)}_{n,k}(1,\ldots,n; \{\ell\})$, is given by the following diagram
\be
\text{FL} = \int_{GL(2)}\;\;\; \vcenter{\hbox{\includegraphics[width=5.5cm]{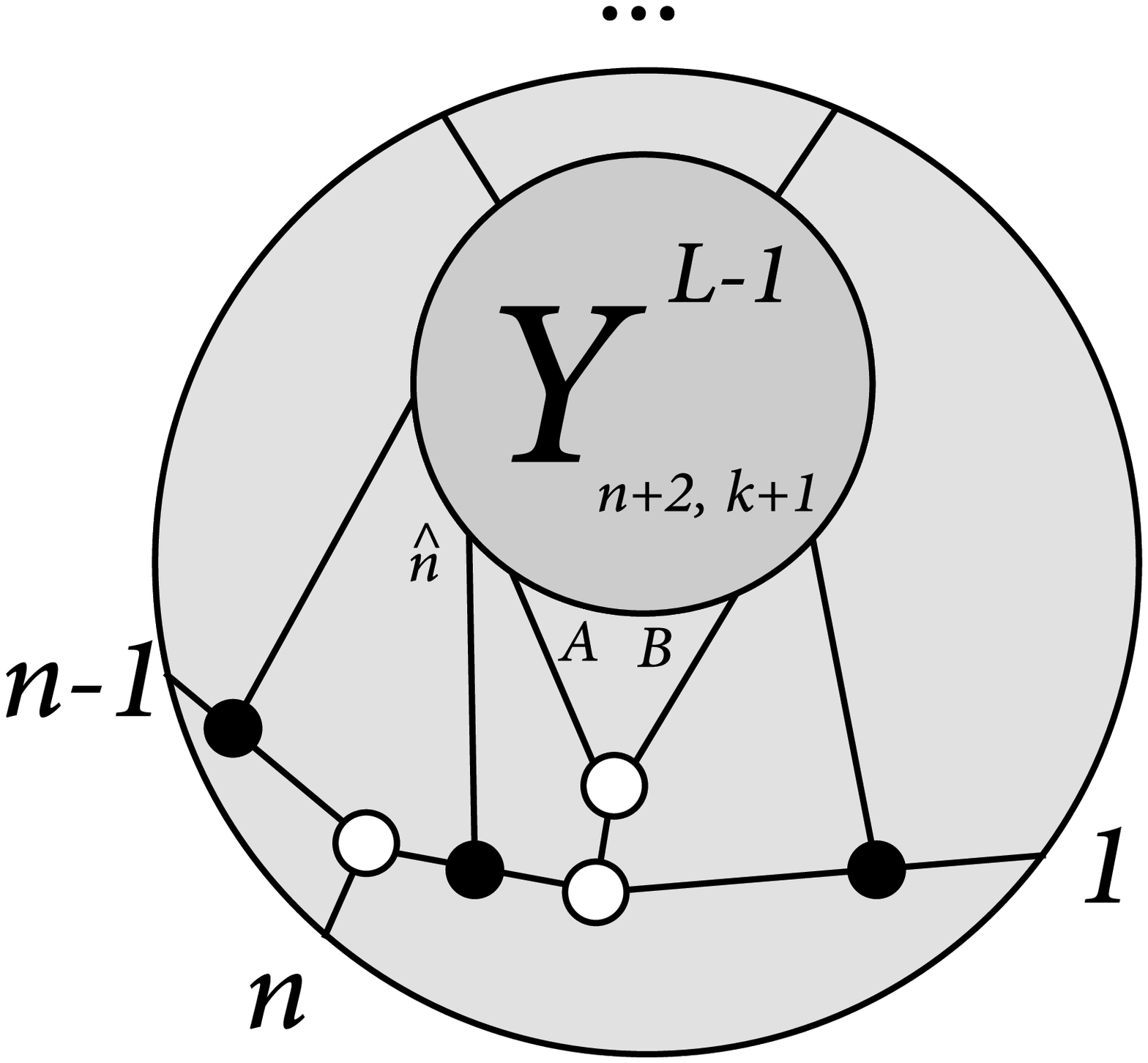}}}~.
\ee\\
The $GL(2)$ integral sign is just there to remind us that there is a $GL(2)$ residue we must take, which we will discuss in a moment. One checks readily that the required shift $\hat{\Z}_n = (n-1,n)\cap (1,A,B)$ in the FL term is expressed in the diagram.

In order to do the $GL(2)$ residue diagrammatically, we first do a new BCFW shift ${\Z_B}\rightarrow\Z_B + w \Z_1$ on the sub-diagram $Y_{n+2,k+1}^{L-1}$. As usual, this contains a boundary (i.e. $w\rightarrow \infty$) term, a factorization channel FL-FAC (i.e. forward limit of factorization channel), and a forward limit FL-FL (i.e. forward limit of forward limit). The boundary term in general does not contribute. Let us first look at the FL-FAC term, which is everything for the FL of one-loop amplitudes, since FL-FL terms do not contribute. We have
\be
\text{FL-FAC} = \sum_{j=3}^{n-1} \int_{GL(2)}\;\;\; \vcenter{\hbox{\includegraphics[width=5cm]{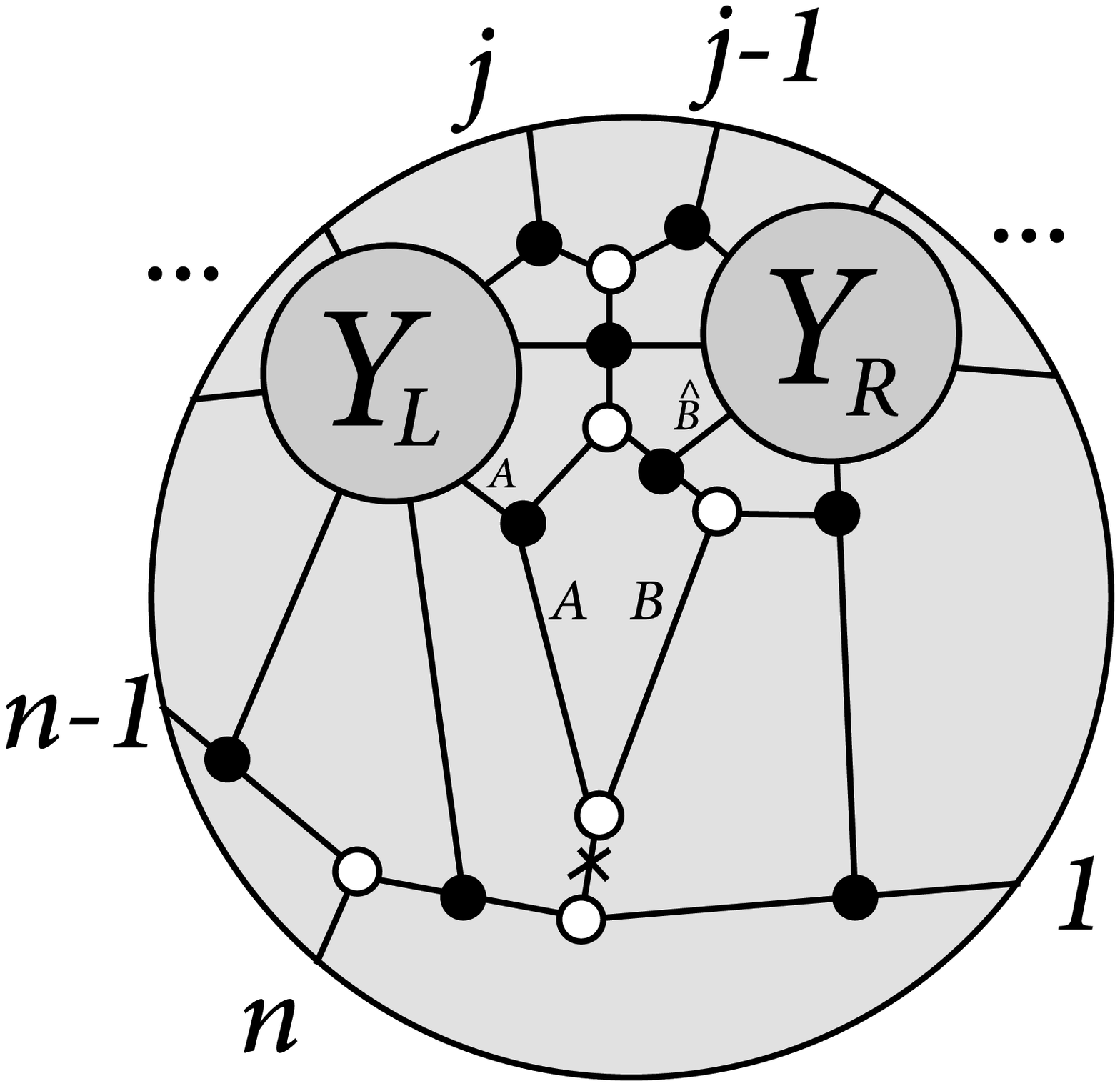}}}
\ee\\
where we sum over all left and right sub-diagrams for which $L_L + L_R = L-1$, $k_L + k_R = k$, and $n_L + n_R= n+4$. The boundary case $\Z_j = \hat{\Z}_n$ is zero after doing the fermionic integrals for $\Z_A,\Z_B$, and so is not included in the summation.

Now recall that the $GL(2)$ residue takes $\Z_A,\Z_B\rightarrow (A,B)\cap (1,n-1,n)$. The point $(A,B)\cap (1,n-1,n)$ can be found on the diagram, and is labeled by a cross. When taking the residue, the line $A$ coming out of the left sub-diagram must be cut and reconnected to the crossed line.

But what about the line $\hat B$ coming out of the right sub-diagram? Surely that must be reconnected as well. A quick look at the diagram shows that $\hat {\Z}_B = (1,B)\cap(j-1,j,A)$. When taking the residue, this becomes $\hat{\Z}_B\rightarrow (A,B)\cap (1,n-1,n)$, which again is the crossed line. So the $\hat B$ line must also be reconnected to the cross. This completes the $GL(2)$ residue. The advantage of using diagrams is that we did not have to do this residue analytically. The final form of the FL-FAC term is thus given by the following\\
\be
\text{FL-FAC} = \sum_{j=3}^{n-1} \;\;\; \vcenter{\hbox{\includegraphics[width=4.5cm]{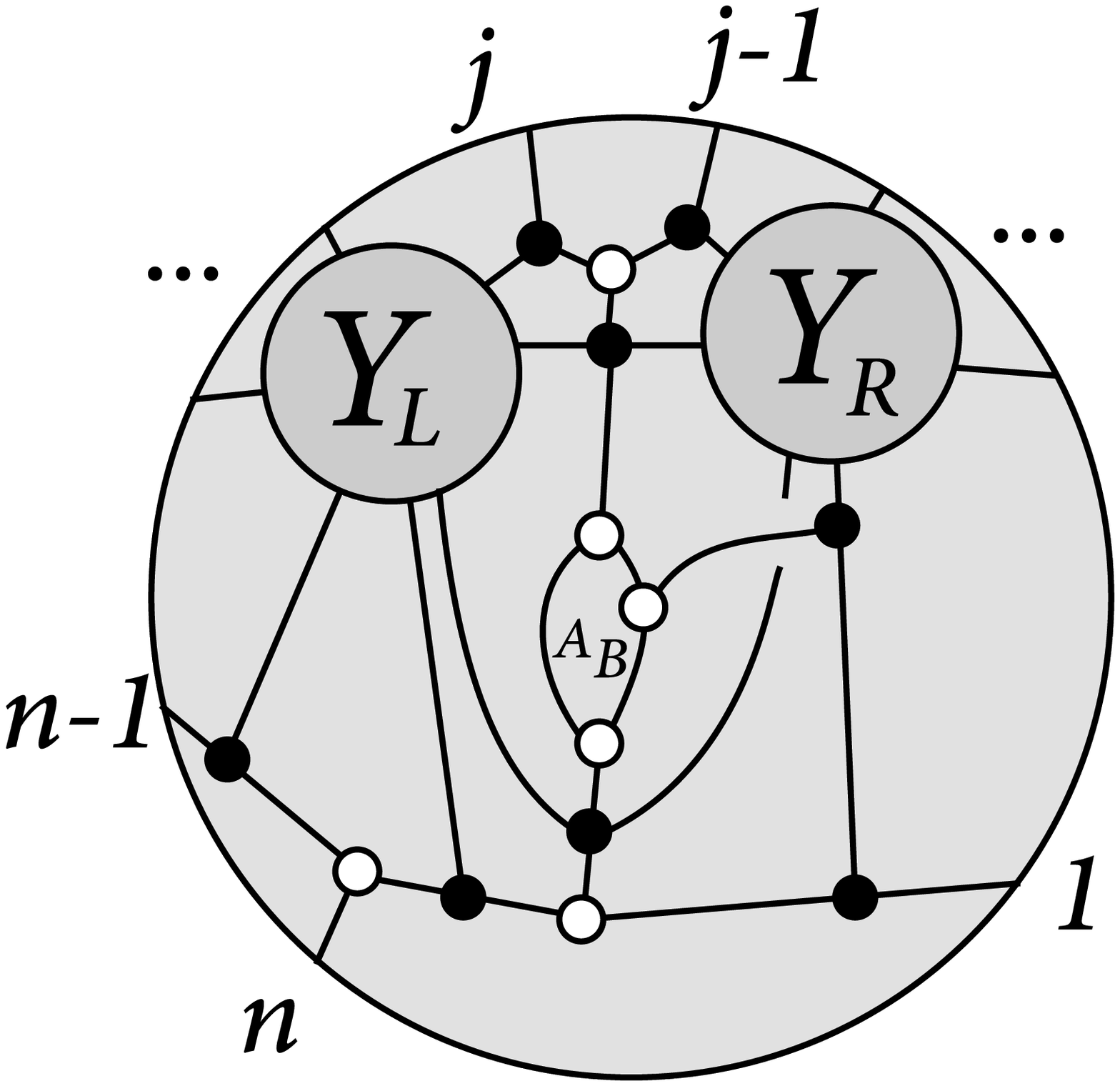}}}\;\;\; = \sum_{j=3}^{n-1} \;\;\; \vcenter{\hbox{\includegraphics[width=4.5cm]{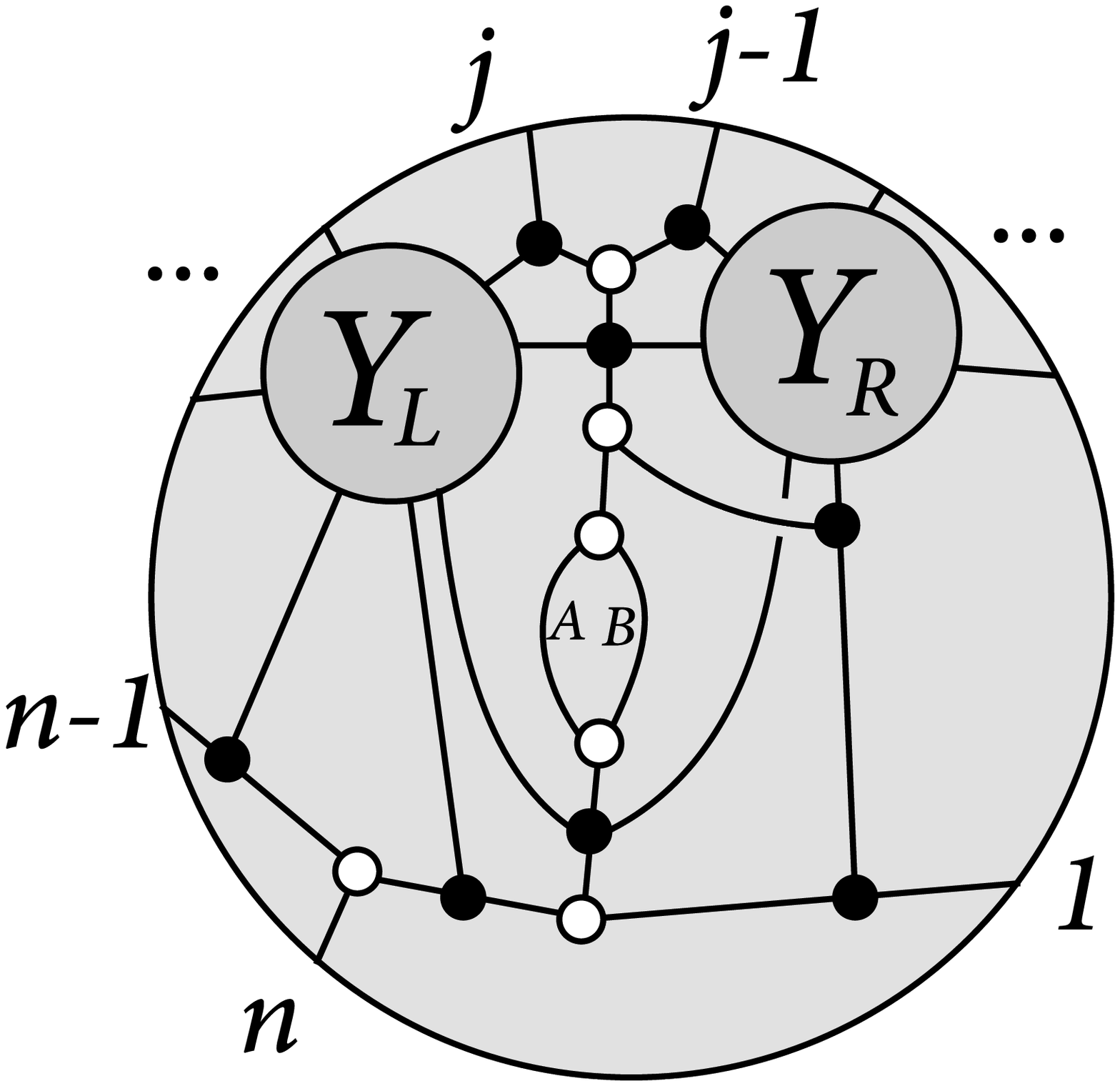}}}
\ee\\
where the second diagram is obtained from the first by some merge and expansion of white vertices.\\

In the boundary case where $j = n{-}1$, we should identify the external lines $j$ and $n{-}1$ as follows

\be
\vcenter{\hbox{\includegraphics[width=4.9cm]{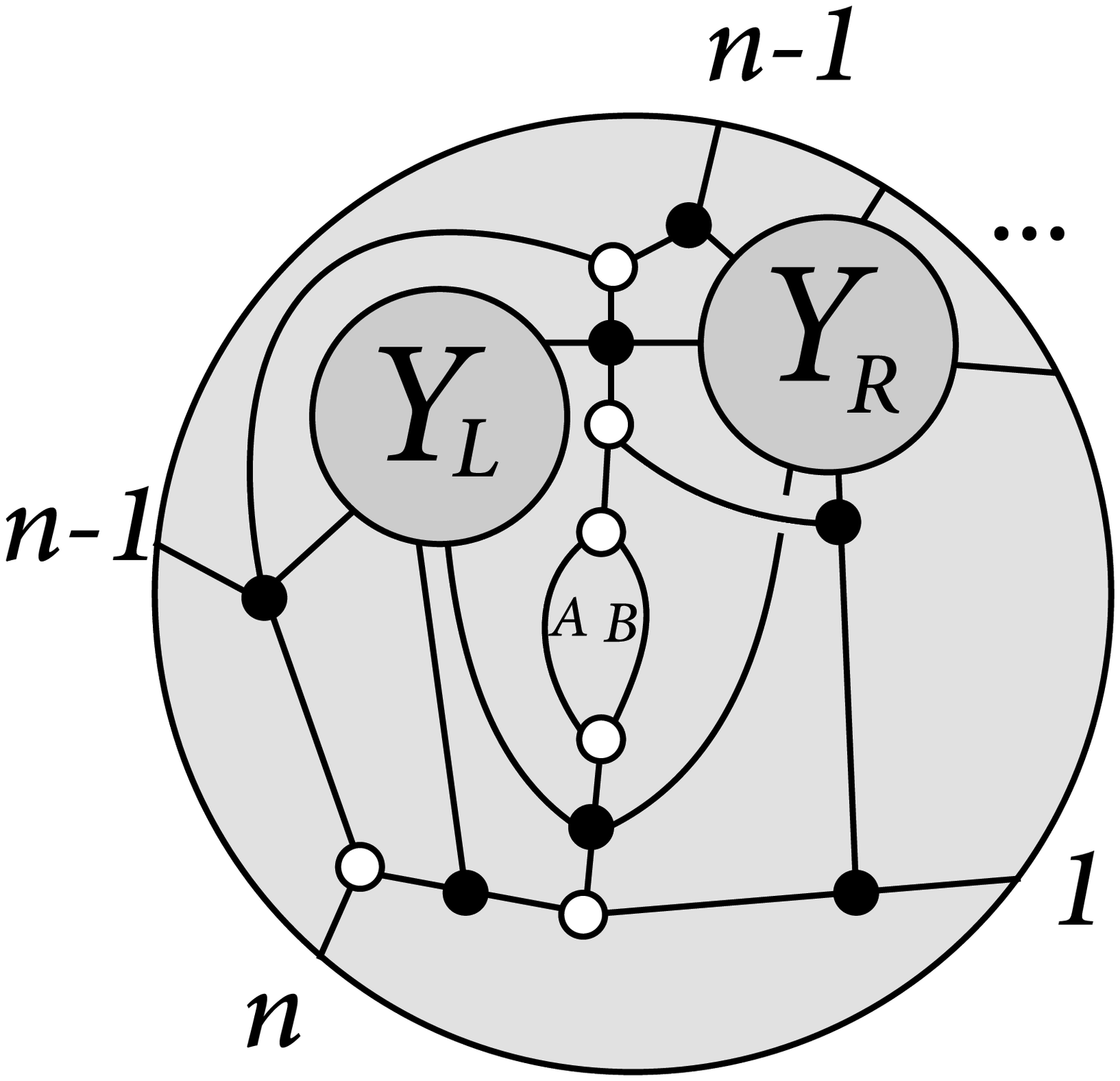}}}
\ee\\
We notice that the process of doing the $GL(2)$ integral introduces one degree of non-planarity in the diagram. In other words, two of the bridges appearing in the diagram intersect. Although this may seem peculiar, we can still do boundary measurements in the usual way. Furthermore, we see that the loop variables $\Z_A,\Z_B$ have been isolated in a bubble-like structure. This will be very convenient for writing down the loop integrand, as we will show in a moment.\\

\subsection{One-loop amplitudes}

Our diagrams at loop level are not only conceptually interesting, but they also serve as a powerful tool for computing loop integrands. It is obvious that the calculation in momentum twistor space is more efficient than in the original space. Besides, an important advantage of our diagrams is that, it bypasses the technical difficulties of performing GL$(2)$ integrals in the forward limit, and one can directly read off the result \textit{algebraically} from the diagrams. We will first derive general one-loop integrands, which yields the Kermit representation~\cite{Bourjaily:2013mma}, and then move to give results for the two-loop case. 
\subsubsection{One-loop MHV amplitudes}

We now derive the BCFW representation of the one-loop $n$-point MHV integrand. In this case, only the B and FL-FAC terms contribute, where the FL-FAC involves a factorization into two MHV trees. It follows that

\be
\text{FL-FAC} = \sum_{j=3}^{n-1} \;\;\;\vcenter{\hbox{\includegraphics[width=4.5cm]{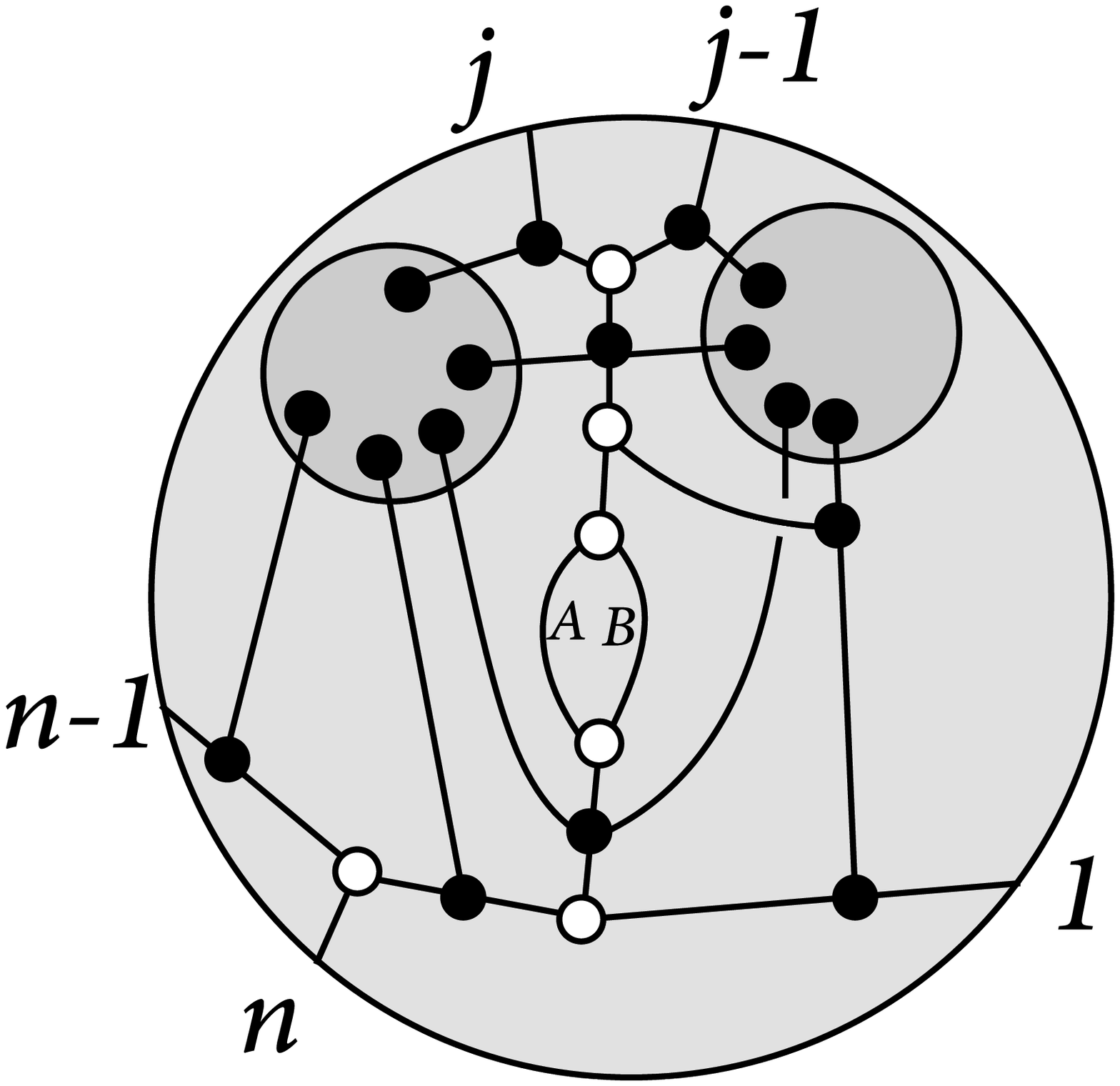}}} = \;\;\;\sum_{j=3}^{n-1}\;\;\; \vcenter{\hbox{\includegraphics[width=4.5cm]{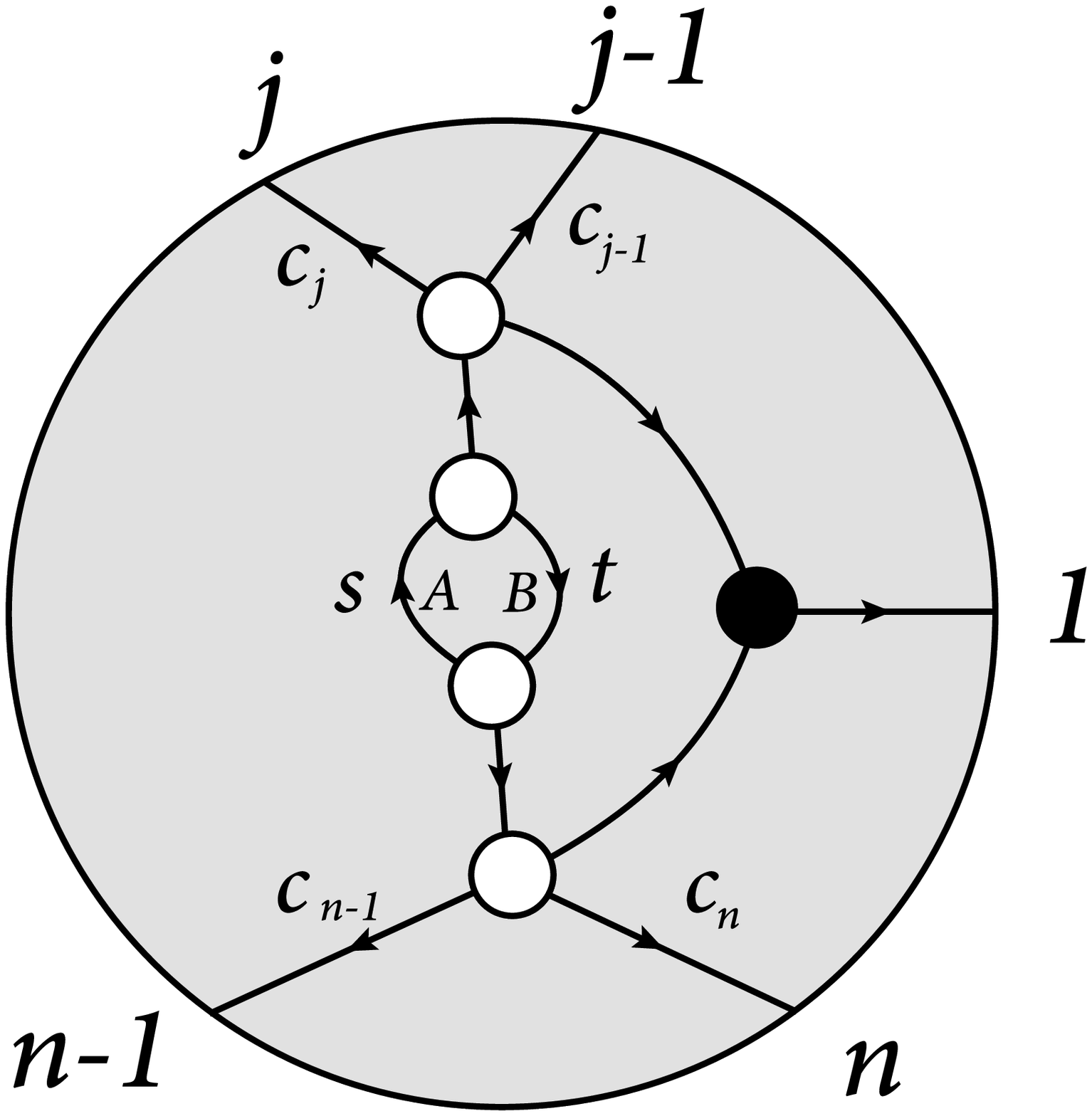}}}
\ee\\

In the boundary case where $j = n{-}1$, we identify the external lines $j$ and $n{-}1$ as follows.
\be
\;\;\; \vcenter{\hbox{\includegraphics[width=4.5cm]{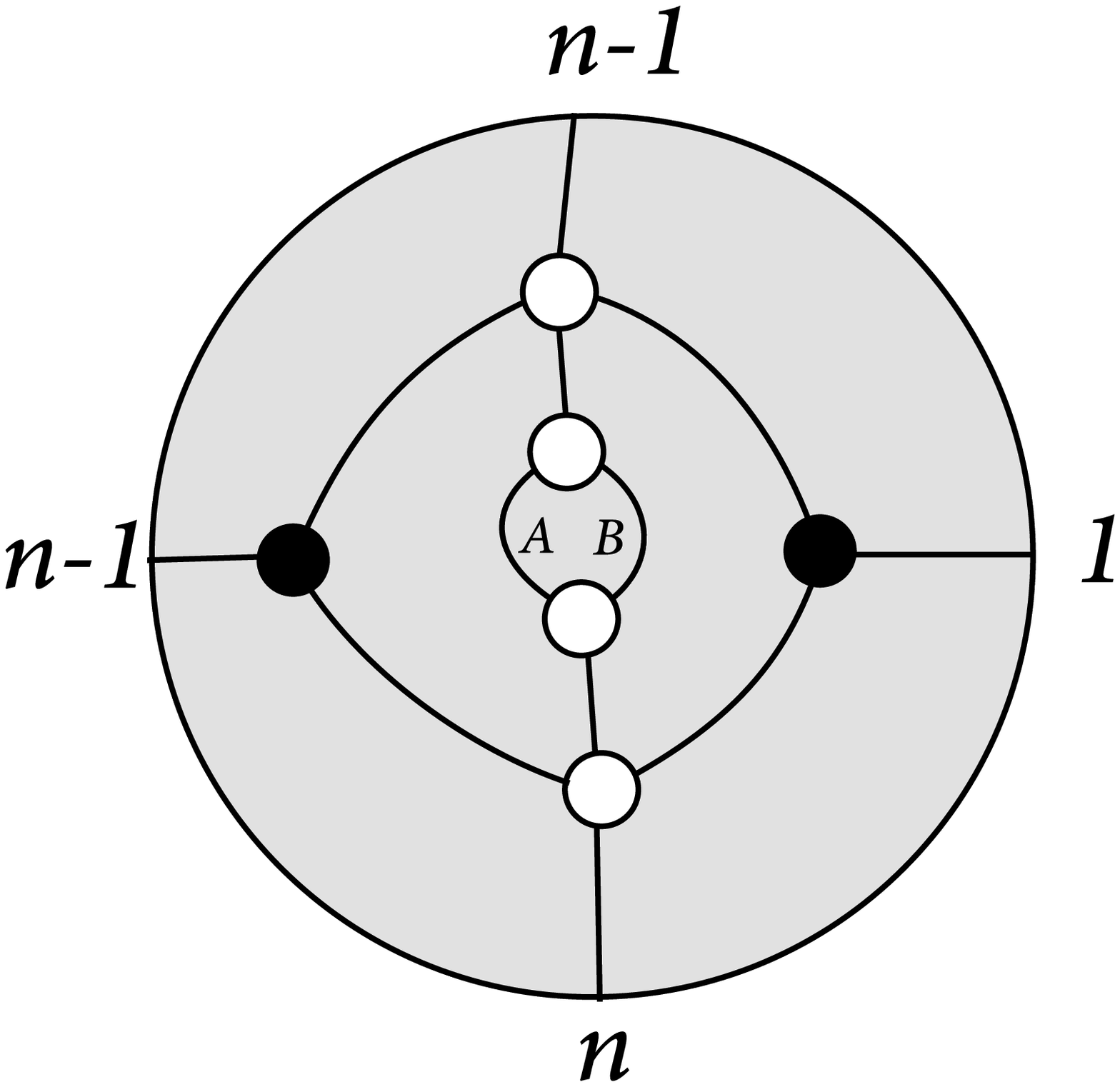}}}
\ee

Any one of these diagrams contains $n_F=7$ faces, and so must involve $n_F-1 = 6$ integration variables. As a general rule, we will always attach two extra $GL(1)$ gauges for every forward limit we take, so this reduces the diagram to a 4-form. But let us keep all 6 variables for now. There are no delta functions since the $k$-charge of this diagram is 0 (i.e. there are no arrows going into the diagram). So this diagram is literally given by
\be
\frac{ds}{d}\frac{dt}{t}\;\frac{d c_{j{-}1}} {c_{j{-}1}} \frac{d c_j}{c_j} \frac{d c_{n{-}1}}{c_{n{-}1}} \frac{d c_n}{c_n}
\ee
By following the arrows in the diagram and doing the usual boundary measurements, we can rewrite $c_1,...,c_4$ in terms of $\Z_A,\Z_B$.
\be
\Z_A + t \Z_B &=& \Z_1 + c_{j-1} \Z_{j-1}+c_j \Z_j \nl
\Z_B + s\Z_A &=& \Z_1 + c_{n-1} \Z_{n-1} + c_n \Z_n
\ee
It follows that
\be
\text{FL-FAC} = \frac{1}{\text{Vol}[GL(1)]^2}\frac{ds}{s}\frac{dt}{t}\;d^4\ell_{AB} \sum_{j=3}^{n-1}  K_{AB}(1,n{-}1,n;1,j{-}1,j)
\ee
where we have abbreviated $d^4\ell_{AB} = \left<AB d^2A\right>\left<AB d^2 B\right>$, and we define the Kermit
\be\label{1loopkerm}
K_{AB}(1,n{-}1,n;1,j{-}1,j)\equiv \frac{\left<AB(1,n{-}1,n)\cap (1,j{-}1,j)\right>^2}{\left<AB1\;n{-}1\right>\left<AB1\;n\right>\left<AB\; n{-}1\;n\right>\left<AB1\;j{-}1\right>\left<AB1\;j\right>\left<AB\; j{-}1\;n\right>}\nl
\ee
We can now gauge fix $s,t\rightarrow 1$ to obtain
\be
\text{FL-FAC} = d^4 \ell_{AB} \sum_{j=3}^{n-1}  K_{AB}(1,n{-}1,n;1,j{-}1,j)
\ee
This last step is universal, so we will often omit writing the $d\log s$ and $d\log t$ factors. Thus the full BCFW recursion for one-loop MHV gives
\be
Y_{n,k=0}^{L=1}(\Z_1,...,\Z_n) &=& \text{B} + \text{FL-FAC}\nl
&=&Y_{n-1,k=0}^{L=1}(1,...,n{-}1)+d^4\ell_{AB}\sum_{j=3}^{n-1}  K_{AB}(1,n-1,n;1,j-1,j)\nl
\ee
Solving this relation in closed form gives the well-known Kermit form of one-loop MHV,
\be
Y_{n,k=0}^{L=1}(\Z_1,...,\Z_n) = d^4\ell_{AB} \sum_{i<j} K_{AB}(1,i{-}1,i;1,j{-}1,j)
\ee

It is well known that each Kermit diagram gives $d\log c_{i{-}1}\,d\log c_i\,d\log c_{j{-}1}\,d\log c_j$, which corresponds to a cell of the one-loop MHV amplituhedron. The Kermit form nicely provides a triangulation of the amplituhedron. 

\subsubsection{General one-loop amplitudes}

In general, to all loop orders the FL-FAC term is given by Kermit terms multiplied by left and right sub-amplitudes. We can see this diagrammatically by doing boundary measurements on the middle portion of the FL-FAC diagram. The steps here are very similar to what we did for the FAC term. 

For one-loop amplitudes, in addition to B and FAC terms, only the FL-FAC term contributes, and both sub-amplitudes are tree amplitudes, which then gives the Kermit representation for one-loop integrand:

\be
\text{FL-FAC} &=& \;\;\;\vcenter{\hbox{\includegraphics[width=7cm]{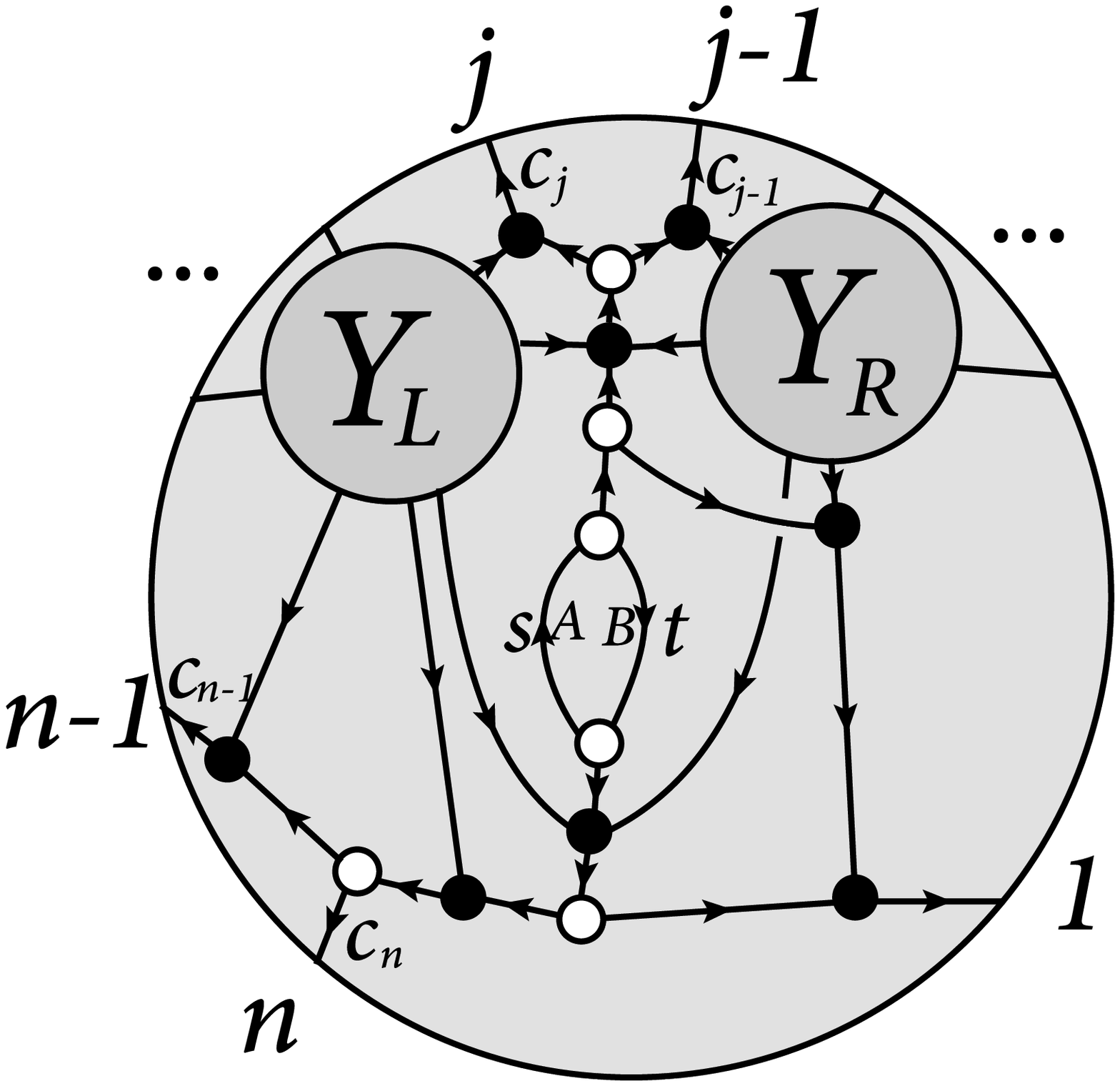}}} \nl
&=&
\sum_{j=3}^{n-1}\frac{dc_{j-1}dc_jdc_{n-1}dc_{n}}{c_{j-1}c_{j}c_{n-1}c_n}\,Y_{n_L,k_L}^{L_L}(\hat{\Z}_j,\Z_j,...,\Z_{n-1},\hat{\Z}_n,\hat{\Z}_{nAB})Y_{n_R,k_R}^{L_R}(\Z_1,...,\Z_{j-1},\hat{\Z}_j,\hat{\Z}_{nAB})\nl
\ee
where the loop variables are given by
\be
\Z_A + t\Z_B &=& \Z_1 + c_{j{-}1}\Z_{j{-}1}+c_j \Z_j\nl
\Z_B + s\Z_A &=& \Z_1+c_{n{-}1}\Z_{n{-}1}+c_n\Z_n
\ee
and $\hat{\Z}_j = (j-1,j)\cap(1,A,B)$, $\hat{\Z}_{nAB}=(A,B)\cap(1,n-1,n)$ and $\hat{\Z}_n = (n-1,n)\cap(1,A,B)$.

It is now clear that the $d\log c$ form just becomes a Kermit, so we get
\be
\text{FL-FAC} &=& 
d^4\ell_{AB}\sum_{j=3}^{n-1}K_{AB}(1,j-1,j;1,n-1,n)\nl
&&\times Y_{n_L,k_L}^{L_L}(\hat{\Z}_j,\Z_j,...,\Z_{n-1},\hat{\Z}_n,\hat{\Z}_{nAB})Y_{n_R,k_R}^{L_R}(\Z_1,...,\Z_{j-1},\hat{\Z}_j,\hat{\Z}_{nAB})
\ee

Before turning to higher loops, here we present an example, the one-loop five-point NMHV integrand, $Y_{5,k=1}^{L=1}$, in full details. The only contributions in this case are FAC and FL-FAC. The FAC term involves only one diagram and is given by the factorization with 4-point 1-loop MHV on the left and 3-point MHV tree on the right.

\be
\text{FAC} = \;\;\;\vcenter{\hbox{\includegraphics[width=5cm]{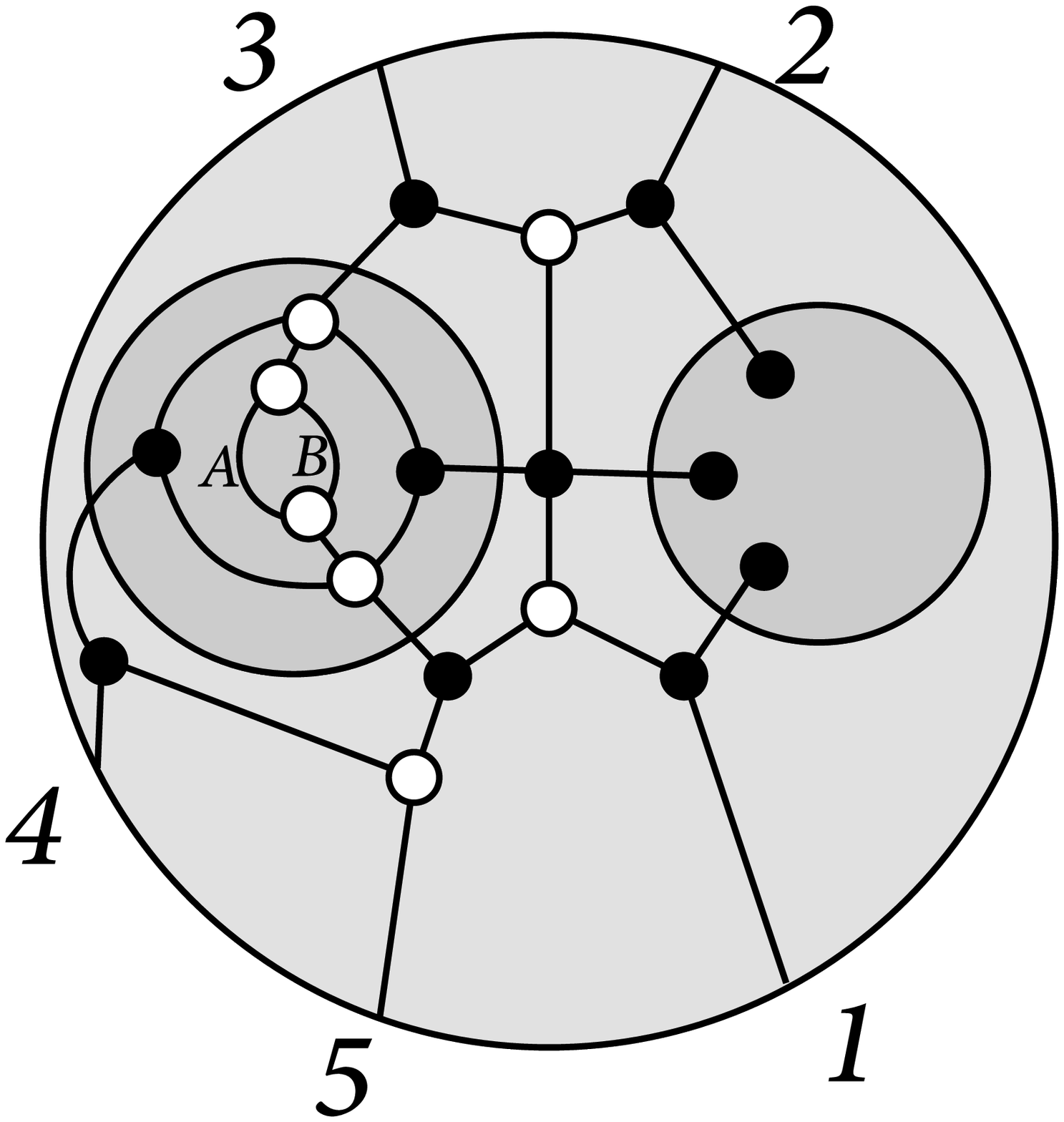}}} \;\;\;= \;\;\; \vcenter{\hbox{\includegraphics[width=5cm]{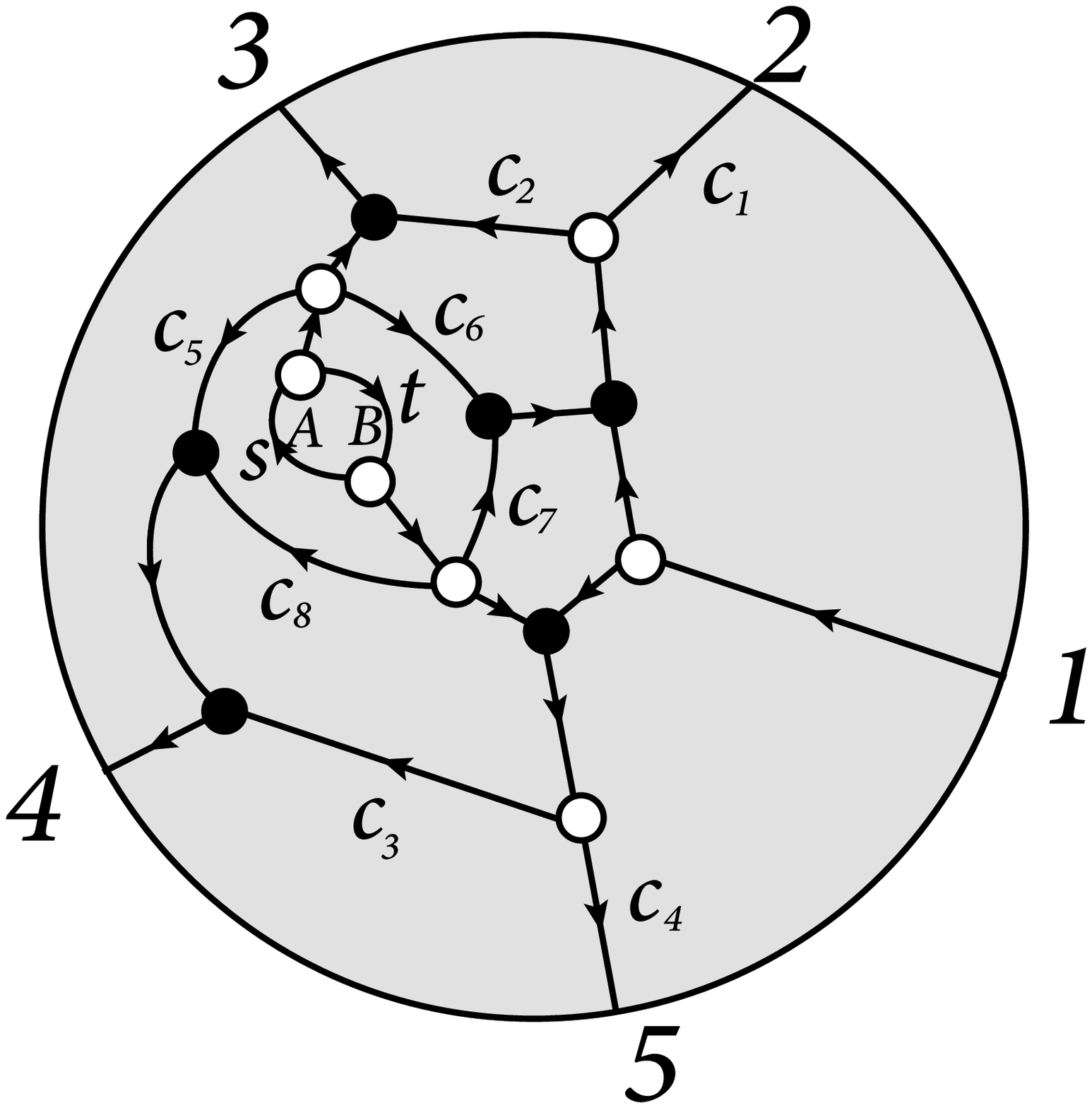}}}
\ee\\

This graph has $n_F=11$ faces, so it should have $n_F -1 = 10$ integration variables. But remember that we mod out the $s,t$ using two $GL(1)$ gauges so we really only have 8 integrations. There is only one arrow going into the diagram, which means that the diagram is NMHV, so there should be one delta function $\delta^{4|4}(...)$. Indeed, the diagram is given by

\be
\text{FAC} = \int\frac{dc_1\;...\; dc_8}{c_1\;...\;c_8}\delta^{4|4}(\Z_1-c_1 \Z_2-c_2 Z_3-c_3 \Z_4-c_4 \Z_5)
\ee

Doing the boundary measurements for the loop variables gives
\be
\Z_A + t \Z_B &=& \Z_3 + c_5 \Z_4 + c_6 (c_1 \Z_2 + c_2 \Z_3)\nl
\Z_B + s \Z_A &=& (c_3 \Z_4 + c_4 \Z_5) + c_7 (c_1 \Z_2 + c_2 \Z_3)+c_8 \Z_4
\ee

On the support of the delta function, we see that
\be
c_1\Z_2 + c_2 \Z_3 &\sim& (23)\cap(145) \equiv \Z_2'\nl
c_3 \Z_4 + c_4 \Z_5 &\sim& (45)\cap(123) \equiv \Z_4'
\ee

After rescaling some of the integration variables we get
\be
\Z_A + t \Z_B &=& \Z_3 + c_5 \Z_4 + c_6 \Z_2'\nl
\Z_B + s \Z_A &=& \Z_4' + c_7 \Z_2'+ c_8 \Z_4
\ee

This looks just like the Kermit, so the $c_5,...,c_8$ part of the form would just give a Kermit. The remaining variables $c_1,...,c_4$ can be integrated trivially over the delta function to yield the R-invariant $[1,2,3,4,5]$. The result is
\be
\text{FAC} = d^4\ell_{AB} K_{AB}(432'; 44'2')[1,2,3,4,5]
\ee
which simplifies to
\be
\text{FAC} = d^4\ell_{AB} \frac{\delta^{0|4}(\eta_1 \left<2345\right>+\eta_2 \left<3451\right>+\eta_3 \left<4512\right>+\eta_4 \left<5123\right>+\eta_5 \left<1234\right>)}{
\left<1245\right>\left<1235\right>\left<AB23\right>\left<AB34\right>\left<AB45\right>\left<AB1(45)\cap(123)\right>}
\ee

Now let us work on the FL-FAC term. There are two diagrams FL-FAC = FL-FAC-1 + FL-FAC-2. The first diagram FL-FAC-1 is the forward limit of a factorization channel with 5-point NMHV tree on the left and 4-point MHV tree on the right.

\be
\text{FL-FAC-1} = \;\;\;\vcenter{\hbox{\includegraphics[width=5cm]{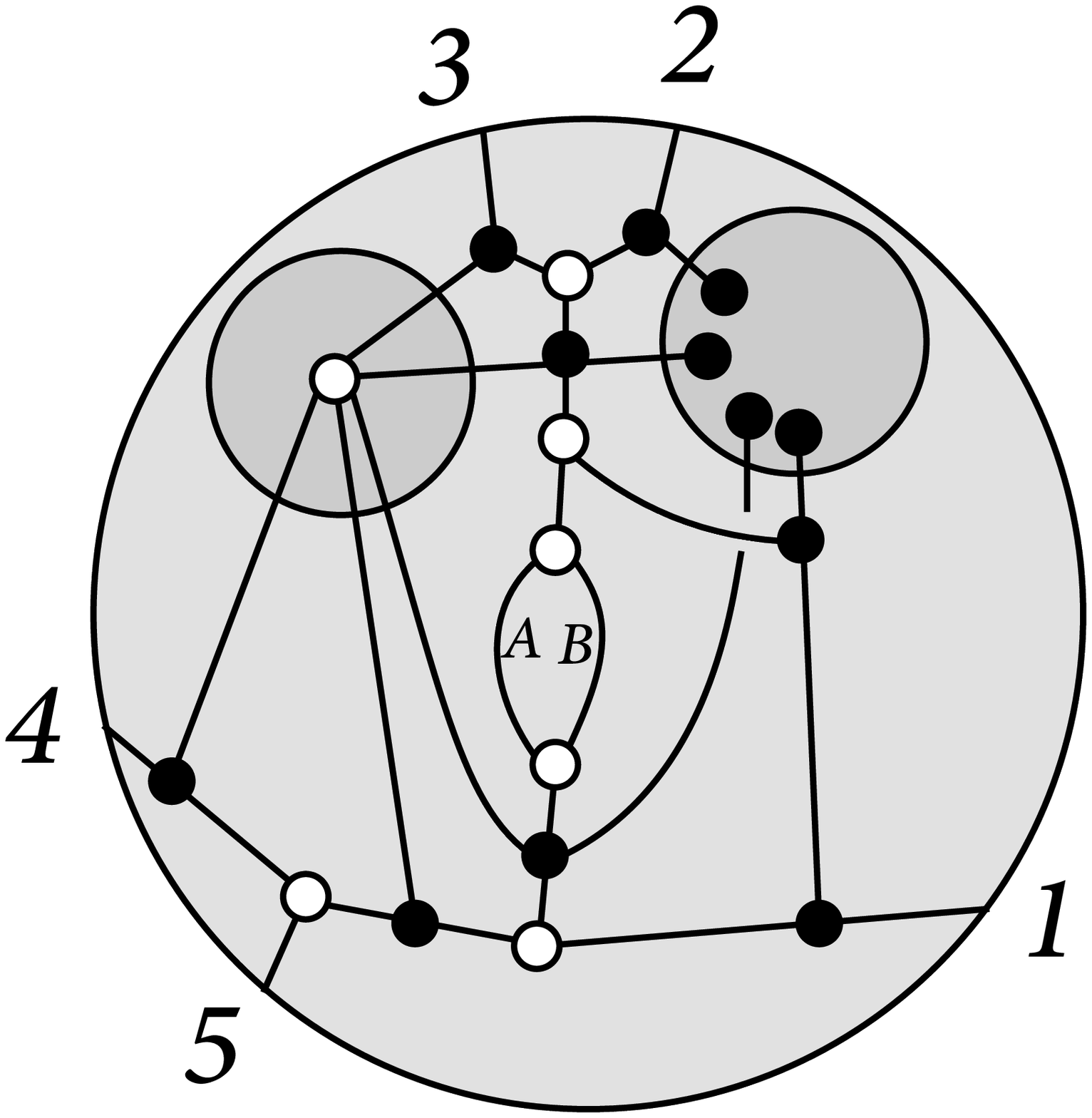}}} \;\;\;= \;\;\; \vcenter{\hbox{\includegraphics[width=5cm]{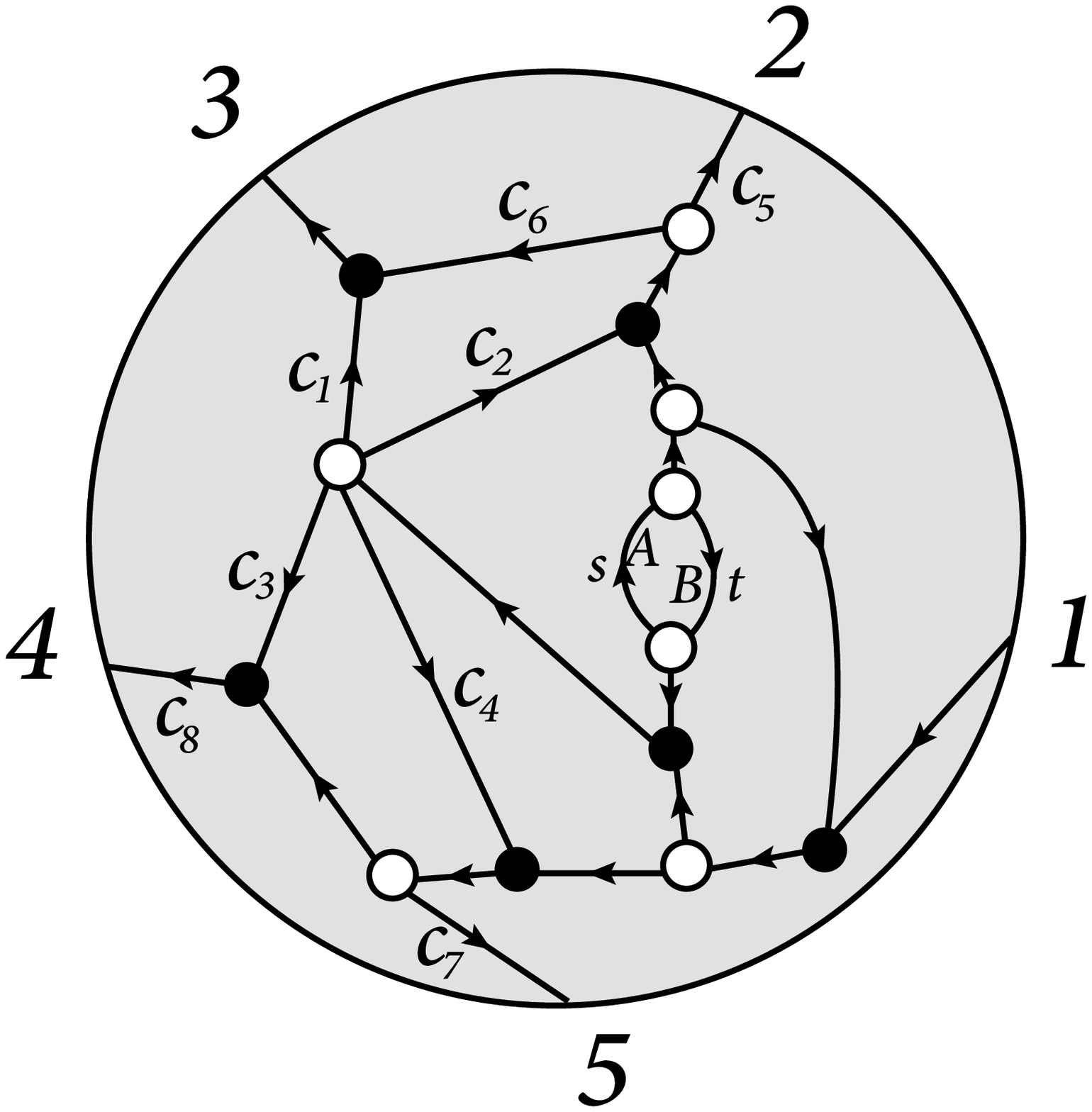}}}\nl
\ee

Again, this diagram has $n_F = 11$ faces, so we should find 8 integration variables and one delta function $\delta^{4|4}(...)$. We find

\be
\text{FL-FAC-1} = \int \frac{dc_1\;...\;dc_8}{c_1\;...\;c_8}\delta^{4|4}((\Z_1 - c_7 \Z_5 - c_8 \Z_4) - \xi)
\ee

The loop variables are given by
\be
\Z_A + s \Z_B &=& c_5 \Z_2 + c_6 \Z_3+ (c_7 \Z_5 + c_8 \Z_4 + \xi) \nl
\Z_B + t \Z_A &=& \xi
\ee

where
\be
\xi = c_1 \Z_3+ c_2 (c_5 \Z_2 + c_6 \Z_3) + c_3 c_8 \Z_4  + c_4 (c_7 \Z_5 + c_8 \Z_4)
\ee

On the support of the delta function, we can rewrite the loop variables as
\be
\Z_A + s \Z_B &=& \Z_1 + c_5 \Z_2 + c_6 \Z_3 \nl
\Z_B + t \Z_A &=& \Z_1 - c_7 \Z_5 - c_8 \Z_4
\ee

Using these new loop variable expressions, we find the following shifts
\be 
\Z_1 - c_7 \Z_5-c_8 \Z_4 &\sim& (AB)\cap (145) \equiv \Z_A'' \nl
c_7 \Z_5 + c_8 \Z_4 &\sim& (45)\cap(1AB) \equiv \Z_4''\nl
c_5\Z_2+c_6 \Z_3 &\sim& (23)\cap (1AB) \equiv \Z_2''
\ee
Substituting these into the delta function and rescaling the integration variables appropriately gives us
\be
\text{FL-FAC-1} = \int \frac{dc_1\;...\;dc_8}{c_1\;...\;c_8}\delta^{4|4}(\Z_A''+c_1\Z_3 + c_2 \Z_2'' + c_3 \Z_4 + c_4 \Z_4'')
\ee

Like before, the loop variables take on the Kermit form, so the $c_5,...,c_8$ part of the form just gives us a Kermit, and the remaining integrals can be integrated over the delta function to give an R-invariant. The result is
\be
\text{FL-FAC-1} &=& d^4\ell_{AB}K_{AB}(123;145)[3,4,A'',4'',2'']\nl
&=& d^4\ell_{AB}\frac{\delta^{0|4}(\eta_1 \left<2345\right>+\eta_2 \left<3451\right>+\eta_3 \left<4512\right>+\eta_4 \left<5123\right>+\eta_5 \left<1234\right>)}{\left<2345\right>\left<AB12\right>\left<AB23\right>\left<1345\right>\left<AB15\right>\left<AB4(15)\cap(234)\right>}\nl
\ee

Finally, the last term FL-FAC-2 is given as the forward limit of the factorization with 4-point MHV tree on the left and 5-point NMHV tree on the right. We just give the result here
\be
\text{FL-FAC-2} = \;\;\;\vcenter{\hbox{\includegraphics[width=5cm]{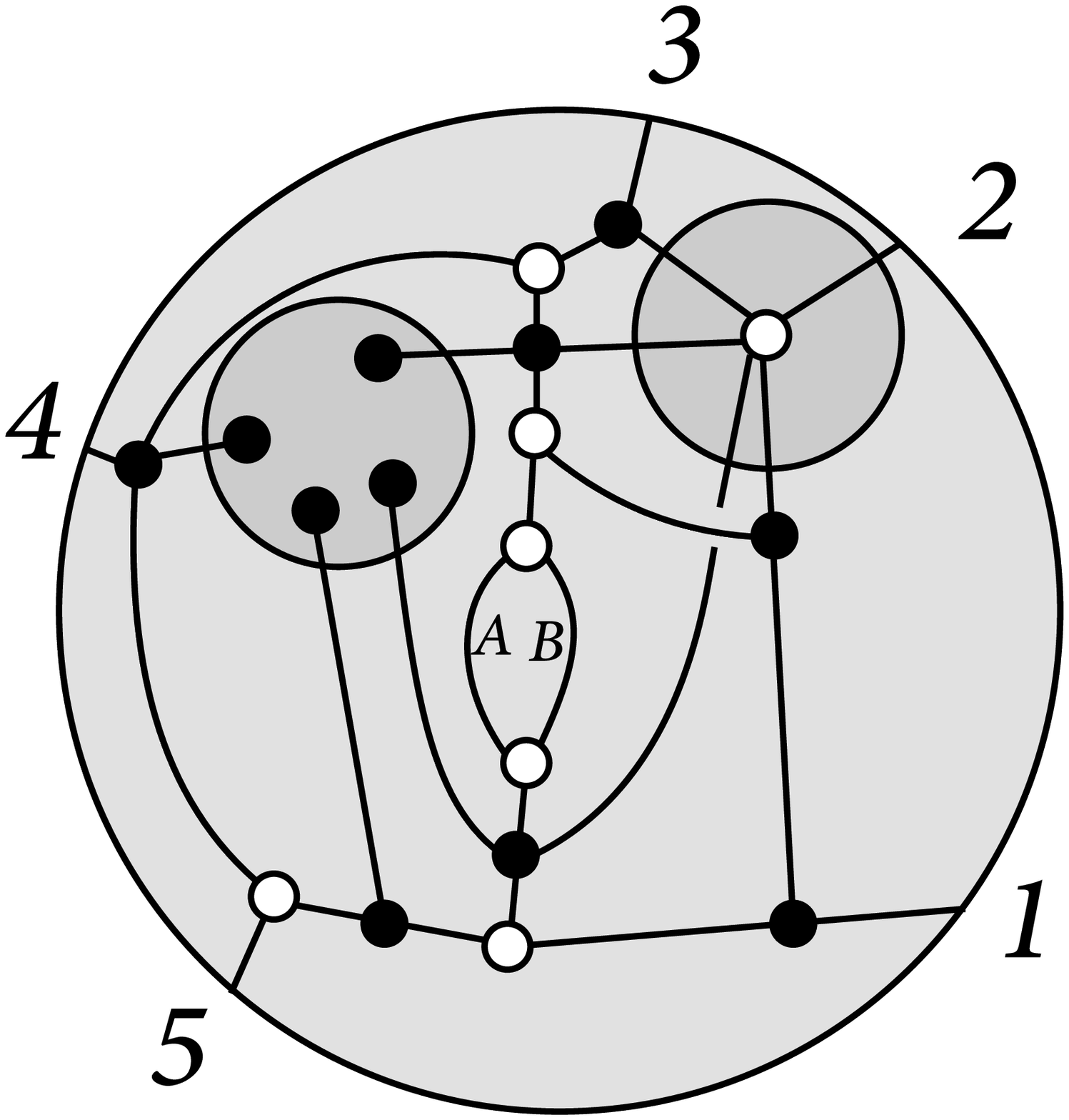}}} \;\;\;= \;\;\; \vcenter{\hbox{\includegraphics[width=5cm]{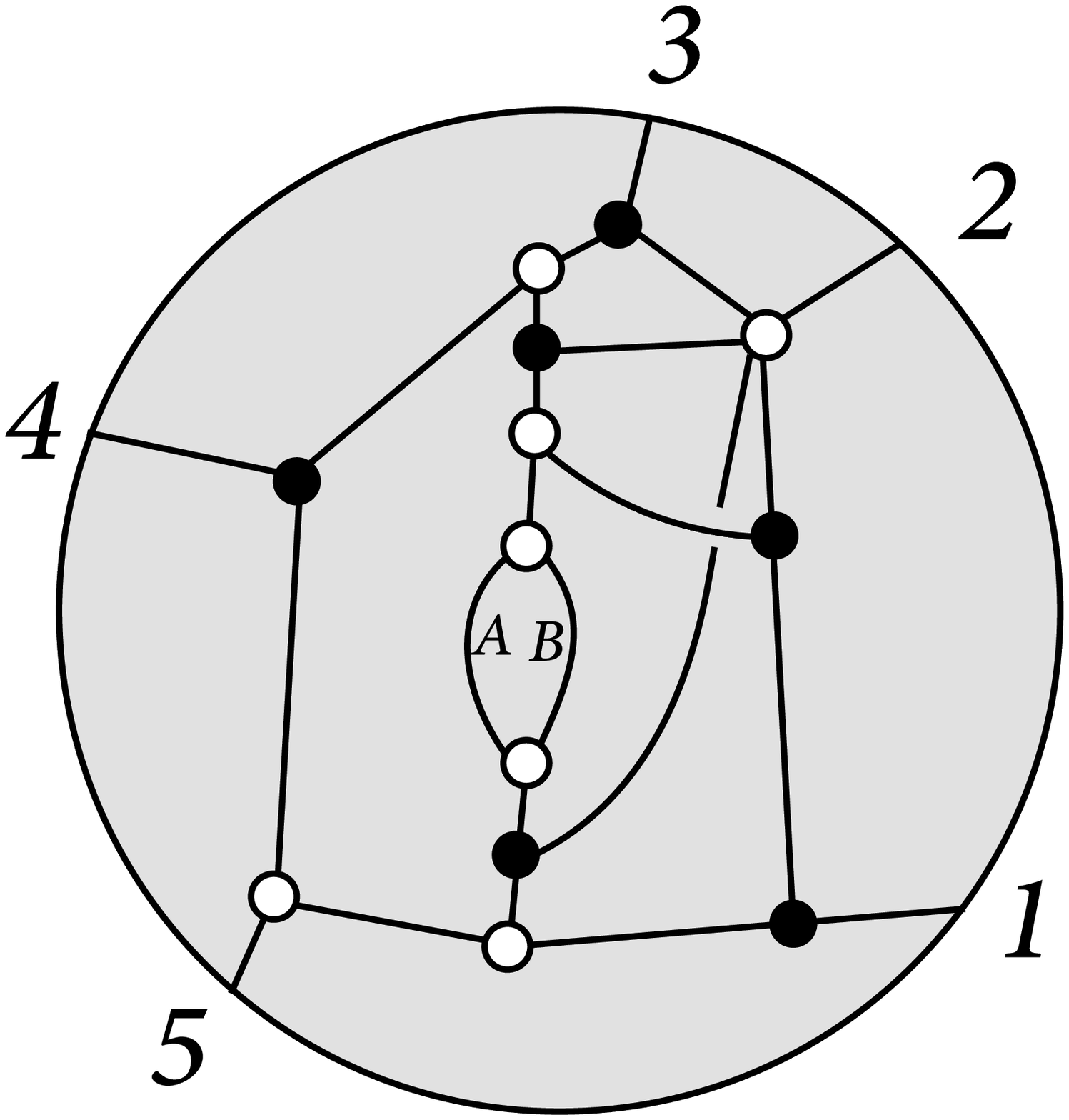}}}\nl
\ee
which is equal to
\be
\text{FL-FAC-2}&=& d^4\ell_{AB}K_{AB}(134;145)[1,2,3,(34)\cap(1AB),(AB)\cap(145)] \nl
&=&d^4\ell_{AB}\frac{\delta^{0|4}(\eta_1 \left<2345\right>+\eta_2 \left<3451\right>+\eta_3 \left<4512\right>+\eta_4 \left<5123\right>+\eta_5 \left<1234\right>)\left<AB14\right>^2}{\left<1234\right>\left<AB12\right>\left<AB34\right>\left<AB45\right>\left<AB15\right>\left<AB1(45)\cap(123)\right>\left<AB4(51)\cap(234)\right>}\nl
\ee

The final result for the 5-point 1-loop NMHV integrand is the sum of all the contributions
\be
Y_{5,k=1}^{L=1}(\Z_1,...,\Z_5) = \text{FAC} + \text{FL-FAC-1}+\text{FL-FAC-2}.
\ee
These diagrams also give the three $C,D$-matrices for three cells of the amplituhedron. It is already non-trivial to see how the three cells provide a triangulation of this one-loop five-point NMHV case. 

\subsection{Two-loop amplitudes}

At higher loops, the only new feature one would encounter is the iteration of taking forward limits, and the general structure is very clear:  there are $L$ bubbles at $L$ loops, and the contributions can be classified as those with $L$ bubbles connected by factorization ``bridges", those with a FL-FL part, \textit{i.e.} two connected bubbles, and the rest, etc. Basically it requires some bookkeepings to work out all diagrams for a given multi-loop amplitudes. 

Here we restrict ourselves to two-loop amplitudes. In addition to B and FAC, which are of the same form as before, the FL-FAC term is also of the same form as the one-loop case. The new, non-trivial contribution is the FL-FL term, which involves yet another $GL(2)$ residue. Let us call the second loop variable $CD$. We begin by drawing the diagram corresponding to the two forward limits:
\be
\text{FL-FL} = \int_{GL(2)}^{AB,CD} \;\vcenter{\hbox{\includegraphics[width=5cm]{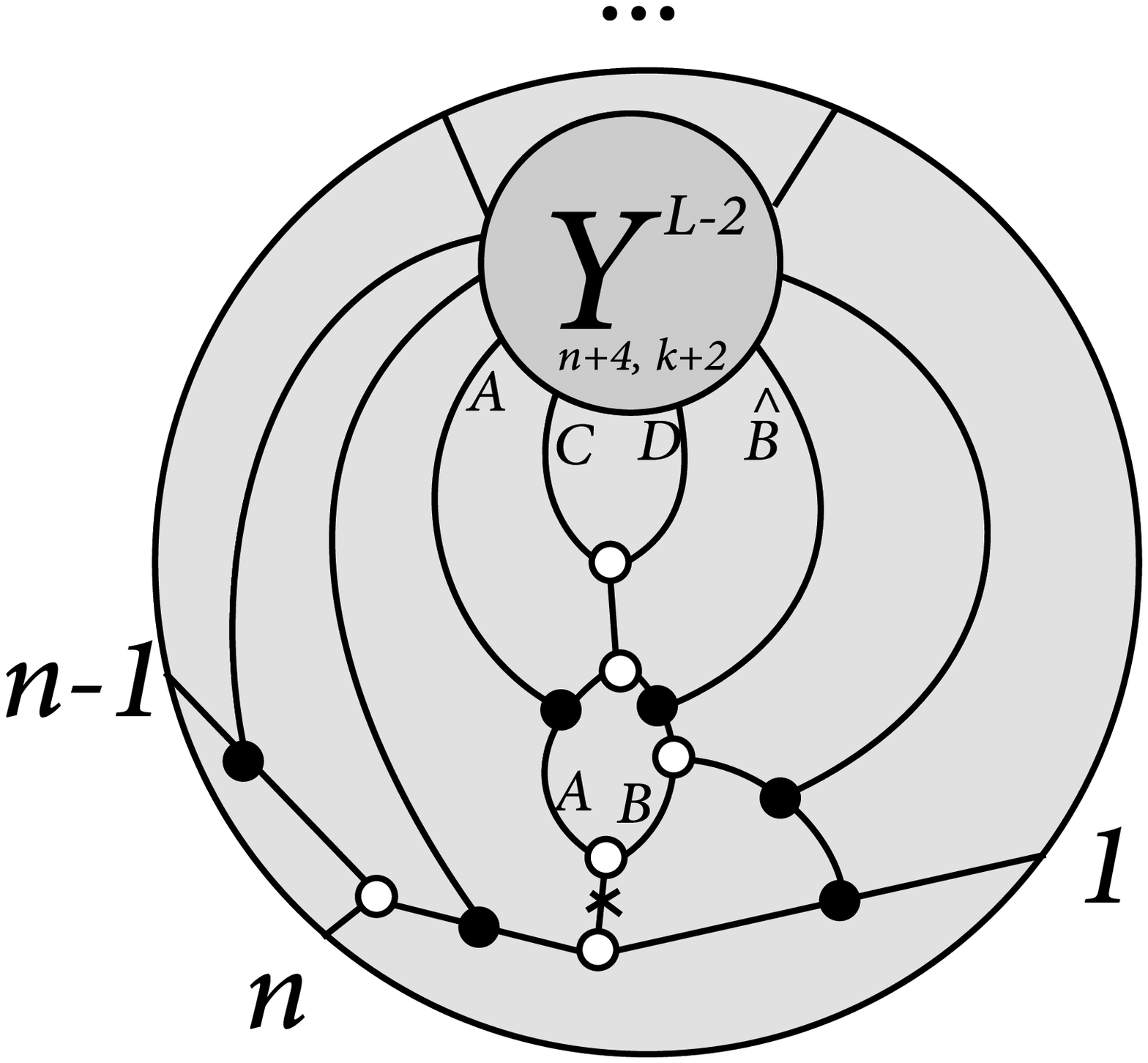}}}\;\; = \int_{GL(2)}^{CD} \;\vcenter{\hbox{\includegraphics[width=5cm]{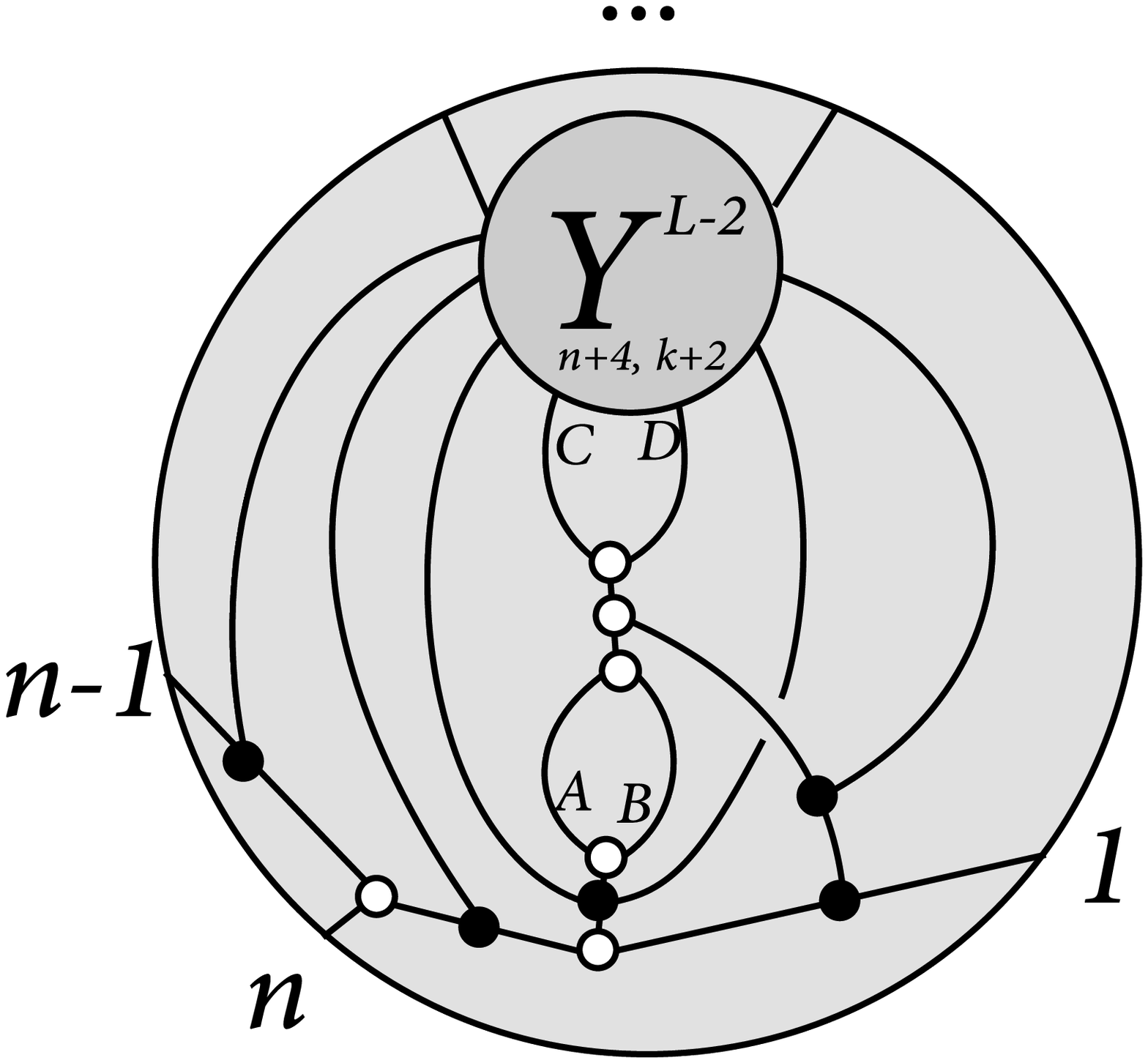}}}\nl
\ee
where the second diagram is obtained by doing the $GL(2)$ integral for $AB$. The procedure here is the same as before. Just reattach the two lines $A$ and $\hat{B}$ coming out of the sub-diagram to the crossed line. \\

Going one step further, we perform yet another BCFW shift on the sub-diagram, and we can concentrate on the FL-FL-FAC term, which is the only new contribution for two-loop amplitudes, and the form again generalizes to FL-FL-FAC term at all loops. We will use the shift $\Z_C\rightarrow \Z_C+w\Z_A$, in which case the $w\rightarrow\infty$ term vanishes in the forward limit of $CD$,
\be\label{FLFLFAC}
\text{FL-FL-FAC} =\sum_{j=2}^{n}\int_{GL(2)}^{CD} \;\vcenter{\hbox{\includegraphics[width=5cm]{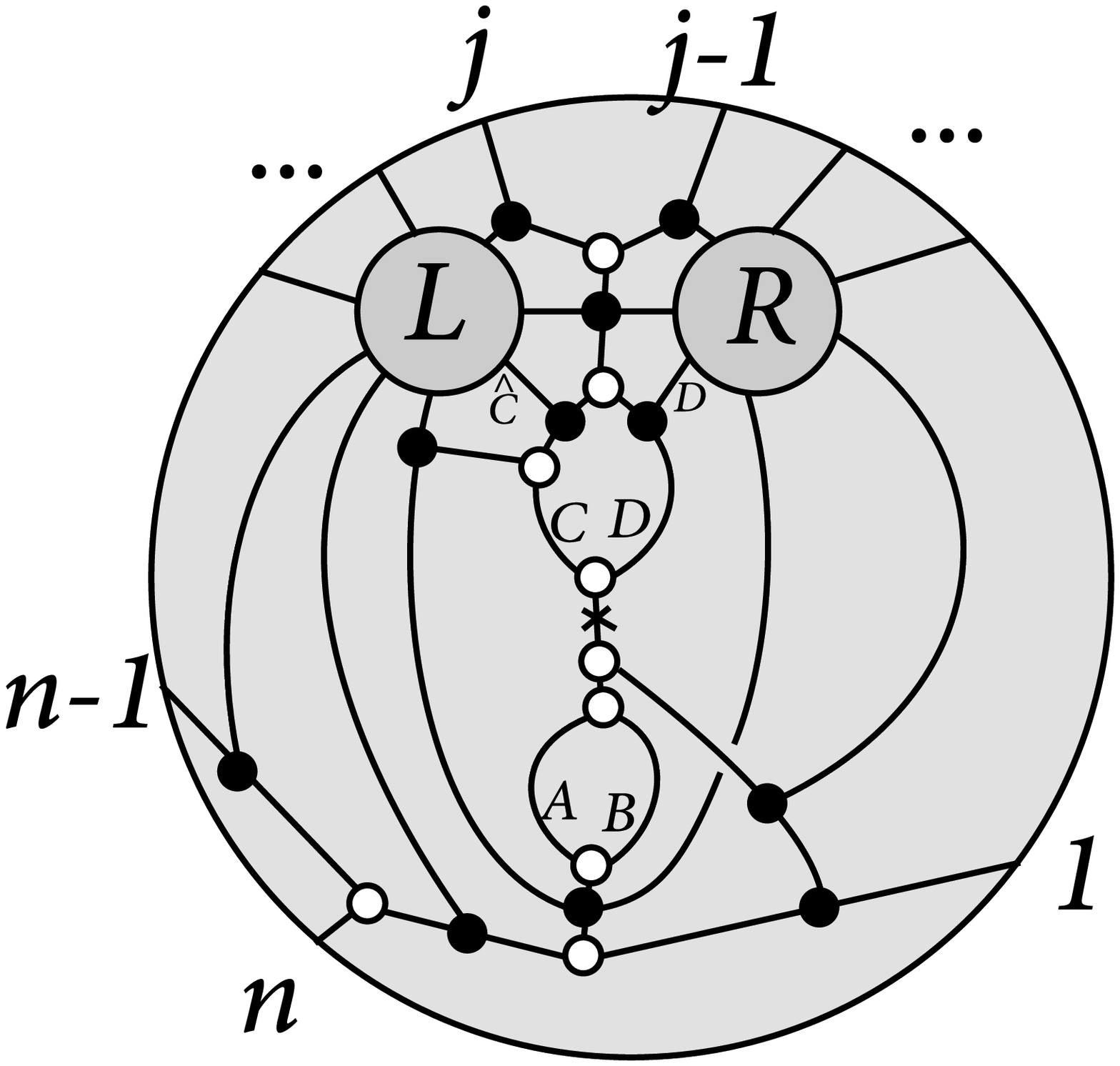}}}\;\; = \sum_{j=2}^{n}\;\vcenter{\hbox{\includegraphics[width=5cm]{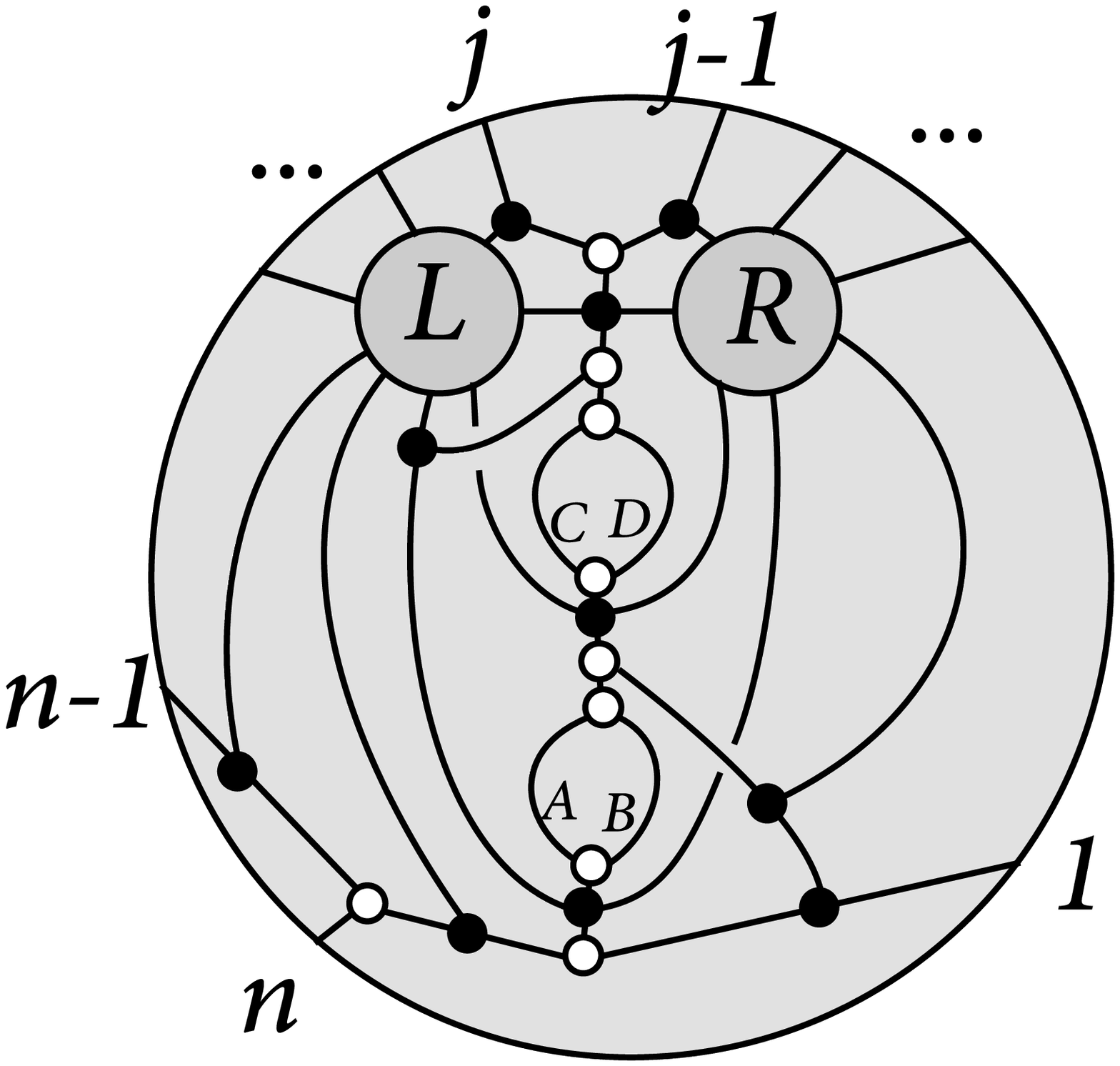}}}~.\nl
\ee\\

In the second diagram we have done the $GL(2)$ integral for $CD$ by reattaching the two lines $\hat{C}$ and $D$ coming out of the two sub-diagrams to the crossed line. Recalling that the forward limit takes $C,D\rightarrow (C,D)\cap (1,A,B)$, we see that the crossed line precisely represents this limit. We note that $k_L{+}k_R=k{+}1$ and $L_L{+}L_R = L{-}2$.\\

\subsubsection{Example: two-loop four-point MHV}

Before turning to general two-loop amplitudes, here we give the simplest example: the four-point integrand. The computation is already quite non-trivial and interesting, which shows all the essential features of our diagrammatic formulation of multi-loop integrands.

The two-loop four-point integrand is given by the forward limit of one-loop six-point NMHV result, which can be worked out similar to the one-loop five-point case above, and includes 16 terms. When applying the forward limit to it using (\ref{FLFLFAC}), we find only 8 terms are non-vanishing. There are 2 terms coming from FL-FAC and 6 terms coming from FL-FL. For each diagram we include also the momentum twistor expression (without the $d^4\ell_{AB}\,d^4\ell_{CD}$ factor), and the corresponding $D_{AB},D_{CD}$ matrices in the amplituhedron which we display in the form\\
\be 
\mathcal{D}^{(2)}\equiv \left( \begin{array}{cccc}
D_{AB}\\
D_{CD} \end{array} \right)
\ee
To write the expressions in a more compact form, we need the following shifted twistors, from either FAC and FL, or two FL's:

\be
\hat{A} &=& (A,B)\cap(1,3,4),\quad \hat{C}' = (C,D)\cap(1,A,B),\nl
\hat{3}' &=& (2,3)\cap(1,A,B),\quad \hat{4}' = (3,4)\cap(1,A,B),\nl
\hat{2} &=& (1,2)\cap(\hat{A},C,D),\quad \hat{3} = (2,3)\cap(\hat{A},C,D), \quad \hat{4} = (3,4)\cap(\hat{A},C,D),
\ee

So far we have not tried to make the cells positive. In general, boundary measurements do not guarantee positivity. In what follows, however, we have adjusted the signs of some of the bridge variables so as to make the cells positive. We put all the 8 variables on the edges of each diagram, and the matrix can be read off from the boundary measurement from $A,B,C,D$ to external legs. In addition, each expression, multiplied by $d^4\ell_{AB}\,d^4\ell_{CD}$, is given by the $d\log$'s of the 8 variables.\\

\begin{flalign}
&\vcenter{\hbox{\includegraphics[width=5cm]{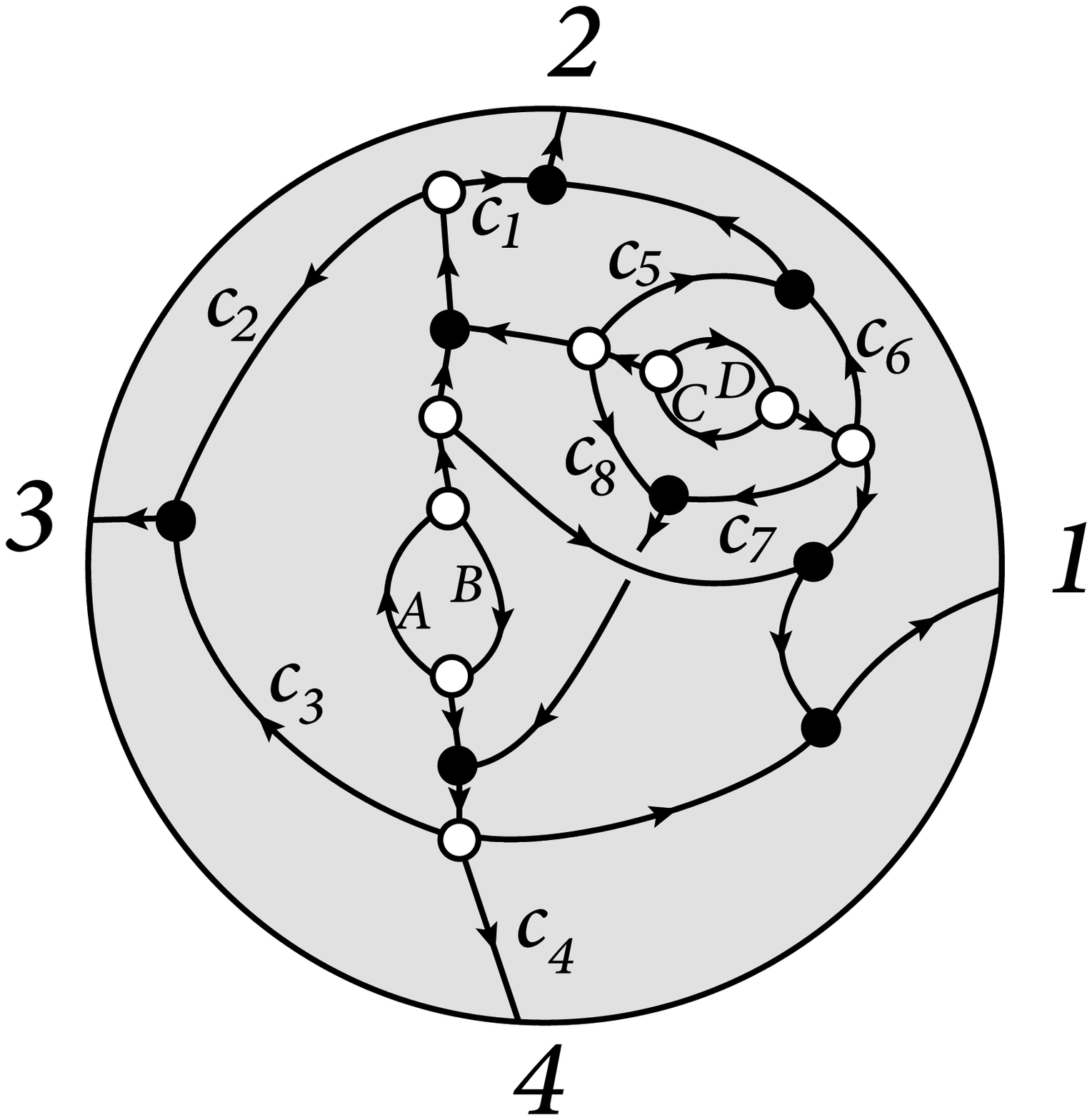}}}\hspace*{1cm}
\left( \begin{array}{cccc}
-1 & -c_1 & -c_2 & 0  \\
1 & 0 & -c_3 & -c_4 \\
c_8 & -c_1-c_5 & -c_2-c_3 c_8 & -c_4 c_8 \\
1+c_7 & c_6 & -c_3 c_7 & -c_4 c_7 
\end{array} \right)&
\end{flalign}\\

\be
\text{FL-FAC-1} = \frac{\la 1234\ra^4\la AB13\ra^2}
{\la AB23\ra \la AB34\ra\la AB14\ra\la CD12\ra\la CD23\ra\la CD\hat{3}'\hat{A}\ra\la CD1\hat{A}\ra}
\ee\\

\begin{flalign}
&\vcenter{\hbox{\includegraphics[width=4.8cm]{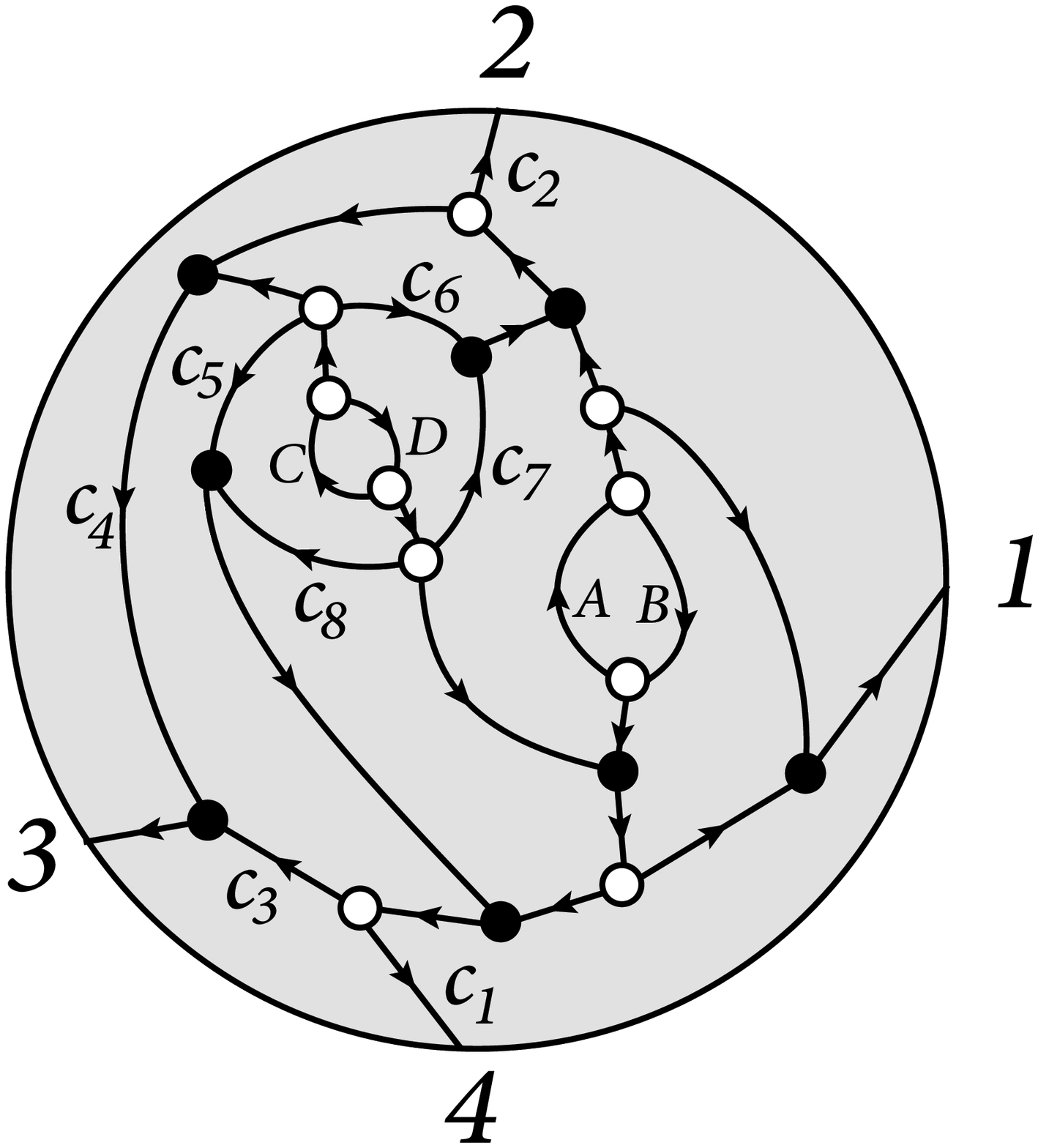}}}\hspace*{1cm}
\left( \begin{array}{cccc}
1 & c_2 & c_4 & 0  \\
-1 & 0 & c_3 & c_1 \\
0 & c_2c_6 & c_4+c_3 c_5+c_4c_6 & c_1 c_5 \\
-1 & -c_2c_7 & c_3- c_4c_7+c_3 c_8 & c_1+c_1 c_8 \end{array} \right)&
\end{flalign}\\

\be
\text{FL-FAC-2} = \frac{\la 1234\ra^4\la AB13\ra^2}
{\la AB12\ra \la AB23\ra\la AB14\ra\la CD23\ra\la CD34\ra\la CD\hat{4}' 1\ra\la CD\hat{3}'\A\ra}
\ee\\

\begin{flalign}
&\vcenter{\hbox{\includegraphics[width=4.8cm]{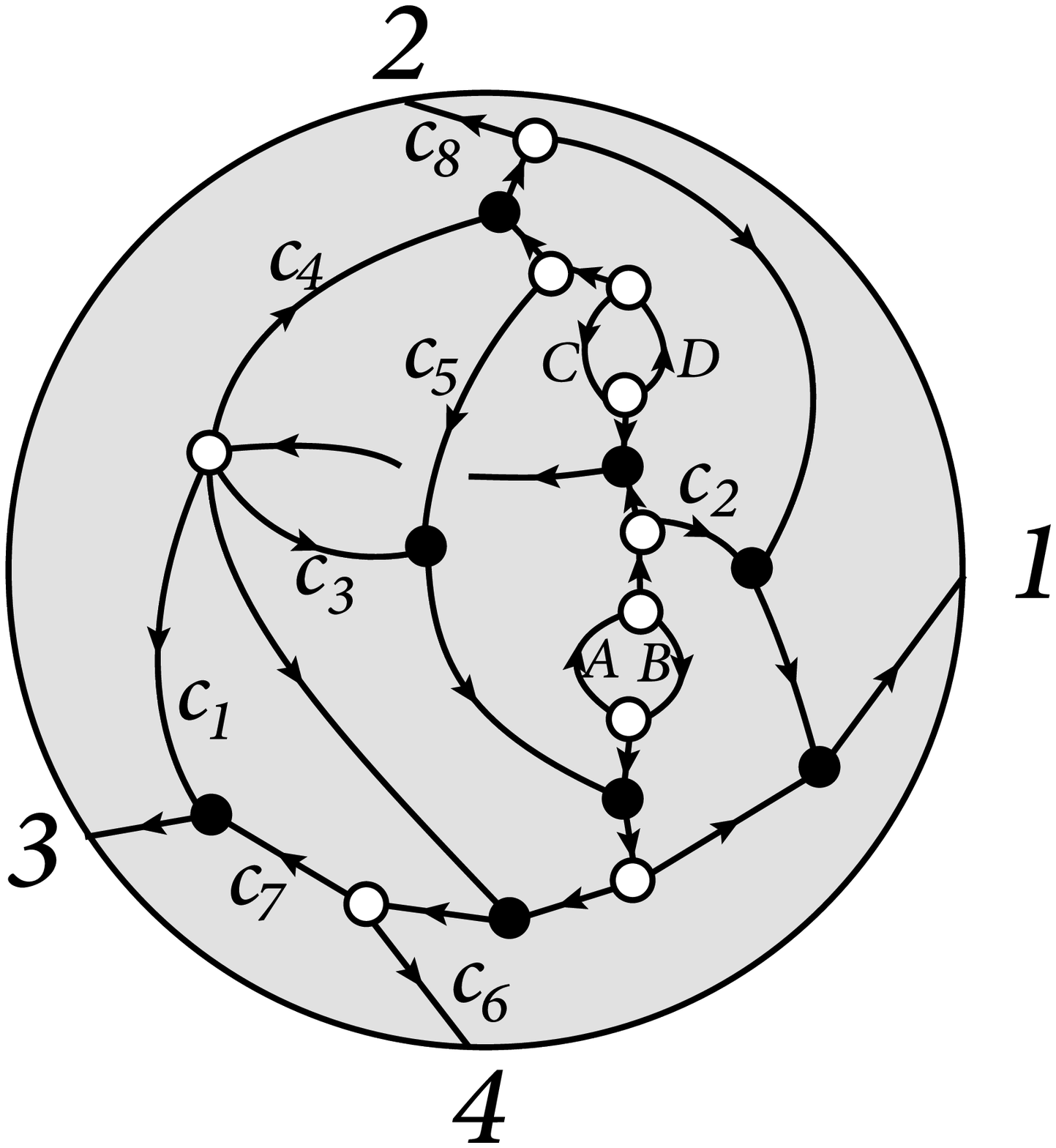}}}\hspace*{0.6cm}
\left( \begin{array}{cccc}
{-}c_2{+}c_3{-}c_4 & -c_4 c_8 & -c_1{-}c_7{-}c_3 c_7 & -c_6{-}c_3 c_6  \\
1 & 0 & -c_7 & -c_6 \\
c_3{-}c_4 & -c_4 c_8 & -c_1{-}c_7{-}c_3c_7 & -c_6{-}c_3 c_6 \\
1{+}c_5 & c_8 & -c_5 c_7 & -c_5 c_6 \end{array} \right)&
\end{flalign}\\

\be
\text{FL-FL-1} =\frac{\la AB(134)\cap(1CD)\ra^2\la\A \hat{C}' 12\ra^2\la 1234\ra^3\la AB34\ra\la CD\A 1\ra^2}
{\la AB14\ra\la ABCD\ra\la CD12\ra\la CD\A 2\ra\la \hat{C}' 134\ra\la \hat{2} \4' \hat{C}'\A\ra\la \hat{2} 3 \A \hat{C}' \ra\la \hat{C}' \hat{2} 34\ra}
\ee\\

\begin{flalign}
&\vcenter{\hbox{\includegraphics[width=4.8cm]{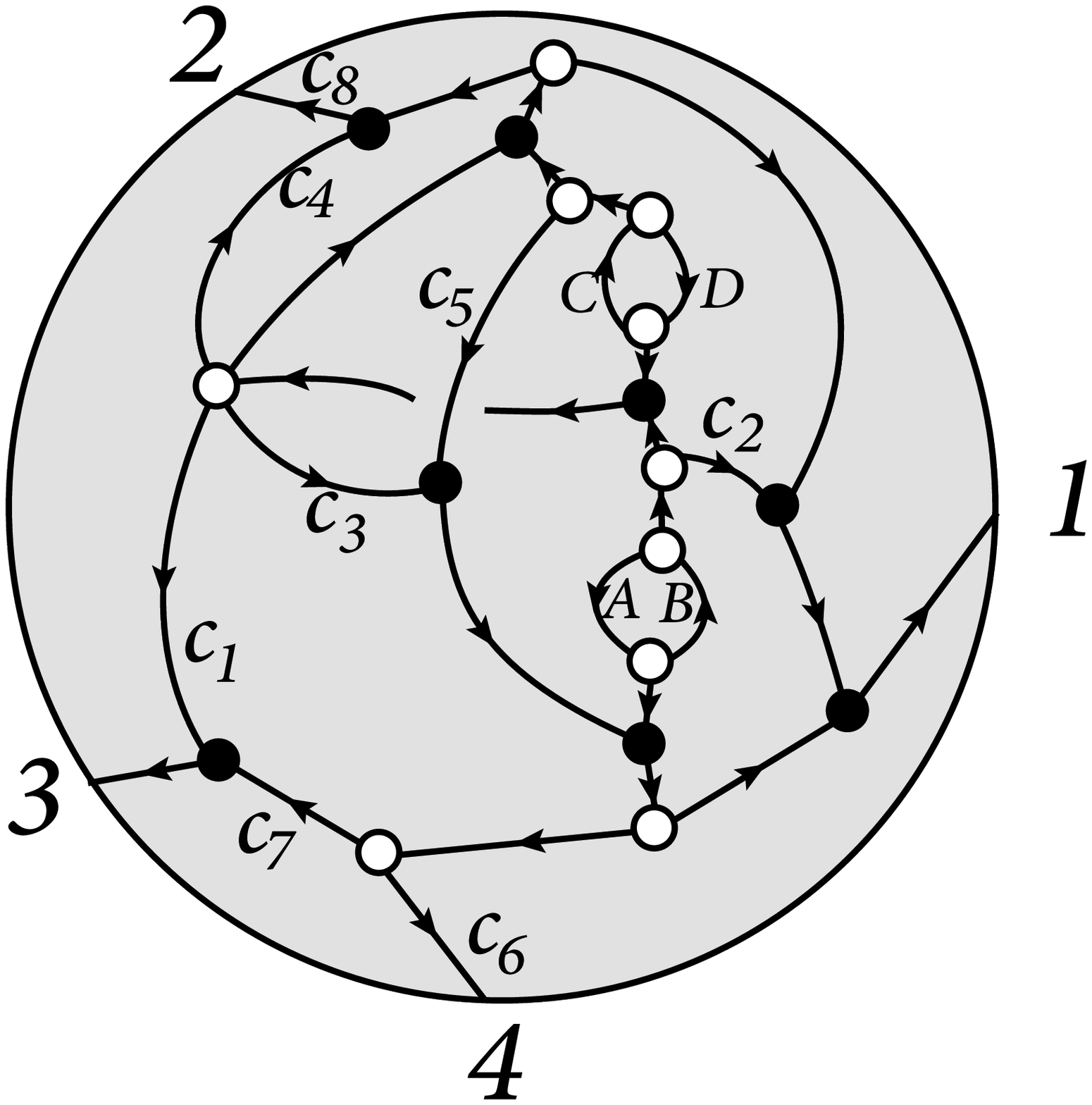}}}\hspace*{1cm}
\left( \begin{array}{cccc}
1 & 0 & -c_7 & -c_6  \\
1+c_2-c_3 & c_8+c_4 c_8 & c_1+c_3 c_7 & c_3c_6 \\
1+c_5 & c_8 & -c_5c_7 & -c_5 c_6 \\
1-c_3 & c_8+c_4 c_8 & c_1+c_3 c_7 & c_3 c_6 \end{array} \right)&
\end{flalign}\\

\be
\text{FL-FL-2} = \frac{\la AB(134)\cap(1CD)\ra^2\la \A \hat{C}' 12\ra\la 1234\ra^3\la AB13\ra^2}
{\la AB14\ra\la AB34\ra\la ABCD\ra\la CD\A 1\ra\la CD12\ra\la \A \hat{C}' 23\ra\la \A \hat{C}' 3 \hat{2}\ra\la \hat{C}' 123\ra}
\ee\\

\begin{flalign}
&\vcenter{\hbox{\includegraphics[width=4.8cm]{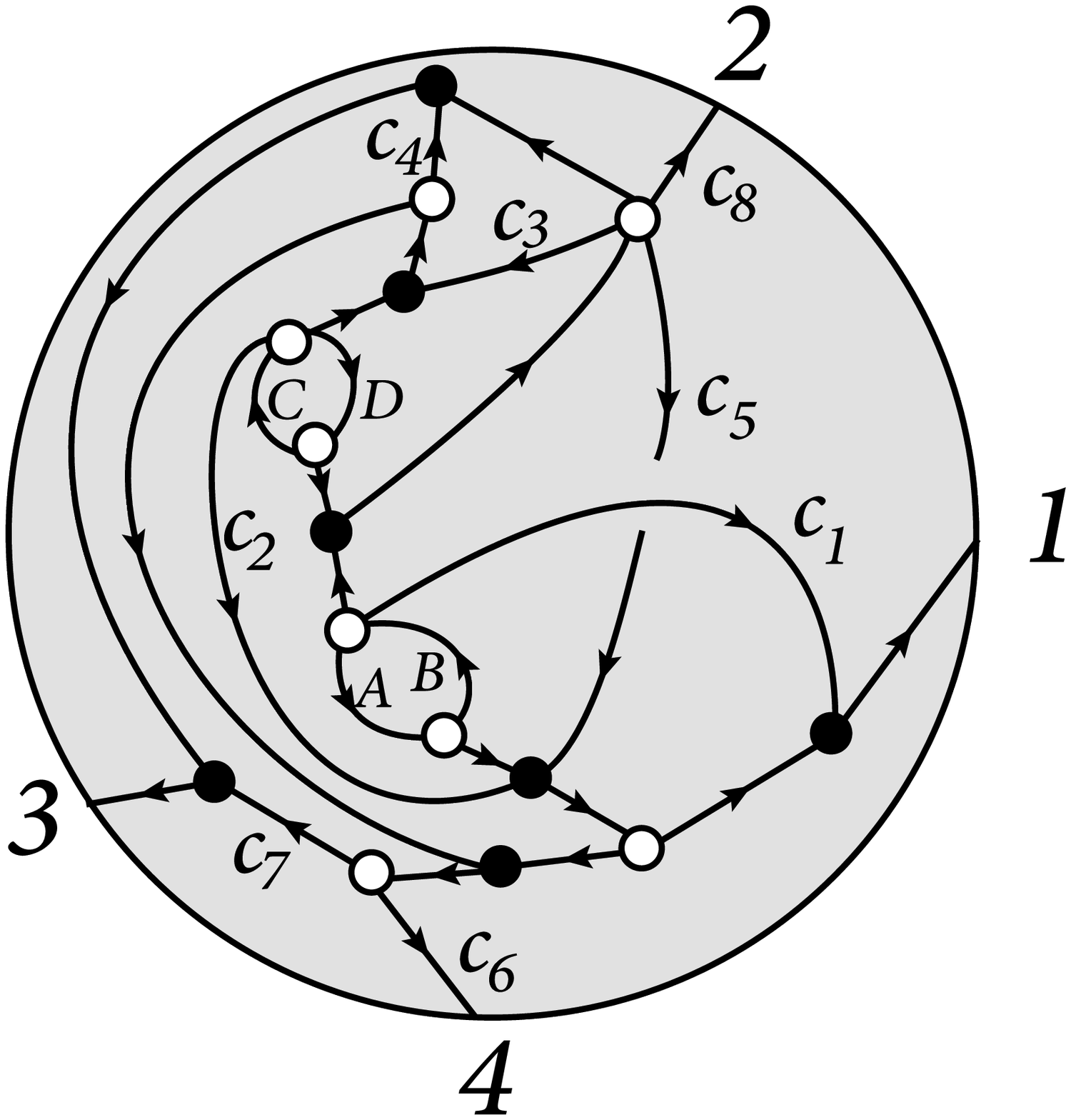}}}\hspace*{0.8cm}
\left( \begin{array}{cccc}
1 & 0 & -c_7 & -c_6  \\
c_1{+}c_5 & c_8 & 1{+}c_3 c_4{+}c_3 c_7{-}c_5 c_7 & c_3c_6{-}c_5c_6 \\
c_2 & 0 & -c_4{-}c_7{-}c_2c_7 & -c_6{-}c_2c_6 \\
c_5 & c_8 & 1{+}c_3 c_4{+}c_3 c_7{-}c_5 c_7 & c_3 c_6{-}c_5 c_6 \end{array} \right)&
\end{flalign}\\

\be
\text{FL-FL-3} = \frac{\la AB(134)\cap(1CD)\ra^2\la AB34\ra^3\la \hat{C}'134\ra\la 1234\ra^3}
{\la AB14\ra\la ABCD\ra\la CD34\ra\la CD\4'\A\ra\la \hat{C}'234\ra\la \hat{4} \hat{C}' \A 2\ra\la \hat{C}' \A 23\ra}
\ee\\

\begin{flalign}
&\vcenter{\hbox{\includegraphics[width=4.8cm]{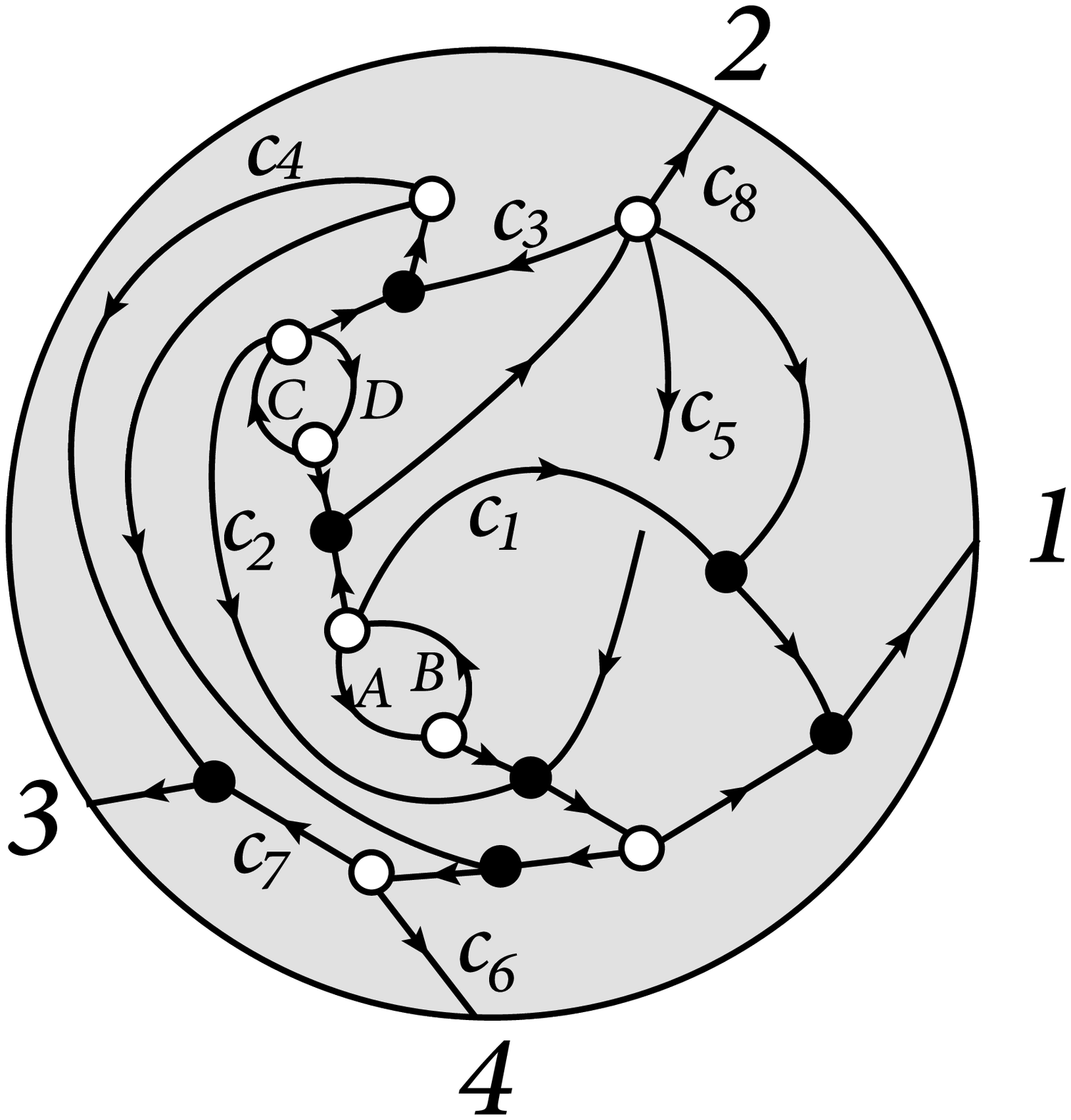}}}\hspace*{0.3cm}
\left( \begin{array}{cccc}
1 & 0 & -c_7 & -c_6  \\
1+c_1+c_5 & c_8 & c_3 c_4+c_3 c_7-c_5 c_7 & c_3c_6-c_5c_6 \\
c_2 & 0 & -c_4-c_7-c_2c_7 & -c_6-c_2c_6 \\
1+c_5 & c_8 & c_3 c_4+c_3 c_7+c_5 c_7 & c_3 c_6-c_5 c_6 \end{array} \right)&
\end{flalign}\\

\be
\text{FL-FL-4} = \frac{\la AB(134)\cap(1CD)\ra^2 \la \hat{C}' 134\ra^2\la \A \hat{4} 12\ra^3}
{\la AB14\ra\la ABCD\ra\la CD\A 3\ra\la CD34\ra\la \hat{C}'\hat{4} 12\ra\la \hat{C}'\hat{4} \A 1\ra\la \hat{C}' \hat{4}\A 2\ra\la \hat{C}' \A 12\ra\la CD\4' 1\ra}\nl
\ee\\

\begin{flalign}
&\vcenter{\hbox{\includegraphics[width=4.8cm]{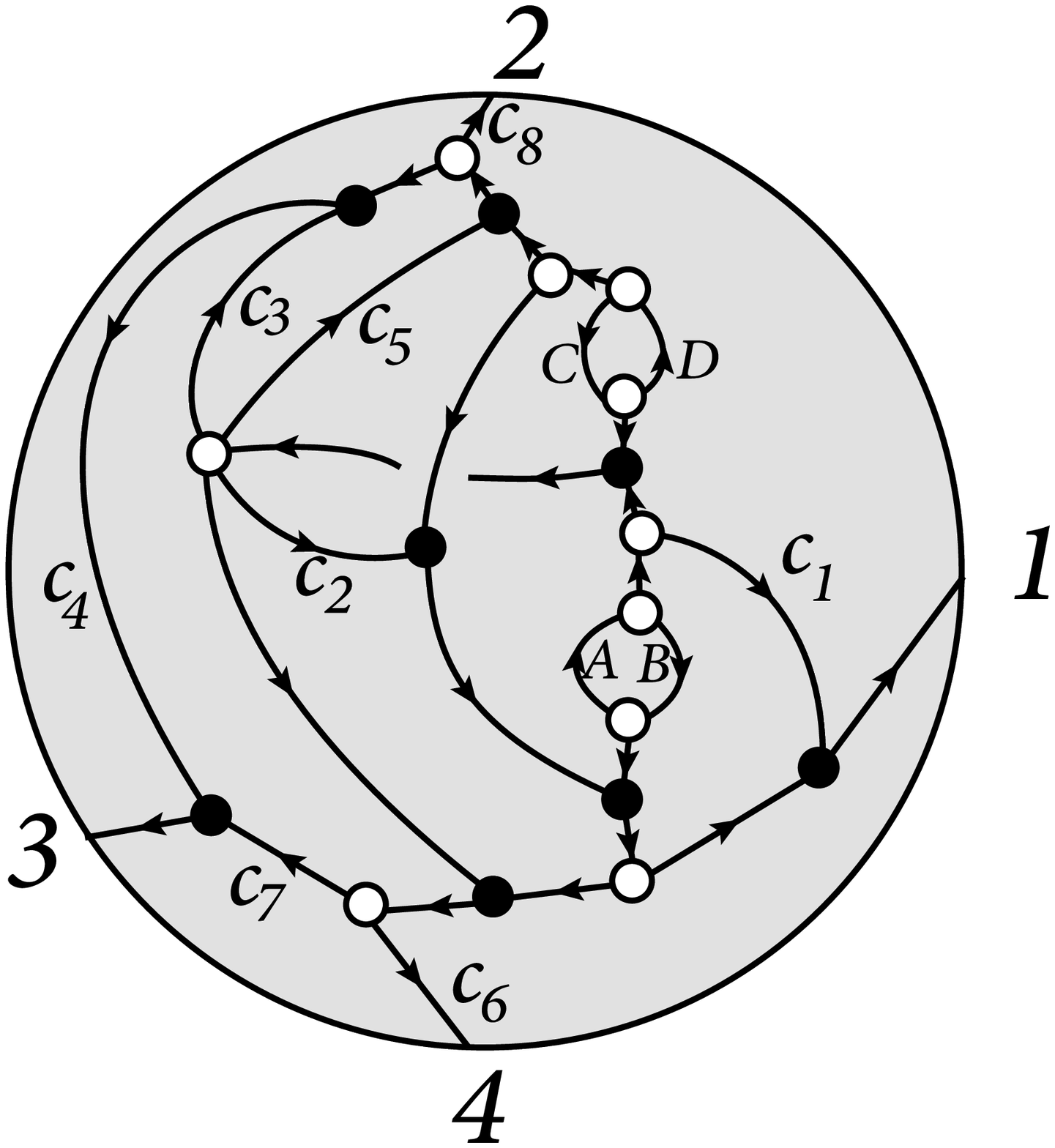}}}\hspace*{0.5cm}
\left( \begin{array}{cccc}
-c_1{+}c_2 & -c_5 c_8 & -c_3c_4{-}c_4c_5{-}c_7{-}c_2c_7 & -c_6{-}c_2c_6  \\
1 & 0 & -c_7 & -c_6 \\
c_2 & -c_5c_8 & -c_3c_4{-}c_4c_5{-}c_7{-}c_2c_7 & -c_6{-}c_2c_6 \\
1 & c_8 & c_4{-}c_7 & - c_6 \end{array} \right)&
\end{flalign}\\

\be
\text{FL-FL-5} = \frac{\la AB(134)\cap(1CD)\ra^2 \la AB34\ra\la \hat{C}' \A 23\ra\la 1234\ra^3}
{\la AB14\ra\la ABCD\ra\la CD\A2\ra\la CD23\ra\la \hat{C}' 134\ra\la \A \hat{3} \hat{C}' \4'\ra\la \hat{C}' 234\ra}
\ee\\

\begin{flalign}
&\vcenter{\hbox{\includegraphics[width=5cm]{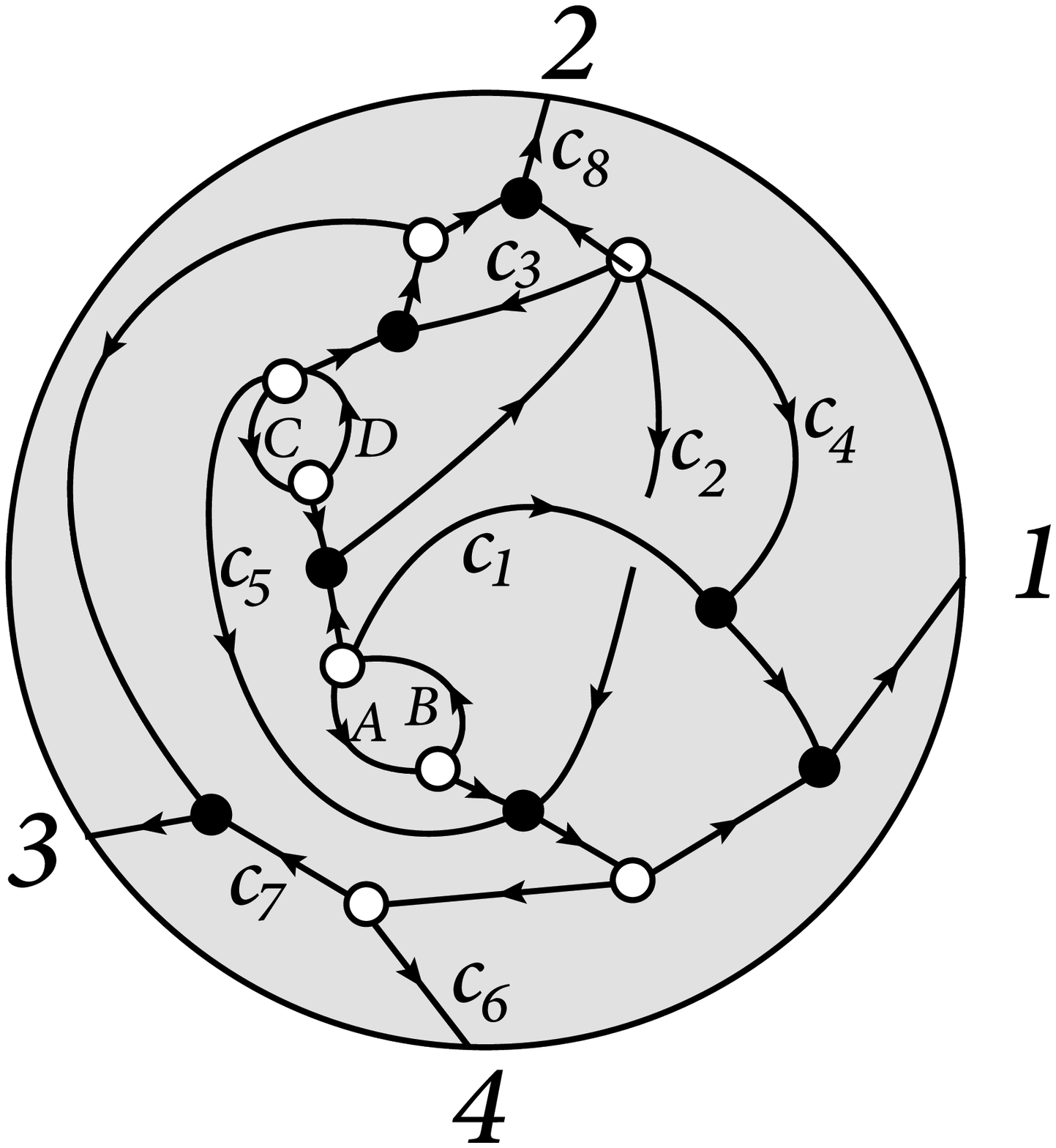}}}\hspace*{1cm}
\left( \begin{array}{cccc}
1 & 0 & -c_7 & -c_6 \\
c_1+c_2+c_4 & c_8+c_3 c_8 & c_3-c_2c_7 & -c_2c_6  \\
c_2+c_4 & c_8+c_3c_8 & c_3-c_2c_7 & -c_2c_6 \\
-c_5 & c_8 & 1+c_5c_7 & c_5c_6 \end{array} \right)&
\end{flalign}\\

\be
\text{FL-FL-6} = \frac{\la AB(134)\cap(1CD)\ra^2\la \hat{C}' \A 23\ra\la \A 123\ra^3}
{\la AB13\ra\la AB14\ra\la AB34\ra\la ABCD\ra\la CD23\ra\la CD3\A\ra\la \hat{C}'123\ra\la \hat{3} \hat{C}'\A 1\ra\la \hat{C}' \A 12\ra}\nl
\ee

The two-loop four-point integrand is the sum of the 8 terms, plus the other 8 terms obtained by $(AB\leftrightarrow CD)$. As one can check numerically, given a set of positive data and a random point inside the amplituhedron, it lies in one and only one of the 8 cells with the above $\mathcal{D}$ matrices, except for points on the boundary of cells. 

\subsubsection{Two-loop MHV and the cells}

Having studied the four-point case, now we present the complete result for any two-loop MHV amplitudes. As discussed above, we have to include FL-FAC and FL-FL-FAC terms, and by solving the recursion with B term, the two type of terms correspond to two types of ``Kermits":
\be
Y_{n,0}^{(2)}=
\frac 1 2\left(\sum_{i,j,k,l} (K^{(2), a}_{i,j; k,l}+ K^{(2), a}_{i,l; j,k}) +\sum_{i,j,k} (K^{(2),b}_{i,j; k}+K^{(2),b}_{j,i; k})+ ( A B \leftrightarrow  C D)\right),
\ee
where the ranges of summation are generically $2\leq i\leq j\leq k\leq l\leq n$ and $2\leq i<j\leq k\leq n$ respectively. We suppressed the momentum-twistor diagrams for these Kermits, since they are of the same form as those for four points. We will concentrate on giving the explicit momentum-twistor expressions and the positive $\mathcal{D}$ matrices for the cells corresponding to these Kermit terms. 

It is straightforward to count the number of terms (prior to symmetrization) for type-a and type-b Kermits, in terms of binomial numbers: $N^{(a)}_n=2\times C_{n,4}=n(n{-}1)(n{-}2)(n{-}3)/12$, and $N^{(b)}_n=2\times (C_{n,3}{-}1)=n(n{-}1)(n{-}2)/3-2$. For example, for $n=4$, what we had above correspond to $K^{(2),a}_{2,3;3,4}$, $K^{(2),a}_{3,4; 3,4}$, and $K^{(2),b}_{2,3;4}$, $K^{(2),b}_{2,4;4}$, $K^{(2),b}_{3,4;4}$, $K^{(2),b}_{3,2;4}$, $K^{(2),b}_{4,2;4}$, $K^{(2),b}_{4,3;4}$. It can be easily checked that the general expressions and $\mathcal{D}$ matrices reduce to the four-point results above.   

The first type of Kermit comes from the forward-limit of factorization terms at one loop, thus is given by the product of two one-loop Kermit, which takes the form in eq.~(\ref{1loopkerm}). 
For the Kermit with loop variable $CD$, its ``reference twistor" $Z$ is defined as $\hat{A}_l \equiv (A B)\cap (1, l{-}1, l)$, and the generic form for $K^{(2),a}_{i,j; k,l}$ and $K^{(2),a}_{i,l; j,k}$ are:
\be\label{2loopKa}
&&K^{(2), a}_{i,j; k,l}=d^4\ell_{AB}\,d^4\ell_{CD}\,K_{A B}(1,k{-}1, k; 1,l{-}1, l)\,K_{C D}(\hat{A}_l, i{-}1, i; \hat{A}_l,j{-}1,j) \,,\nl &&K^{(2), a}_{i,l;j,k}=d^4\ell_{AB}\,d^4\ell_{CD}\,K_{A B}(1,i{-}1, i; 1,l{-}1,l)\,K_{C D}(\hat{A}_l, j{-}1, j; \hat{A}_l,k{-}1,k)\,,
\ee

There are boundary terms when $j=k$ for $K_{i,j; k,l}$, as well as when $i=j$ and/or $k=l$ for $K_{i,l;j,k}$,  which are given by the following replacement:
\be
j=k: \, j\to \hat{j}'\equiv (j{-}1\,j)\cap (1, A, B)\,; \quad i=j:\, j{-}1 \to \hat{j}', \,\quad k=l:\, k\to \hat{k}'
\,. \nl
\ee

By choosing positive coordinates in the diagrams, one can make the associated $\mathcal{D}$ matrices positive. Here we list explicitly, for type-a Kermits,  these $\mathcal{D}$ matrices which represent the cells of the amplituhedron   In the summation, there are two sets of type-a Kermit terms. Let us look at the first sets, where we need to consider generic case, as well as ``boundary" cases. Note that we only display non-zero columns in the matrices below.  

For $1<i-1<i<j-1<j<k-1<k<l-1<l$:
\be
\mathcal{D}_{i,j;k,l}^{(2),a} = 
\bordermatrix{
&1     &i-1     &i     &j-1	     &j      &k-1       &k     &l-1       &l\cr
&1     &0       &0     &0        &0      &0         &0     &-c_{l-1}  &-c_l\cr
&1     &0       &0     &0        &0      &c_{k-1}   &c_k   &0         &0\cr
&1     &c_{i-1} &c_i   &0        &0    &0         &0     &-c_{l-1}  &-c_l\cr
&-1    &0       &0     &c_{j-1}  &c_j     &0         &0     &c_{l-1}   &c_l
}
\ee

For $1<i-1<i<j-1<j=k<l-1<l$:
\be
\mathcal{D}_{i,j;k,l}^{(2),a} = 
\bordermatrix{
&1     &i-1     &i       &k-1       &k     &l-1       &l\cr
&1     &0       &0       &0         &0     &-c_{l-1}  &-c_l\cr
&1     &0       &0       &c_{k-1}   &c_k   &0         &0\cr
&1     &c_{i-1} &c_i     &0         &0     &-c_{l-1}  &-c_l\cr
&-1    &0       &0       &c_{j-1}c_{k-1}+c_j c_{k-1}         &c_j c_k     &c_{l-1}   &c_l
}
\ee

All other boundary cases can be obtained from one of the matrices above by "merging". For example, suppose we look at the second matrix where $j=k$, and suppose in addition that we want $i=k-1$. To get this, we merge columns $i$ and $k-1$ into one column by adding them component-wise. The resulting matrix will still be positive.\\

Similarly we give the matrices for the second sets, together with boundary cases. For $1<i-1<i<j-1<j<k-1<k<l-1<l$:

\be
\mathcal{D}_{i,l;j,k}^{(2),a} = 
\bordermatrix{
&1     &i-1     &i     &j-1	     &j      &k-1       &k     &l-1       &l\cr
&1     &0       &0     &0        &0      &0         &0     &-c_{l-1}  &-c_l\cr
&1     &c_{i-1} &c_i   &0        &0      &0         &0     &0         &0\cr
&1     &0       &0     &c_{j-1}  &c_j    &0         &0     &-c_{l-1}  &-c_l\cr
&-1    &0       &0     &0        &0      &c_{k-1}   &c_k   &c_{l-1}   &c_l
}
\ee

For $1<i-1<i<j-1<j<k-1<k=l$:
\be
\mathcal{D}_{i,l;j,k}^{(2),a} = 
\bordermatrix{
&1     &i-1     &i     &j-1	     &j      &k-1       &k     \cr
&1     &0       &0     &0        &0      &-c_{l-1}  &-c_l     \cr
&1     &c_{i-1} &c_i   &0        &0      &0         &0     \cr
&1     &0       &0     &c_{j-1}  &c_j    &-c_{l-1}  &-c_l     \cr
&-1    &0       &0     &0        &0      &c_{l-1}c_{k-1}+c_{l-1}c_k+c_{l-1}   &c_l c_k+c_l   
}
\ee

For $1<i-1<i=j<k-1<k<l-1<l$:
\be
\mathcal{D}_{i,l;j,k}^{(2),a} = 
\bordermatrix{
&1     &i-1     &i     &k-1	     &k      &l-1       &l     \cr
&1     &0       &0     &0        &0      &-c_{l-1}  &-c_l     \cr
&1     &c_{i-1} &c_i   &0        &0      &0         &0     \cr
&1     &c_{j-1}c_{i-1}       &c_j c_i+c_{j-1}c_i     &0  &0    &-c_{l-1}  &-c_l     \cr
&-1    &0        &0      &c_{k-1}   &c_k   &c_{l-1}   &c_l
}
\ee

For $1<i-1<i=j<k-1<k=l$:
\be
\mathcal{D}_{i,l;j,k}^{(2),a} = 
\bordermatrix{
&1     &i-1     &i     &l-1       &l     \cr
&1     &0       &0     &-c_{l-1}  &-c_l     \cr
&1     &c_{i-1} &c_i   &0         &0     \cr
&1     &c_{j-1}c_{i-1} &c_j c_i+c_{j-1}c_i   &-c_{l-1}  &-c_l     \cr
&-1    &0       &0     &c_{l-1}c_{k-1}+c_{l-1}c_k+c_{l-1}   &c_l c_k+c_l   
}
\ee
Again all other boundary cases can be obtained from one of the matrices above by "merging".\\

The second type of Kermit comes from the forward-limit of forward-limit terms at one-loop, and has the generic form: \be\label{2loopKb}
K^{(2),b}_{i,j; k}&=&\frac{\l C D d^2 C\r\l C D d^2 D\r\l \hat{A}_k \hat{C}'\,i{-}1\,i\r^2\l \hat{A}_k\,\hat{i}_k\,j{-}1\,j\r^3}{\l C D \hat{A}_k\, i{-}1\r\l C D \hat{A}_k\, i\r\l C D\,i{-}1\, i\r\l \hat{A}_k \hat{C}'\,\hat{i}_k\, j{-}1\r\l \hat{A}_k \hat{C}'\,\hat{i}_k\, j\r\l \hat{A}_k \hat{C}'\,j{-}1\, j\r\l \hat{C}'\,\hat{i}_k\,j{-}1\, j\r}\nl
&&\times \frac{ \l A B d^2 A\r\l A B d^2 B\r \l A B (1, k{-}1, k)\cap (1,C, D)\r^2 }{\l A B C D\r\l A B\,1\,k{-}1\r\l A B\,1\,k\r\l A B\,k{-}1\,k\r}\,,
\ee
where we have defined $\hat{i}_k\equiv (i{-}1\, i)\cap (\hat{A}_k, C, D)$, $\hat{A}_k\equiv (A B)\cap (1, k{-}1, k)$ and also $\hat{C}'=(C D)\cap (1, A, B)$. Note that in the four-point case above, we have omitted the subscript of $k=4$ of $\hat{A}_k$. The only boundary case is when $j=k$, and we replace $j\to \hat{j}'\equiv (j{-}1 j)\cap (1 A B)$. There is a similar formula for $K^{(2),b}_{j,i,k}$ with $i\leftrightarrow j$.

Now we turn to the corresponding cells, and list the positive $\mathcal{D}$ matrices.  For the first set of the type-b Kermits, we consider generic and boundary cases.  For $1<i-1<i<j-1<j<k-1<k$:

\be
\mathcal{D}_{i,j;k}^{(2),b} = 
\bordermatrix{
&1     &i-1     &i     &j-1	     &j      &k-1       &k     \cr
&1     &0       &0     &0        &0      &-c_{k-1}  &-c_{k} \cr
&-a+c_1&c_{i-1} &c_i   &c_{j-1}  &c_j    &a c_{k-1} &a c_k \cr
&1     &c_{i-1} &c_i   &0        &0      &-c_{k-1}  &-c_k   \cr
&-a    &c_{i-1} &c_i   &c_{j-1}  &c_j    &a c_{k-1} &a c_k  \cr
}
\ee

For $1<i-1<i<j-1<j=k$:
\be
\mathcal{D}_{i,j;k}^{(2),b} = 
\bordermatrix{
&1     &i-1     &i      &k-1       &k     \cr
&1     &0       &0      &-c_{k-1}  &-c_{k} \cr
&-a+c_1&c_{i-1} &c_i    &c_{j-1}+a c_{k-1}+c_j c_{k-1} &a c_k+c_j c_k \cr
&1     &c_{i-1} &c_i    &-c_{k-1}  &-c_k   \cr
&-a    &c_{i-1} &c_i    &c_{j-1}+a c_{k-1}+c_j c_{k-1} &a c_k+c_j c_k  \cr
}
\ee

For the second set, there are also generic and boundary cases. For $1<i-1<i<j-1<j<k-1<k$:

\be
\mathcal{D}_{j,i;k}^{(2),b} = 
\bordermatrix{
&1     &i-1     &i     &j-1	     &j      &k-1       &k     \cr
&-a-c_1&-c_{i-1} &-c_i   &-c_{j-1}  &-c_j    &a c_{k-1} &a c_k \cr
&1     &0       &0     &0        &0      &-c_{k-1}  &-c_{k} \cr
&-a    &-c_{i-1} &-c_i   &-c_{j-1}  &-c_j    &a c_{k-1} &a c_k  \cr
&1     &0 &0           &-c_{j-1}   &-c_{j}  &-c_{k-1} &-c_k  \cr
}
\ee

For $1<i-1<i<j-1<j=k$:
\be
\mathcal{D}_{j,i;k}^{(2),b} = 
\bordermatrix{
&1     &i-1     &i      &k-1       &k     \cr
&-a-c_1&-c_{i-1}&-c_i   &ac_{k-1}-c_j c_{k-1}-c_{j-1}c_{k-1} &ac_k-c_j c_k \cr
&1     &0       &0      &-c_{k-1}  &-c_{k} \cr
&-a    &-c_{i-1} &-c_i    &ac_{k-1}-c_j c_{k-1}-c_{j-1}c_{k-1} &a c_k-c_j c_k  \cr
&1     &0 &0    &-c_{k-1}-c_j c_{k-1}-c_{j-1}c_{k-1}  &-c_k-c_j c_k   \cr
}
\ee

It is not obvious but one can check that both $K^a$ and $K^b$ are given by the $8$ $d\log$'s of the variables in the corresponding $\mathcal{D}$ matrix.

\subsubsection{General two-loop amplitudes}

The last result from our diagrams we will present is the BCFW representation of all two-loop amplitudes, which as we discussed, includes B and FAC terms in the same form as the tree-level case, and FL-FAC and FL-FL-FAC terms which we write down now. 

The FL-FAC term is again identical to the one-loop case.  We sum over left L and right R sub-amplitudes for which $k_L + k_R = k$ (prior to symmetrization). 
\be
\sum_{i=3}^{n-1}K_{1;i,n}^{(1),AB}Y_L(\hat{i}',i,i{+}1,...,n{-}1,\hat{n}',\hat{A}_n)Y_R(\hat{i}',\hat{A}_n,1,2,...,i{-}1)\,.
\ee

For the FL-FL-FAC term, we sum over all left L and right R sub-amplitudes for which $k_L {+} k_R = k{+}1$, and to be as explicit as possible we discuss three cases. 

In the special case where $k_L = 0$, we factorize the right sub-amplitude R into RL and RR (by doing the BCFW shift $D$ to $(B1)\cap(ACD)$ on R) and sum over all RL and RR for which $k_{RL}{ +} k_{RR} = k_R{-}1 = k$. This gives us the term
\be
\sum_{2\le j < i \le \hat{n}'}K_{i,j;n}^{(2),b}\; Y_{RL}(\hat{j}_n,j,j{+}1,...,i{-}1,\hat{i}_n)Y_{RR}(\hat{j}_n,\hat{C}'',\hat{A}_n,1,...,j{-}1)
\ee
where $\hat{j}_n = (j{-}1,j)\cap(\hat{A}_n CD)$, $\hat{i}_n = (i{-}1,i)\cap(\hat{A}_n CD)$, $\hat{C}''=(\hat{C}' \hat{A}_n)\cap (\hat{i}_n,j{-}1,j)$.

In another case where $k_R = 0$, we factorize L into LL and LR (by shifting $(AC)\cap(D,i{-}1,i)$ on L) and sum over contributions for which $k_{LL}{+}k_{LR} = k_L {-} 1 = k$. This gives us the term
\be
\sum_{2\le i<j\le \hat{n}'} K_{i,j;n}^{(2),b} Y_{LL}(\hat{j}_n,j,j{+}1,...,n{-}1,\hat{n}',\A_n,\hat{C}'')Y_{LR}(\hat{i}_n,i,i{+}1,...,j{-}1,\hat{j}_n)
\ee
where $\hat{n}'=(n{-}1,n)\cap(1AB)$.

In the final case where $k_R,k_L>0$, we factorize both L and R using the same shifts as above and sum over all LL LR RL RR for which $k_{LL}{+}k_{LR} = k_L {-} 1$ and $k_{RL} {+} k_{RR} = k_R{-}1$. Formally, we have L $=$ L-B $+$ L-FAC and R $=$ R-B $+$ R-FAC so that L $\times$ R = L-FAC $\times$ R $+$ L-B $\times$ R-FAC $+$ L-B $\times$ R-B, which is equivalent to
\be
\sum_{2\le i<j\le \hat{n}'} K_{i,j;n}^{(2),b} Y_{LL}(\hat{j}_n,j,j{+}1,...,n{-}1,\hat{n}',\A_n,\hat{C}'')Y_{LR}(\hat{i}_n,i,i{+}1,...,j{-}1,\hat{j}_n)Y_R(\hat{i}_n,\hat{C}',\hat{A}_n,1,...,i{-}1)\nl
+\sum_{2\le j < i \le \hat{n}'}K_{i,j;n}^{(2),b}\; Y_L(\hat{i}_n,i,i{+}1,...,n{-}1,\hat{n}',\hat{A}_n)Y_{RL}(\hat{j}_n,j,j{+}1,...,i{-}1,\hat{i}_n)Y_{RR}(\hat{j}_n,\hat{C}'',\hat{A}_n,1,...,j{-}1)\nl
\ee
The first term is the L-FAC $\times$ R. The second term is the L-B $\times$ R-FAC. In principle we should also include L-B $\times$ R-B, but this does not contribute since there are not enough $C$ and $D$ fermionic delta functions in the forward limit.

The three cases can be combined into a single formula, which gives the FL-FL-FAC term for any two-loop N$^k$MHV integrand prior to symmetrization
\be 
\text{FL-FL-FAC}=\hspace{12cm}\nl
\sum_{\substack{2\le i<j\le \hat{n}'\\k_1{+}k_2{+}k_3=k}} K_{i,j;n}^{(2),b} Y_1(\hat{j}_n,j,j{+}1,...,n{-}1,\hat{n}',\A_n,\hat{C}'')Y_2(\hat{i}_n,i,i{+}1,...,j{-}1,\hat{j}_n)Y_3(\hat{i}_n,\hat{C}',\hat{A}_n,1,...,i{-}1)\nl
+\sum_{\substack{2\le j < i \le \hat{n}'\\k_1{+}k_2{+}k_3=k}}K_{i,j;n}^{(2),b}\; Y_1(\hat{i}_n,i,i{+}1,...,n{-}1,\hat{n}',\hat{A}_n)Y_2(\hat{j}_n,j,j{+}1,...,i{-}1,\hat{i}_n)Y_3(\hat{j}_n,\hat{C}'',\hat{A}_n,1,...,j{-}1)\nl
\ee
where in each of the two terms we sum over all sub-amplitudes $Y_{1,2,3}$ for which $k_1+k_2+k_3=k$. We have thus derived an {\it algebraic} recursion relation for all two-loop amplitudes. Note that at higher loops, FL-FL-FAC term is not enough, since we need additional contributions coming from forward limit terms of L and R which we did not include.

To complete this section, we note that there is a small subtlety regarding the fermionic components of $\hat{C}'=(C,D)\cap(1,A,B)$. Naively, one might expand $\eta_{\hat{C}'}=\eta_{C}\left<D1AB\right>-\eta_{D}\left<C1AB\right>$, but this is wrong because $\eta_C$ and $\eta_D$ have already been integrated out. The other expansion of $\eta_{\hat{C}'}$ in terms of $\eta_1,\eta_A,\eta_B$ also does not make sense since $\eta_A$ and $\eta_B$ have been integrated out. By tracing back to the forward limit calculation, we find that there is a fermionic delta function whose support gives
\be
\eta_{\hat{C}'}&=&\eta_{C}\left<D1AB\right>-\eta_{D}\left<C1AB\right>\\
&=&-\frac{\eta_{j{-}1}\la j,\hat{A}_n,\hat{C}',\hat{i}_n\ra+\eta_j\la \hat{A}_n ,\hat{C}',\hat{i}_n,j{-}1\ra+\eta_{\hat{A}_n}\la \hat{C}',\hat{i}_n,j{-}1,j\ra+\eta_{\hat{i}_n}\la j{-}1,j,\hat{A}_n,\hat{C}'\ra}{\la \hat{i}_n,j{-}1,j,\hat{A}_n\ra}\nl
\ee
This gets rid of any dependence on fermionic components of loop momentum super-twistors. Of course, the bosonic components of $\hat{C}'$ are defined in the usual way.

\section{Conclusions and Discussions}

In this paper, we propose ``momentum-twistor diagrams" as a new diagrammatic representation of all-loop amplitudes/Wilson loops in planar $\mathcal{N}=4$ SYM. Formulated as on-shell diagrams in momentum twistor space, the diagrams manifest the dual superconformal symmetry, and naturally give factorization and forward-limit contributions for the all-loop integrand. Compared to the original on-shell diagrams in momentum space, such contributions are represented in a very different fashion, as we discussed in detail; and it is much more efficient to determine and evaluate the new diagrams in practice, which we have demonstrated through various calculations including all two-loop amplitudes. Similar to the fact that on-shell diagrams can be associated with factorizations and forward-limits of amplitudes in momentum space, the momentum-twistor diagrams are naturally related to such properties of the Wilson loops, which are discussed in details in~\cite{simonWL}.

Our diagrams are closely related to other types of interesting diagrams. One can take the geometric dual of our diagrams,  or ``region diagrams", where each trivalent black/white vertex is associated with a black/white triangle, and internal/external legs correspond to edges of the triangles, with the external $n$ edges forming the polygon contour of the Wilson loops. At least at tree-level, the result is the polygon triangulated by black and white triangles, and the merge of trivalent vertices correspond to merge of triangles into polygon regions, such that we can make the region diagram ``bi-partite". It remains an open question to see how the loop diagrams are represented in this dual picture. We can also modify the on-shell diagrams slightly to obtain ``CSW-like diagrams", which evaluate to the MHV-vertex expansion in momentum-twistor space~\cite{Bullimore:2010pj}. The reference twistor $*$ can be represented by an additional leg from a fixed puncture in the diagram, and the basic R-invariant or ``propagator", $[*,a,b,c,d]$, is given by the diagram (\ref{Rinv}) with leg $e$ replaced by the leg $*$. We can similarly write down all tree-level higher-$k$ diagrams, which gives the product of such R-invariants with expected shifted twistors~\cite{Bullimore:2010pj}, and it may be interesting to work out such diagrams at loop-level as well. This construction may provide new connections between on-shell diagrams and the Feynman diagrams (in an axial gauge)~\cite{Bullimore:2010pj, masonskinner}, for Wilson loops in momentum twistor space.

Perhaps more interestingly, the new diagrams are very useful for studying the geometry of the amplituhedron for the all-loop integrands. Given any such diagram, one can extract $C,D$ matrices associated with it and certain positive coordinates, which in turn gives a cell of the full amplituhedron. There is strong evidence that the collection of all such diagrams for any loop integrand exactly gives a triangulation of the corresponding amplituhedron, which is very non-trivial to obtain otherwise. Other known representation of the same integrand, such as MHV-vertex expansion~\cite{Bullimore:2010pj}, or local integral expansion~\cite{ArkaniHamed:2010gh}, generally do not correspond to any triangulations of the amplituhedron. Given the valuable data provided by the diagrammatic formulations, it would be extremely interesting to understand how they triangulate the amplituhedron. In fact, beyond one-loop MHV case, it seems any BCFW representation does not give the most natural triangulation~\cite{Arkani-Hamed:2013kca}, and geometrically it is intriguing why these diagrams do not overlap with each other inside the amplituhedron, already for the 8 cells of the two-loop four-point case. We expect that our diagrams to be vital for exploring the rich structures of the amplituhedron to all loops. 

In this aspect, it would be highly desirable to completely understand how to obtain $C,D$-matrices, to arbitrary loop order, from iterating forward-limits, or equivalently, how to write the forward limit as an matrix operation acting on the $C,D$-matrices. We suspect that there may be interesting combinatoric structures behind momentum-twistor diagrams at loop level, generalizing the permutations for the reduced diagrams in both original and momentum-twistor spaces. Relatedly, since each diagram by itself is Yangian invariant, it would be very interesting to study individual diagrams, as opposed to the full integrand, at loop level. Of course they are cells of a single object, the amplituhedron, but without understanding its geometry completely, can we say anything about the origin of individual diagrams? It would be highly desirable, beyond trees and one-loop MHV, to associate the diagrams with residues of some generalized Grassmannian integrals which depend on external and loop variables. In particular, such an understanding can shed more light on how to relate different BCFW representations to each other, as well as to other forms such as the local form~\cite{ArkaniHamed:2010gh}. We plan to address these questions in the future. 

Furthermore, given this diagrammatic representation for the all-loop integrand, it is natural to ask if one can understand some of their fascinating properties better, such as the Yangian-invariance as positive diffeomorphism, the positivity of the rational integrand, and structures of multi-loop amplitudes integrated from the integrand. One novel and unexpected feature of the diagrams is the appearance of non-planarity at loop level. Generally speaking, every forward limit operation gives rise to one degree of non-planarity. Although the diagrammatic origin of this feature is clear (\textit{i.e.} doing the GL(2) integral by reconnecting lines on the diagram), its physical interpretation is still unclear. Are there some combinatorics behind the non-planarity? Are there extensions of our diagrams which are related to non-planar amplitudes, or even some non-planar extension of the amplituhedron?

It is natural to ask if these diagrams can be extended to other Yangian invariant theories like ABJM. However, doing so requires a proper construction of momentum twistor variables in three dimensions. Given the similarities of planar integrands of $\mathcal{N}=4$ SYM and of $\mathcal{N}=1,2$ SYM in momentum-twistor space, one may also ask if the momentum-twistor diagrams can be applied to those theories as well. Finally, could the on-shell diagrams in momentum-twistor space be generalized to study off-shell quantities, such as correlation functions of planar SYM? After all, in the light-like limits, correlators of half-BPS operators become equivalent to the amplitudes/Wilson loops~\cite{Alday:2010zy,Eden:2010zz}.

\section*{Acknowledgements}

We are grateful to N.~Arkani-Hamed for numerous discussions and encouragements, and to T.~Bargheer, J.~Bourjaily, S.~Caron-Huot, Y.~t.~Huang, J.~Trnka for helpful comments. Y. B would also like to thank T.~Lam for helpful discussions. The work of S. H is supported by Zurich Financial Services Membership and the Ambrose Monell Foundation. The work of Y. B is supported by Natural Sciences and Engineering Research Council of Canada PGS M and the Department of Physics, Princeton University.




\end{document}